\pdfoutput=1
 \documentclass[12pt,a4paper]{article}
\usepackage[hmargin=2.5cm,vmargin=3cm]{geometry}
\usepackage[english]{babel}
\usepackage[utf8]{inputenc}

\usepackage[bbgreekl]{mathbbol}
\usepackage{dsfont} 
\usepackage{amsmath,amssymb,amsfonts,amsthm}
\usepackage{mathrsfs}
\usepackage{caption}
\usepackage{subcaption}

\usepackage{bbold}
\DeclareSymbolFontAlphabet{\mathbb}{AMSb}
\DeclareSymbolFontAlphabet{\mathbbl}{bbold}

\numberwithin{equation}{section}
\usepackage[bottom]{footmisc}
\usepackage{cite}

\usepackage{xcolor}
\usepackage[
  bookmarksnumbered,
  linktocpage=true,
  colorlinks=true,
  citecolor={green!50!black}
]{hyperref}

\usepackage{hyperref}

\usepackage{tikz}
\usetikzlibrary{arrows,calc}

\DeclareMathAlphabet{\mathpzc}{OT1}{pzc}{m}{it}

\usepackage{verbatim}

\newcommand{\ZZ}{\mathbb{Z}}

\newcommand{\RR}{\mathbb{R}}
\newcommand{\CC}{\mathbb{C}}
\newcommand{\dd}{\mathrm{d}}
\newcommand{\ex}{\mathrm{e}}

\newcommand{\spindle}{\mathbbl{\Sigma}}

\newcommand{\Morb}{\mathbb{M}}

\newcommand{\flp}{\mathfrak{p}}

\newcommand{\evol}{\mathds{V}} 
\newcommand{\Fext}{F} 

\newcommand{\be}{\begin{equation}} 
\newcommand{\ee}{\end{equation}}
\newcommand{\bea}{\begin{equation} \begin{aligned}} 
\newcommand{\eea}{\end{aligned} \end{equation}}

\newcommand{\fan}{d}
\newcommand{\fix}{n}

\begin{document}

\begin{titlepage}

\vskip 2.5cm

\begin{center}

\today

\vskip 2.5cm

{\Large \bf Equivariant localization and holography}

\vskip 2cm
{Dario Martelli$^{\mathrm{a,b}}$ and Alberto Zaffaroni$^{\mathrm{c,d}}$}

\vskip 0.8cm

${}^{\mathrm{a}}$\textit{Dipartimento di Matematica ``Giuseppe Peano'', Universit\`a di Torino,\\
Via Carlo Alberto 10, 10123 Torino, Italy}

\vskip 0.2cm

${}^{\mathrm{b}}$\textit{INFN, Sezione di Torino,  Via Pietro Giuria 1, 10125 Torino, Italy}

\vskip 0.2cm

${}^{\mathrm{c}}$\textit{Dipartimento di Fisica, Universit\`a di Milano-Bicocca\\
Piazza della Scienza 3, 20126 Milano, Italy}

\vskip 0.2cm

${}^{\mathrm{d}}$\textit{INFN, sezione di Milano-Bicocca, Piazza della Scienza 3, 20126 Milano, Italy}

\end{center}

\vskip 2 cm

\begin{abstract}
\noindent  
We discuss the theory of equivariant localization focussing on applications relevant for   holography. We consider 
geometries comprising  compact and non-compact  toric orbifolds, as well as more general non-compact  toric 
Calabi-Yau singularities. A key object in our constructions is the \emph{equivariant volume}, for which we describe two methods of evaluation: the Berline-Vergne fixed-point formula and the Molien-Weyl formula, supplemented by the Jeffrey-Kirwan prescription.
 We present two applications in supersymmetric field theories. Firstly, we describe a method for 
 integrating the anomaly polynomial of SCFTs on compact toric orbifolds.
 Secondly, we discuss equivariant orbifold indices that are expected to play a key role in the computation
 of supersymmetric partition functions. In the context of  supergravity, we propose that the equivariant volume can be used to characterise universally the geometry of a 
  large class of supersymmetric solutions. As an illustration, we employ equivariant localization  to prove the factorization in gravitational blocks of various supergravity free energies, recovering previous results as well as obtaining generalizations.

\end{abstract}

\end{titlepage}

\tableofcontents

\section{Introduction}

The theory of equivariant co-homology is a rich mathematical subject, with numerous applications in diverse areas of geometry and mathematical physics. 
A remarkable consequence
 of this theory is that quite generally, on manifolds endowed with a Hamiltonian action of a Lie group, there exist \emph{localization formulas} that express 
certain integrals in terms of contributions arising from the fixed point sets of the group action, thus simplifying enormously their evaluation. 
 A prime example of this feature is the classic result of  Duistermaat-Heckmann \cite{Duistermaat}.
More generally, such integrals involve  equivariant characteristic classes and 
 may arise in the evaluation of equivariant 
 indices of transversally elliptic operators  \cite{berline1982classes,ATIYAH19841}. See for instance the review articles \cite{VERGNE2,vergne2006applications},
 or \cite{Pestun:2016qko} for a discussion perhaps more accessible to physicists. In this paper we will need  an extension of these results to the setup of 
 orbifolds \cite{Vergne1996EQUIVARIANTIF}, including spindles and other orbifolds with singularities in co-dimension less than two, that are not so often considered
 in the physics literature but have  recently received attention in the context of holography.

A very general framework in which the ideas of equivariant localization  are realized  is that of symplectic toric geometry. In this case one starts with a symplectic manifold $(M_{2m},\omega)$ in
 dimension $2m$, equipped with an effective Hamiltonian action of the real torus $\mathbb{T}^m$.  The image of the associated moment maps  
is a convex rational  polytope ${\cal P}\subset \mathbb{Q}^m$ \cite{Delzant}.
Together with the   angular coordinates on  the torus, $\phi_i\sim \phi_i+2 \pi$, the moment maps $y_i$ can be used as  ``symplectic coordinates'', endowing the manifold with 
 a natural coordinate system $(y_i,\phi_i)$ in which  any metric can then be written in a canonical form  in terms of  combinatorial data of the polytope  \cite{Guillemin1994KaehlerSO}.
In the toric setting the fixed points sets of the $\mathbb{T}^m$ action are always isolated singularities, so that  the localization formulas  take the form of sums over fixed points. 
The applications of the  localization theorems in this context range from the quantization of symplectic manifolds to algorithms for computing the volumes of  polytopes and counting integral points. See e. g. 
 \cite{Barvinok,10.2307/41062}. The extensions to symplectic toric orbifolds was discussed in  \cite{Lerman:1995aaa}
and that  to non-compact toric cones in   \cite{Moraes1997MomentMO}.

In theoretical physics  equivariant localization came to the fore 
in the work of Nekrasov as a technique for calculating  partition functions counting instantons in supersymmetric field theories 
\cite{nekrasov2002seibergwitten}. See e.g. \cite{Tachikawa:2014dja} for a review. Applications of toric geometry motivated 
by the AdS/CFT correspondence were first discussed in \cite{Martelli:2004wu} and further developed in \cite{Martelli:2005tp}, where the volume functional of toric Sasakian manifolds was shown to be extremized by Sasaki-Einstein metrics.
The extension to the more general equivariant setting and the relation to fixed point theorems and the  index-character of the associated Calabi-Yau cone singularities was explored in \cite{Martelli:2006yb}.
From the viewpoint of holography, these results can be used to  compute  the volume and other properties of Sasaki-Einstein manifolds, without explicit knowledge of the metric, from which in turn one can  infer properties of the dual field theories. 
Subsequent  developments  in geometry include results about:  toric Sasaki-Einstein metrics  \cite{Gauntlett:2006vf,Futaki:2006cc}, 
  K\"ahler-Einstein metrics \cite{Li:2016qwl}, extremal Sasaki metrics \cite{vanCoevering:2012dg,Boyer:2016yae},  conformally K\"ahler 
Einstein-Maxwell metrics \cite{futaki2017volume}.

Following on the steps of \cite{Martelli:2005tp,Martelli:2006yb} we take holography as a motivation for uncovering  precise mathematical relationships between geometry and supersymmetric field theories. In particular,  we provide
new evidence that toric geometry and equivariant localization are well-suited mathematical frameworks for this purpose.
In the context of supergravity our aim is to develop a universal approach to study the geometry underlying supersymmetric solutions  based on  \emph{extremization problems} analogous to those formulated in \cite{Martelli:2005tp}  and 
 \cite{Couzens:2018wnk}. We will argue that the functionals to be extremized  can be calculated in each case  using the technique 
of equivariant localization, generalizing the results appeared in \cite{Martelli:2006yb,Gabella:2009ni,Gabella:2011sg,BenettiGenolini:2019jdz}.\footnote{For  applications of localization to the calculation of the supergravity path integral see for example \cite{Dabholkar:2010uh,Dabholkar:2014wpa,deWit:2018dix,Jeon:2018kec}.}
We will consider, in particular, the novel supergravity constructions featuring compact spaces with conical singularities, like spindles and other orbifolds \cite{Ferrero:2020laf,Ferrero:2020twa,Hosseini:2021fge,Boido:2021szx,Ferrero:2021wvk,Ferrero:2021ovq,Couzens:2021rlk,Faedo:2021nub,Ferrero:2021etw,Giri:2021xta,Couzens:2021cpk,Suh:2022olh,Arav:2022lzo,Couzens:2022yiv,Suh:2022pkg,Suh:2023xse,Amariti:2023mpg,Cheung:2022ilc,Faedo:2022rqx,Couzens:2022lvg,FFM2}. 
These solutions  imply that we should work in the orbifold setting from the outset and indeed from our results we will obtain a direct localization proof 
of the factorization in gravitational blocks \cite{Hosseini:2019iad} 
recently discussed in the literature \cite{Hosseini:2021fge,Faedo:2021nub,Faedo:2022rqx,Boido:2022mbe}.

In the context of supersymmetric field theory our motivation is that of extending to the orbifold realm some tools that are well-established for field theories compactified on smooth  manifolds.
Specifically, working with orbifold equivariant co-homology, 
we wish to put on  a sounder mathematical footing the technique of  integration of the  anomaly polynomials of even-dimensional SCFTs compactified on orbifolds. 
Furthermore, following
\cite{Inglese:2023wky}, where a new  index for three-dimensional ${\cal N}=2$  theories was computed exploiting the spindle index-character, we present a discussion of 
 equivariant orbifold indices, that we expect to be key building blocks for  computing supersymmetric partition functions of SQFTs on orbifolds, extending 
 Pestun's \cite{Pestun:2007rz} approach to supersymmetric localization.  When the underlying space is (the resolution of) 
 a non-compact Calabi-Yau singularity the 
 same objects have been employed previously to compute Hilbert series of 
  the moduli spaces of supersymmetric field theories \cite{Benvenuti:2006qr,Martelli:2006vh,Butti:2006au}. We expect that new insights can be gained from the study 
  of equivariant indices and their relation to the equivariant volume.   In this paper, however, we restrict our localization techniques to the evaluation of classical geometrical objects.
We now summarize the structure of this  paper.

In section \ref{sec:equivol} we recall the symplectic geometry description of compact toric orbifolds, following 
\cite{Guillemin1994KaehlerSO,Lerman:1995aaa,abreu2000kahler,Abreu:2001to}. 
We introduce  the \emph{equivariant volume}  of symplectic toric orbifolds, that is our main object of interest. We discuss two alternative methods for 
evaluating this, namely the Berline-Vergne fixed point formula and the Molien-Weyl integral formula, that exploits the presentations of the toric orbifolds as symplectic quotients, based on the Lerman and Tolman's generalization \cite{Lerman:1995aaa} of Delzant's construction \cite{Delzant}. Although the topics covered in this section are mainly not original, we consider  an
 extension to  non-compact toric orbifolds, 
 which leads to localization formulas for odd-dimensional orbifolds, arising as the base of complex cones thus generalizing the results of \cite{Martelli:2006yb,Boyer:2017jwa}  for the Sasakian volume. 
 Aspects of the  equivariant volume of non-compact  symplectic toric manifolds were recently studied in  \cite{Nekrasov:2021ked,Cassia:2022lfj}.
In this section we also discuss the \emph{equivariant orbifold index} of the Dolbeault complex, twisted by a holomorphic line orbi-bundle. In the toric setting, 
 the geometric interpretation of this object is that of counting integer lattice points inside a convex integral polytope (or polyhedral cone, in the non-compact case), corresponding to sections of the line bundle.
  It  can then be thought of as the quantum  (or K-theoretical) version of the equivariant volume, which is  recovered in  a limit in which the lattice spacing goes to zero.
  Besides this close relationship with the equivariant volume,  equivariant orbifold indices are expected to provide fundamental building blocks for the construction of supersymmetric partition functions defined on orbifolds 
   \cite{Inglese:2023wky}, with applications to black hole microstrate counting. We therefore present some   examples of these indices in appendix \ref{app:character}.

In section \ref{sec:examples} we discuss in detail a number of examples of toric orbifolds and their associated equivariant volumes. We start with the complex projective line $\mathbb{WP}^1_{[n_1,n_2]}$, also known as the spindle.
This is the simplest compact toric orbifold, which arises in  complex dimension one. It  has a prominent role in several  recent supergravity constructions corresponding to 
various wrapped  branes 
\cite{Ferrero:2020laf,Ferrero:2020twa,Hosseini:2021fge,Boido:2021szx,Ferrero:2021wvk,Ferrero:2021ovq,Couzens:2021rlk,Faedo:2021nub,Ferrero:2021etw,Giri:2021xta,Couzens:2021cpk,Suh:2022olh,Arav:2022lzo,Couzens:2022yiv,Suh:2022pkg,Suh:2023xse,Amariti:2023mpg}. 
Moving to complex dimension two, we consider generic compact toric orbifolds, described by a rational convex polyhedron 
with an arbitrary number of vertices. We then consider in more detail triangles, namely weighted projective spaces $\mathbb{WP}^2_{[n_1,n_2,n_3]}$ and their quotients by discrete groups, as well as quadrilaterals. Examples of supergravity solutions comprising Hermitian (non-K\"ahler) metrics on quadrilateral toric orbifolds are discussed in \cite{Cheung:2022ilc,Faedo:2022rqx,Couzens:2022lvg,FFM2}.
As a warm-up for section \ref{sec:CY},   we also describe the pecularities of the  non-compact case in a few explicit examples.

Section \ref{anomal:sec} concerns the application of equivariant localization to the calculation of the anomaly polynomial of 4d and 6d SCFTs compactified on various orbifolds. Using this approach we prove 
the localized form of the anomaly polynomial for theories compactified on the spindle  \cite{Ferrero:2020laf,Boido:2021szx,Hosseini:2021fge,Ferrero:2021wvk},
 as well as on general four dimensional toric orbifolds, for which one example was considered in \cite{Cheung:2022ilc}. Our approach leads to a 
 uniform derivation for SCFTs compactified on different orbifolds and explains the localized form of the integrated anomaly polynomials 
 that was previous observed in examples. 

Finally, in section \ref{sec:CY} we discuss the application of equivariant localization in the context of supergravity. 
Firstly, we show that the results of \cite{Martelli:2006yb} on the localized form of the Sasakian volume are immediately recovered from the equivariant volume of the associated non-compact Calabi-Yau singularity.
Furthermore, we prove that also the \emph{master volume} introduced in \cite{Gauntlett:2018dpc}, in the context of GK geometry \cite{Gauntlett:2007ts}, can be extracted from the equivariant volume and as a corollary we derive  the factorization in gravitational blocks of the  supergravity free energies for D3 and M5 branes wrapped on the spindle, reproducing  the results of \cite{Boido:2022mbe} in the toric case.  We then propose analogous constructions for other branes wrapped on the spindle from which, in each case, we can extract the localized form of the gravitational blocks, which was anticipated in  \cite{Faedo:2021nub}.
This leads us to propose that equivariant localization is the  
common thread of the geometry of supersymmetric supergravity solutions, at least in setups with a holographic interpretation. 
In particular, we believe that the equivariant volume should be a key object for  setting up  extremal problems characterising supersymmetric geometries, in different supergravity theories. 

In section \ref{sec:discuss} we discuss our findings and indicate some directions for future work. The paper contains three appendices with some complementary material.

\section{The equivariant volume}
\label{sec:equivol}

\subsection{The symplectic geometry description of toric orbifolds}
\label{sec:geo}

In this section we review the geometry of   symplectic toric orbifolds following the general formalism developed in \cite{Guillemin1994KaehlerSO,abreu2000kahler,Abreu:2001to,Lerman:1995aaa}. 
We emphasise that although it appears that we are relying on symplectic geometry, we will be interested in computing topological quantities that do not depend on the existence of an integrable symplectic structures and 
 it should be possible to reformulate our computations entirely in terms of complex geometry. 
 In particular, our results are applicable also to situations in which one is interested in metrics that are not compatible with a symplectic structure, such as the  Hermitian (non-K\"ahler)  metrics 
on toric orbifolds constructed in \cite{Cheung:2022ilc,Faedo:2022rqx,Couzens:2022lvg,FFM2}.

We consider a symplectic toric orbifold $(\Morb_{2m},\omega)$ in dimension $2m$ equipped with an effective Hamiltonian action of the real torus $\mathbb{T}^m=\mathbb{R}^m/2\pi \mathbb{Z}^m$. We introduce symplectic coordinates $(y_i,\phi_i)$ with $i=1,\ldots,m$ where $\phi_i$ are angular coordinates on the the torus, $\phi_i\sim \phi_i+2 \pi$. In terms of these coordinates the symplectic form is given by\footnote{We adopt the Einstein summation convention for the indices (two repeated indices imply the sum). However, for sake of clarity, in some formulas we will  write  explicitly the  sums over the indices.}
\be \omega = \dd y_i \wedge \dd \phi_i\, .\ee
By a generalization of Delzant's theorem \cite{Delzant}, compact symplectic toric orbifolds are classified by labelled polytopes which are the image of $\Morb_{2m}$ under the moment maps $y_i$ associated with the toric action \cite{Lerman:1995aaa}. The image of $\Morb_{2m}$  is simply obtained by forgetting the angular coordinates $\phi_i$ and it is a rational simple convex polytope $\mathcal{P}$ in $\mathbb{R}^m$. We can describe it by introducing a set of linear functions\footnote{We use intercheangably the notation $y$ and $y_i$ to denote a point in $\RR^m$.}
\be\label{linear}  l_{a} (y) = y_i v_i^a - \lambda_{a}\, ,\qquad a=1, \ldots, \fan \, ,\ee
where $v^a$ are vectors in $\mathbb{R}^m$. 
The convex polytope   is the subset of $\mathbb{R}^m$ defined by
\be\label{poly} \mathcal{P}=\{ y\in \mathbb{R}^m \, : \, l_{a} (y) \ge 0\} \, \qquad a=1, \ldots, \fan \, .\ee
 The linear equations \be  \qquad  l_a(y)=0 \ee define the  facets $\mathcal{F}_a$ of the polytope. We denote with $\fan$ the number of facets of $\mathcal{P}$. The condition that $\mathcal{P}$ be rational is equivalent to the fact  that the vectors $v^a$ have integer entries. The condition that $\mathcal{P}$ be simple requires that each vertex $p$ lies at the intersection of precisely $m$ facets 
\be l_{a_1}(p) =l_{a_2}(p) = \dots = l_{a_m}(p) = 0 \, ,\ee
and the corresponding $m$ vectors $\{v^{a_1},\dots,v^{a_m}\}$ are a basis for  $\mathbb{R}^m$. The polytope comes equipped with a {\it label} for each facet, a positive integer $n_a$ such that the structure group of every point in the inverse image of $\mathcal{F}_a$ is $\mathbb{Z}_{n_a}$. 

This construction exhibits $\Morb_{2m}$ as a torus fibration over the polytope $\mathcal{P}$ and generalises the familiar construction for toric varieties. As in the latter
 case, the torus fibration is non-degenerate in the interior of $\mathcal{P}$. A particular one-cycle in  $\mathbb{T}^m$, determined by the vector $v^a$, collapses at the facet $\mathcal{F}_a$. Thus each facet $\mathcal{F}_a$ defines a symplectic subspace of $\Morb_{2m}$ of real codimension two. In the complex case this becomes a divisor\footnote{They are called ramification divisors, while 
 $\hat D_a = n_a D_a$ are called branched divisors, with multiplicity (or ramification index) $n_a$.} which we will denote $D_a$.  Similarly, at the intersection of $q$ facets, more cycles in $\mathbb{T}^m$ degenerates and we have a symplectic subspace of $\Morb_{2m}$ of real codimension $2q$. The intersection of $m$ facets is a vertex of the polytope and it corresponds to a fixed point of the $\mathbb{T}^m$ action. We denote with $\fix$ the number of such fixed points. In particular, we can give an alternative definition of the polytope $\mathcal{P}$ as  the convex hull of the images of the  fixed points of $\Morb_{2m}$.

In the context of toric varieties, the vectors $v^a$ define what is called the fan of $\Morb_{2m}$. We will use the same terminology but the reader should be aware that we are dealing with a generalization of the concept which allows for more general type of orbifold singularities. For toric varieties, the vectors $v^a$ are primitive, while in the case of generic symplectic toric orbifolds they are not. We can always define for each $v^a$ a primitive vector $\hat v^a$ and a positive integer $n_a$ such that $v^a= n_a \hat v^a$.\footnote{Notice that we are using opposite convections with respect to \cite{Faedo:2022rqx}!}
The integer $n_a$ is precisely the label of the facet $\mathcal{F}_a$ defined above. In particular, each symplectic divisor $D_a$  has a local $\mathbb{Z}_{n_a}$ singularity. This cannot happen for toric varieties which are normal and have no singularities of complex co-dimension less than two. Notice also that the local singularity at the fixed point given by the intersection of the $m$ facets $\mathcal{F}_{a_i}$, with $i=1\, \ldots m$, has order $d=|\det (v^{a_1},\ldots, v^{a_m})|$. $\Morb_{2m}$ is a smooth manifold 
if and only if all the labels $n_a$ are one and
for each vertex $|\det (v^{a_1},\ldots, v^{a_m})|=1$.
 
We want to equip $\Morb_{2m}$ with a compatible complex structure. Any $\mathbb{T}^m$-invariant K\"ahler metric on $\Morb_{2m}$ is of the form \cite{abreu2000kahler}
\be \dd s^2 = G_{ij}(y) \dd y_i \dd y_j + G^{ij}(y) \dd \phi_i \dd \phi_j\, ,\ee
where $G_{ij}$ is determined by a symplectic potential $G(y)$ as
\be G_{ij}=\frac{\partial^2 G}{\partial y_i \partial y_j} \, \ee 
and $G^{ij}=(G^{-1})_{ij}$ is the inverse matrix. Holomorphic coordinates are given by \be\label{complex} z_i =x_i + i \phi_i\, , \qquad x_i =\frac{\partial G}{\partial y_i}  \, .\ee 
The existence of a symplectic potential is equivalent to the integrability  of the complex structure. 

The canonical metric on $\Morb_{2m}$ is given by  \cite{Guillemin1994KaehlerSO} 
\be\label{symppot} G(y)=\frac12 \sum_{a=1}^\fan l_{a} \log l_{a} \, , \ee
so that
\be\label{metric} G_{ij}=\frac{\partial^2 G}{\partial y_i \partial y_j} = \frac12 \sum_{a=1}^\fan \frac{v_i^{a}  v_j^{a}}{l_{a}} \, . \ee

Notice that $G_{ij}$ has poles at the facets but the metric is smooth up to orbifold singularities. The inverse matrix $G^{ij}$ has rank $m-1$ at the facets indicating that  a one cycle in $\mathbb{T}^m$ is degenerate. With a change of coordinates we obtain a smooth orbifold metric on $\mathcal{F}_a$. The structure of poles of \eqref{symppot} is compatible with the degeneration of $\mathbb{T}^m$ at the faces of $\mathcal{P}$ and it is precisely designed to obtain an orbifold metric on $\Morb_{2m}$.  The most general K\"ahler metric on $\Morb_{2m}$ is discussed in  \cite{Abreu:2001to} and it is obtained by replacing the symplectic potential $G(y)$ for the canonical metric with $G(y)+h(y)$, where $h(y)$ is smooth on the whole $\mathcal{P}$. The topological quantities we will discuss in this paper do not depend on the metric and we can safely set $h(y)=0$. In our applications to holography we will encounter metrics that are not K\"ahler and not even symplectic, but the underlying spaces are in fact symplectic toric orbifolds and we can therefore use the symplectic coordinates and the canonical metric to compute topological quantities  that ultimately will not depend on the metric.

Each facet $\mathcal{F}_a$ defines a $\mathbb{T}^m$-invariant divisor $D_a$ and an associated line bundle $L_a$. These objects will be important for our construction so we will spend some time discussing their properties. The first Chern class of $L_a$ has been explicitly computed in  \cite{Guillemin1994KaehlerSO,Abreu:2001to}\footnote{We believe that there is a minus sign error in equation (6.17) in \cite{Guillemin1994KaehlerSO}  and we have therefore changed the sign here.}
\be c_1(L_a) = - \frac{i}{2\pi} \left [ \partial \bar\partial \log l_a \right ] \, ,\ee
where with $[\alpha]$ we denote the co-homology class of the differential form $\alpha$. An explicit representative is 
 given by\footnote{Writing  as in \eqref{complex}, $z_i=x_i+ i \phi_i$ where $x_i =\frac{\partial G}{\partial y_i}$, we have $\partial = \frac{1}{2}\sum_{i=1}^2\left( \dd x_i+ i \dd \phi_i \right) \left( \frac{\partial}{\partial x_i} - i \frac{\partial}{\partial \phi_i}\right)$ and, therefore, for a torus invariant function $f$ that only depends on $y$  
\begin{equation}
\partial \bar \partial f = -\frac{i}{2}\frac{\partial^2 f}{\partial x_i \partial x_j} \dd x_i \wedge \dd\phi_j = -\frac{i}{2}\dd \left(\frac{\partial f}{ \partial x_j}\dd \phi_j\right ) =-\frac{i}{2}\dd \left(G^{kj} \frac{\partial f}{ \partial y_k}\dd \phi_j\right ) \, .
\end{equation}
With some abuse of language, we will often denote with $c_1(L)$ an explicit representative of the corresponding co-homology class.
}
\be\label{CC} c_1(L_a) = \dd \left ( \mu_a^i\dd \phi_i \right ) \, ,\ee
where
\begin{equation}
\label{momaps}
\mu_a^i =\mu_a^i  (y) =  -\frac{1}{4\pi }\frac{G^{ij}v_j^a}{l_a}\, . 
\end{equation}
Notice that $\mu_a^i$ can be seen as a moment map for the torus action
\begin{equation}\label{Chern}
i_{\partial_{\phi_i}} c_1(L_a) = -\dd  \mu_a^i \, .
\end{equation}
From $G^{ij}G_{jk} =\delta^i_k$
we find
\be\label{ide} \sum_{a=1}^\fan \mu_{a}^i v_k^{a} = -\frac{\delta_k^i}{2\pi} \, .\ee

We can relate the Chern classes of the divisors to the K\"ahler form $\omega $  as follows. From
\be\label{yy}
G_{ij} y_j  = \frac12 \sum_a \frac{v_i^{a}  v_j^{a} y_j}{l_{a}} = \frac12 \sum_a \frac{v_i^{a}( l_{a}+\lambda_{a} ) }{l_{a}} = \frac12 \left ( \sum_a \frac{ \lambda_{a} v^{a}_i}{l_{a}} + \sum_a v_i^{a}\right ) \, ,
\ee
we obtain
\be y_i = - 2\pi  \sum_a \lambda_a \mu_i^a +\frac12\sum_a  G^{ij}  v_j^{a} \, ,\ee
so that
\be\label{omega} \omega = \dd y_i \wedge \dd \phi_i = - 2\pi  \sum_a \lambda_a \dd (\mu_i^a \dd \phi_i) +\frac12  \dd ( \sum_a G^{ij}  v_j^{a} \dd \phi_i) \, .\ee
Using the degeneration of $G^{ij}$  at the facets, one can check that $\sum_a G^{ij}  v_j^{a} \dd \phi_i$ is a well-defined one-form, so that the last term on the right hand side of \eqref{omega} is exact. Therefore we obtain the important relation
(see Theorem 6.3 in  \cite{Guillemin1994KaehlerSO}) 
\begin{equation}
\label{kform}
\frac{[\omega]}{2\pi} = -\sum_a \lambda_a c_1(L_a) \, .
\end{equation}
Notice that this equation holds only in co-homology. 
We see that the parameters $\lambda_a$ defining the shape of the polytopes through \eqref{poly} are parameterizing the K\"ahler moduli of the symplectic orbifold.

There are actually only $\fan -m$ independent line bundles. Indeed, using \eqref{CC} and \eqref{ide}, we find $m$ relations among the Chern classes
\be \label{chernlin} \sum_{a=1}^\fan v^a_i c_1(L_a) =0 \, ,\ee
that translate into the familiar $m$ linear relations among the divisors
\be \label{diveq} \sum_{a=1}^\fan v^a_i D_a =0 \, .\ee
These relations imply that, if $\fan$ is the number of vectors in the fan, there are $\fan -m$ independent ($2m-2$)-cycles in homology and therefore only $\fan -m$  independent K\"ahler parameters. The $d$ parameters $\lambda_a$ are associated with the $\mathbb{T}^m$-invariant divisors and provide an overparameterization of the 
K\"ahler moduli.
 
In this paper we will also consider non-compact cases, in particular  non-compact Calabi-Yau cones. In this case, the polytope is replaced by a rational convex polyhedral cone.
Calabi-Yau cones have typically singularities worse than orbifold. This happens when more than $m$ facets intersect at a vertex. To use the general formalism of this section we will perform a partial resolution to have only orbifold singularities. In applications to holography, we will also encounter polytopes and polyhedral cones
 that are \emph{not convex}. We will obtain results by performing a suitable extrapolation from the convex case.

\subsection{The definition of the equivariant volume} \label{sec:equiv}

In this section we define the equivariant volume. In order to simplify the exposition, in this and the next two sections we  assume  that $\Morb_{2m}$ is compact. 
We will discuss the subtleties and necessary modifications for the non-compact case in section \ref{sec:non-comp}.

We want to work equivariantly in the $\mathbb{T}^m$ action on $\Morb_{2m}$, which is generated by the $m$ vector fields $\partial_{\phi_i}$. Correspondingly we introduce  $m$ equivariant parameters $\epsilon_i$, with $i=1,\ldots m$ and the vector field $\xi= \epsilon_i \partial_{\phi_i}$.
We can introduce a Hamiltonian $H=\epsilon_i y_i$ for this vector field
\be i_\xi \omega =  - \dd H \, ,\ee
and define an equivariant K\"ahler form
\be \label{omegaeq} \omega^{\mathbb{T}} = \omega + 2\pi H = \dd ( y_i \dd \phi_i) + 2\pi \epsilon_i y_i \, .\ee
We similarly introduce equivariant Chern classes for the line bundles $L_a$
\be \label{Cherneq} c_1^{\mathbb{T}}(L_a) = c_1(L_a) + 2\pi \epsilon_i \mu_i^a = \dd ( \mu_i^a \dd \phi_i) + 2\pi \epsilon_i \mu_i^a\, \, ,\ee
using the moment maps $\mu^a$.  All these forms  are equivariantly closed
\be  (\dd + 2\pi  i_{\xi}) \omega^{\mathbb{T}}=0 \, ,\qquad (\dd + 2\pi  i_{\xi}) c_1^{\mathbb{T}}(L_a) =0 \, ,\ee
with respect to the $\mathbb{T}^m$ action. Notice that these expressions are formal linear combinations of forms of different degrees. 

Our main object of interest is the equivariant volume of $\Morb_{2m}$ \cite{Duistermaat},
\be\label{evolH} \evol(\lambda_a,\epsilon_i) = \frac{1}{(2 \pi)^m}\int_{\Morb_{2m}} \ex^{-H}  \frac{\omega^m}{m!} \, ,\ee
which is sometimes referred in the literature to as ``symplectic volume''
 or ``equivariant symplectic volume''.

We can  write the equivariant volume as
\be \label{bbb} \evol(\lambda_a,\epsilon_i) = (-1)^m \int_{\Morb_{2m}} \ex^{ -H- \frac{ \omega}{2\pi}} = (-1)^m\int_{\Morb_{2m}} \ex^{- \epsilon_i y_i - \frac{\omega}{2\pi}} \, ,\ee
since the integrand over $\Morb_{2m}$ receives contribution only from the piece of degree $2m$ in the expansion of the exponential. In terms of the equivariant forms  \eqref{omegaeq} and \eqref{Cherneq} we have
\be\label{evol} \evol(\lambda_a,\epsilon_i) = (-1)^m \int_{\Morb_{2m}} \ex^{-\frac{\omega^{\mathbb{T}}}{2\pi}} = (-1)^m \int_{\Morb_{2m}} \ex^{\sum_a \lambda_a c_1^{\mathbb{T}}(L_a)} \, . \ee
Notice also that the two integrands in \eqref{evol} are not equal but they differ by an equivariantly exact form by \eqref{kform} and the integrals are therefore equal. Notice that the last inequality strictly holds only if $\Morb_{2m}$ is compact. We will return to this point in section \ref{sec:non-comp}.

A geometrical interpretation of the equivariant volume of compact orbifolds is that it is the generating functional for the integrals of the equivariant Chern classes
\be \label{cohoexp}\evol(\lambda_a,\epsilon_i) = (-1)^m \sum_p \frac{1}{p!} \sum_{a_1,\ldots , a_p=1}^\fan \lambda_{a_1}\cdots \lambda_{a_p} \int_{\Morb_{2m}} c_1^{\mathbb{T}}(L_{a_1}) \cdots c_1^{\mathbb{T}}(L_{a_p}) \, .\ee
The equivariant intersection numbers
\be \label{bbbbb}D_{a_1\ldots  a_p} =\int_{\Morb_{2m}} c_1^{\mathbb{T}}(L_{a_1}) \cdots c_1^{\mathbb{T}}(L_{a_p})=(-1)^m \frac{\partial^p \evol(\lambda_a,\epsilon_i) } {\partial \lambda_{a_1}\cdots \partial \lambda_{a_p}} \Big |_{\lambda_a=0} \, \ee
are polynomials in $\epsilon_i$ and they are topological in nature. Notice that the integrand in \eqref{bbbbb} is a formal linear combination of forms of various degree and the integral
selects the piece of degree $2m$. In particular, the equivariant intersection numbers  are  different from zero for all $p\ge m$. 
 
The expression in \eqref{evolH} can be easily reduced to an integral over the polytope $\mathcal{P}$ by performing the angular integrations
\be\label{polyvol} \evol(\lambda_a,\epsilon_i) = \frac{1}{(2 \pi)^m}\int_{\Morb_{2m}} \ex^{-H}  \frac{\omega^m}{m!}  =\int_{\mathcal{P}} \ex^{-H}  \dd y_1\ldots \dd y_m= \int_{\mathcal{P}} \ex^{-\epsilon_i y_i}  \dd y_1\ldots \dd y_m \, .\ee
Thus we have another interpretation of the equivariant volume as the volume of the polytope $\mathcal{P}$ with measure $\ex^{-H}$. 
Integrals of this type and their relation to equivariant localization are discussed in \cite{10.2307/41062}. 

The equivariant volume satisfies some interesting identities. From the expression \eqref{evol}
\be \label{def2} \evol(\lambda_a,\epsilon_i) = (-1)^m \int_{\Morb_{2m}} \ex^{\sum_a \lambda_a \left (c_1(L_a) +2\pi \epsilon_i \mu_i^a\right )} \ee
and the identities \eqref{chernlin} and \eqref{ide} we find that the identity 
\be \label{gauge} \evol(\lambda_a +  \beta_j v^a_j, \epsilon_i) = \ex^{-\beta_j\epsilon_j} \evol(\lambda_a, \epsilon_i) \, \ee
holds, for arbitrary $\beta_i\in\mathbb{R}^m$. This formula reflects the fact that only $\fan -m$ parameters $\lambda_a$ are geometrically independent. A closely related and useful identity can be obtained by taking derivatives of \eqref{def2} and using again \eqref{chernlin} and \eqref{ide} 
\be \label{1der} \sum_{a=1}^\fan v^a_i \frac{\partial \evol}{\partial \lambda_a}   = -\epsilon_i \evol \, .\ee
Similarly we have
\begin{equation}
\label{nderid}
  \sum_{a_1,\dots, a_q=1}^\fan v^{a_1}_{i_1} \dots v^{a_q}_{i_q}  \frac{\partial^n {\evol}}{\partial \lambda_{a_1} \dots \partial \lambda_{a_q}} =(-1)^q\epsilon_{i_1} \dots  \epsilon_{i_q}  \evol   \, . 
\end{equation}

In the next sections we review two different methods to evaluate $\evol(\lambda_a,\epsilon_i)$, the fixed point and the Molien-Weyl formula. They are both discussed in the literature. Here we adapt them to our notations and we discuss the relations among them.

\subsection{The fixed point formula}\label{sec:fixed}

The equivariant volume can be computed using a fixed point formula. This can be obtained from \eqref{evolH} using the Duistermaat-Heckman theorem \cite{Duistermaat}, the localization formula for equivariant co-homology \cite{berline2003heat,meinrenken1996symplectic}, or a direct evaluation of \eqref{polyvol} \cite{10.2307/41062,Barvinok}. Here we use the localization approach and, for completeness, in Appendix 
\ref{fixptform4d_appendix}
we report an explicit proof for $m=2$.

Consider a symplectic toric orbifold $\Morb_{2m}$ with the properties discussed in section \ref{sec:geo}. The torus action $\mathbb{T}^m$ acts on $\Morb_{2m}$ with  $\fix$ isolated fixed points corresponding to the vertices of the polytope $\mathcal{P}$. Consider an equivariantly closed form $\alpha^{\mathbb{T}}$ on $\Morb_{2m}$. The  equivariant localization theorem for orbifolds applied to our situation states that 
\begin{equation}
\int_{\Morb_{2m}} \alpha^{\mathbb{T}} = \sum_{A=1}^\fix  \frac{\alpha^{\mathbb{T}}|_{y_A}}{d_A\,  e^{\mathbb{T}}|_{y_A}}\, ,
\label{toric-equiv-integral0}
\end{equation}
where   the sum is over the fixed points $y_A$ of  the $\mathbb{T}^m$ action, $e^{\mathbb{T}}$ is the equivariant Euler class of the tangent bundle at $y_A$ 
 and $d_A$ are the orders of the orbifold singularity
 at the fixed point $y_A$.  
In particular, 
applying this  theorem to the computation of the  equivariant volume \eqref{evol} gives
\be  \evol(\lambda_a,\epsilon_i) =(-1)^m \sum_{A=1}^\fix  \frac{ \ex^{- 
\frac{1}{2\pi}\omega^{\mathbb{T}} |_{y_A}
 } } {d_A\,  e^{\mathbb{T}}|_{y_A}} =(-1)^m \sum_{A=1}^\fix  \frac{ \ex^{ \sum_{a=1}^\fan \lambda_a c_1^{\mathbb{T}}(L_a) |_{y_A}  } } {d_A\, e^{\mathbb{T}}|_{y_A}}\, .\ee

To evaluate the localization formula we need to compute the restriction $c_1^{\mathbb{T}}(L_a) |_{y_A}$ of the equivariant Chern classes of $L_a$ to the $A$-th fixed point.
The fixed point $y_A$ is  defined by $m$ linear equations 
\be l_{a_{1}}=\ldots=l_{a_{m}}=0 \, \ee
for a choice of $m$ intersecting facets associated with the vectors $\{v^{a_1},\ldots ,v^{a_m}\}$. The order of the orbifold singularity is given by
\be d_A = |\det (v^{a_1}, \ldots ,v^{a_m})| \, .\ee
We will denote the $A$-th fixed point also with the $m$-tuples of indices $A=(a_1,\ldots, a_m)$ defining the vertex. 
In the neighbourhood of the fixed point $A=(a_1,\ldots, a_m)$
\be G_{ij} =  \frac12 \frac{v_i^{a_1}  v_j^{a_1}}{l_{a_1}} + \frac12  \frac{v_i^{a_2}  v_j^{a_2}}{l_{a_2}} +\ldots + \frac12 \frac{v_i^{a_m}  v_j^{a_m}}{l_{a_m}}+\ldots \, ,\ee
up to regular pieces. This can be inverted to give
\be\label{facetsGinv} G^{ij} = \frac{2}{d_A^2}\left( (u_{A}^{a_1})_i (u_{A}^{a_1})_j l_{a_1}  + (u_{A}^{a_2})_i (u_{A}^{a_2})_j l_{a_2}+\ldots  + (u_{A}^{a_m})_i (u_{A}^{a_m})_j l_{a_m}  \right ) +\ldots \, ,\ee
where the  vectors $u^{a_i}_A$ are constructed by taking the wedge product of the $m-1$ vectors $v^{a_j}$ with $j\ne i$. The sign ambiguity is determined by requiring $u^{a_i}_A \cdot v^{a_i} = d_A$. The vectors $u^{a_i}_A$  have integer entries and satisfy
\be u^{a_i}_A\cdot v^{a_j} = d_A \delta_{ij}\, \ee as well as
 \be \label{uv}v^{a_1}_i (u^{a_1}_{A})_j+v^{a_2}_i (u^{a_2}_{A})_j+\ldots +v^{a_m}_i (u^{a_m}_{A})_j=d_A \delta_{ij}\, .\ee
 They have a dual geometrical interpretation as the inward normal vectors to the facets of the cone $(v^{a_1}, \ldots ,v^{a_m})$ in the fan or, equivalently, as integer vectors along the $m$ edges of the polytope $\mathcal{P}$ meeting at $y_A$. Notice that the $m$-tuplet of vectors $u^{a_i}_A$ depends on the choice of vertex $A$.
 
The restriction of the moment maps \eqref{momaps} to the fixed points then gives
\be  \mu^{a}_i \Big |_{y_A} = \begin{cases} -\frac{1}{2\pi}  \frac{ u_i^{a}}{d_{A}}  \, \qquad &{\rm if} \,  a\in A  \\
0 \, \qquad &{\rm if} \, a\notin A \end{cases}
\ee
and, therefore
  \be\label{rest3}  c^{\mathbb{T}}_1(L_a) \Big |_{y_A} =   \begin{cases} -\frac{ \epsilon_i u_i^{a}}{d_{A}} \, \qquad &{\rm if}\,  a\in A  \\
0 \, \qquad &{\rm if} \,  a\notin A \end{cases}\, .\ee
Notice that the restriction of $c^{\mathbb{T}}_1(L_a)$ to a fixed point is not zero only if the point belongs to the facet $\mathcal{F}_a$.

Finally the tangent bundle at a fixed point splits as a direct sum of the $m$ line bundles $L_{a_i}$ associated with $A=(a_1,\ldots, a_m)$ 
and so we have \cite{Martelli:2006yb}
\be e^{\mathbb{T}}|_{y_A} =\prod_{a_i \in A} c^{\mathbb{T}}_1(L_{a_i}) |_{y_A} = (-1)^m \prod_{i=1}^m \frac{\epsilon \cdot u^{a_i}_A}{d_{A}} \, .\ee 

Now the fixed point formula gives
\be \label{fp} \evol(\lambda_a,\epsilon_i)=\sum_{A=(a_1,\ldots, a_m)} \frac{\ex^{- \sum_{i=1}^m \lambda_{a_i}  (\frac{\epsilon \cdot u^{a_i}_A}{d_{A}})}}{d_{A}\prod_{i=1}^m \frac{\epsilon \cdot u^{a_i}_A}{d_{A}} } \, ,\ee
where $A$ runs over the $n$ vertices of the polytope. This formula can be also obtained by computing the volume of the polytope $\mathcal{P}$, as discussed, for example, in \cite{10.2307/41062}.

We note on passing that the identities \eqref{gauge}, \eqref{1der} and \eqref{nderid} can be also derived from the fixed point formula.
For example, 
\be \label{gauge2} 
\evol(\lambda_a +  \sum_{j=1}^m \beta_j v^a_j, \epsilon_i) =  \!\!\!\!\!\! \sum_{A=(a_1,\ldots, a_m)} \!\!\!\!\!\! \ex^{-\sum_{i,j,k=1}^m\frac{\beta_j v^{a_i}_j (u^{a_i}_A)_k \epsilon_k}{d_A}    } \frac{\ex^{- \sum_{i=1}^m \lambda_{a_i}  (\frac{\epsilon \cdot u^{a_i}_A}{d_{A}})}}{d_{A}\prod_{i=1}^m \frac{\epsilon \cdot u^{a_i}_A}{d_{A}} } = \ex^{-\sum_{i=1}^m\beta_i\epsilon_i} \evol(\lambda_a, \epsilon_i) \, ,\ee
where we explicitly wrote all the sums to explain the derivation. In particular, we used \eqref{uv} to perform the intermediate summation over $i$. The proof of \eqref{1der} and \eqref{nderid} is similar.

\subsection{The Molien-Weyl formula}

The equivariant volume can be also computed using a Molien-Weyl integral formula following \cite{Nekrasov:2021ked,Cassia:2022lfj}.

The orbifold $\Morb_{2m}$ can be  realized as a symplectic quotient. Consider the space $\mathbb{C}^\fan$, where $\fan$ is the number of vectors in the fan, and
the subgroup $K$ of $\mathbb{T}^d$ of elements of the form
\be \left ( \ex^{2\pi i Q_1}\, , \ldots , \, \ex^{2\pi i Q_d}\, \right ) \in \mathbb{T}^d \, ,\ee
where
\be \label{charges} \sum_{a=1}^d Q_a v^a_i \in \mathbb{Z} \, ,\qquad i=1,\ldots, d \, . \ee
Then $\Morb_{2m}$ is the symplectic reduction \cite{Abreu:2001to,Lerman:1995aaa}
\be \Morb_{2m} = \mathbb{C}^d//K \, ,\ee
generalizing a familiar result in toric geometry. 
Notice that, in general, the group $K$ has a continuous part $U(1)^{\fan-m}$ 
\be Q_a = \sum_{A=1}^{\fan - m} Q_a^A \alpha_A \, ,\qquad \alpha_A\in\mathbb{R} \ee
which can be expressed in terms of the  GLSM charges $Q_a^A\in \mathbb{Z}$, with $A=1,\ldots \fan -m$, satisfying
\be \label{chargesGLSM} \sum_{a=1}^d Q_a^A v^a_i =0 \, , \ee
as well as a discrete part of more complicated characterization.

Consider first the case where there is no discrete part in the quotient and $\Morb_{2m}=\mathbb{C}^\fan//U(1)^{\fan-m}$. The authors of \cite{Nekrasov:2021ked,Cassia:2022lfj} derived a formula for the equivariant volume as a function of K\"ahler parameters $t_A$ of the symplectic reduction and the equivariant parameters of $\mathbb{C}^\fan$. This is 
a Molien-Weyl formula obtained by averaging the equivariant volume of  $ \mathbb{C}^\fan$ with respect to the $U(1)^{\fan-m}$ action. The formula reads
\be\label{molien}  \evol_{MW}(t_A,\bar\epsilon_a)= \int \prod_{A=1}^{\fan - m} \frac{\dd\phi_A}{2\pi i} \frac{\ex^{ \sum_A \phi_A t_A}}{\prod_{a=1}^\fan (\bar\epsilon_a +\sum_A \phi_A Q^A_a)} \, .\ee
A particular contour of integration should be used depending on the direction of the vector of K\"ahler parameters $t_A$. A  prescription based on the Jeffrey-Kirwan (JK) residue \cite{JeffreyKirwan} has been proposed in \cite{Nekrasov:2021ked,Cassia:2022lfj}. Formula \eqref{molien} should be also divided  by the order of the discrete group corresponding to the torsion part of $K$ when this is present. We will discuss subtleties related to the choice of contour and discrete groups when illustrating examples.

Notice that $\evol_{MW}$ depends  on $\fan - m$ K\"ahler parameters $t_A$, which is the right number of geometrically inequivalent parameters for $\Morb_{2m}$.
 It also depends  on $\fan$ equivariant parameters $\bar\epsilon_a$ associated to the $\mathbb{T}^\fan$ action on the ambient space, which is larger than the $m$ parameters that we expect for $\Morb_{2m}$. However, $\fan - m$ equivariant parameters can be eliminated by shifting the integration variables. Indeed,  up to an exponential factor,  the Molien-Weyl integral is invariant under the shift
\be\label{gaugeN} \evol_{MW}(t_A, \bar\epsilon_a + \sum_A Q^A_a \eta_A) =\ex^{-\sum_A \eta_A t_A}  \evol_{MW}(t_A,\bar\epsilon_a) \, ,\ee
for $\eta_A\in\mathbb{R}$.

We thus have two expressions for the equivariant volume, $\evol(\epsilon_i,\lambda_a)$, depending on $m$ equivariant parameters $\epsilon_i$ and $\fan$ K\"ahler parameters $\lambda_a$
but with the "gauge" invariance \eqref{gauge}, and $\evol_{MW}(\bar\epsilon_a,t_A)$, depending on $\fan$ equivariant parameters $\bar\epsilon_a$ and $\fan - m$ K\"ahler parameters $t_A$
but with the "gauge" invariance \eqref{gaugeN}. The two expressions must agree with a suitable mapping between parameters. The two set of parameters can be related by comparing gauge invariant quantities. Based on  several examples, we find 
evidence that this mapping is given by
\be\label{gaugeinv} t_A = -\sum_a \lambda_a Q_a^A \, , \qquad \epsilon_i =\sum_a v_i^a \bar\epsilon_a \, ,\ee
and  we conjecture that the two expressions for the equivariant volume  agree up to a multiplicative factor when expressed in terms of the over-redundant variables $(\lambda_a, \bar\epsilon_a)$, namely\footnote{It would be interesting 
to give a formal proof of (\ref{conj0}). This   would require a careful analysis of all the residues in (\ref{molien}), which is a hard problem, and we do not attempt such a proof here.}
 \be\label{conj0} \evol_{MW}(t_A=  -\sum_a \lambda_a Q_a^A, \bar\epsilon_a) = \ex^{\sum_a \lambda_a \bar\epsilon_a} \evol(\lambda_a, \epsilon_i =\sum_a v_i^a \bar\epsilon_a) \, .\ee

As a consistency check of this formula we can apply to both sides the operator $\sum_a v^a_i \frac{\partial }{\partial \lambda_a}$.
Since the GLSM charges satify \eqref{chargesGLSM} and $\evol_{MW}$ is a function of $t_A = -\sum_a \lambda_a Q_a^A$, we obtain zero on the left hand side. 
On the right hand side we obtain
\be  \ex^{\sum_a \lambda_a \bar \epsilon_a} \left ( \sum_a v^a_i \bar \epsilon_a \evol +  \sum_a v^a_i \frac{\partial \evol}{\partial \lambda_a} \right ) \, , \ee
which is also zero because of the identity \eqref{1der} and the identification \eqref{gaugeinv}. 

\subsection{Remarks on the non-compact case}\label{sec:non-comp}

Most of the previous results hold also for non-compact orbifolds, but with some important differences. We still define the equivariant volume as
\be\label{evolHnc} \evol(\lambda_a,\epsilon_i) = \frac{1}{(2 \pi)^m}\int_{\Morb_{2m}} \ex^{-H}  \frac{\omega^m}{m!}  = (-1)^m \int_{\Morb_{2m}} \ex^{-\frac{\omega^{\mathbb{T}}}{2\pi}} \, , \ee  
which we can also write as an integral over a polyhedron $\mathcal{P}$
\be\label{polyvol2} \evol(\lambda_a,\epsilon_i) =  \int_{\mathcal{P}} \ex^{-\epsilon_i y_i}  \dd y_1\ldots \dd y_m \, .\ee
Since $\Morb_{2m}$ and $\mathcal{P}$ are non-compact, we need to assume that the integrals are convergent. This happens for the examples we will be interested in, where $\mathcal{P}$ is asymptotically a cone. In this case the Hamiltonian $H=\epsilon_i y_i$ acts as a convergence factor, at least if the vector $\epsilon$ lies inside the cone.

The fixed point formula and the Molien formula hold under general conditions also for non-compact orbifolds, and we can even use them as an operative definition for the equivariant volume. For example, the fixed point formula only assumes that there are isolated fixed points in $\mathcal{P}$ and that there are no contributions from infinity.  Notice that the identities \eqref{gauge}, \eqref{1der} and \eqref{nderid}, which follow from the fixed point formula, are still valid.

However, the co-homological interpretation \eqref{cohoexp} fails. In particular, the two integrals in \eqref{evol} are not equal in general. The second term in \eqref{omega}
\be\label{omega0} \omega = \dd y_i \wedge \dd \phi_i = - 2\pi  \sum_a \lambda_a c_1(L_a) +\frac12  \dd ( \sum_a G^{ij}  v_j^{a} \dd \phi_i) \, ,\ee
 is still exact, but it cannot be ignored in a integral over non-compact orbifolds.\footnote{Notice that this does not affect the argument for the fixed point formula. The contribution of $\omega^{\mathbb{T}}$ to the fixed point is still $\label{omega00} \omega^{\mathbb{T}}  = - 2\pi  \sum_a \lambda_a c^{\mathbb{T}}_1(L_a)$,
since the extra term $\frac12  \sum_a G^{ij}  v_j^{a}$ vanishes at the fixed point by \eqref{facetsGinv}.}
 As a result, the equivariant volume can be still formally expanded in power series of $\lambda_a$ but the coefficients cannot be straightforwardly interpreted as generalized intersection numbers as in \eqref{cohoexp}. In particular, while in the compact case, the power series of $\lambda_a$ starts at order $m$ and the coefficients are polynomials in $\epsilon_i$, in the non-compact setting the power series has all coefficients different from zero and these are in general rational function of $\epsilon_i$, as it follows from the fixed point formula.    We will discussed explicit examples of non-compact orbifolds in sections \ref{sec:nnoncompact} and \ref{sec:CY}.

\subsection{Relation to the equivariant orbifold index}\label{sec:eqindex}

In this section we briefly discuss how the equivariant volume arises as a limit of an index character, following the logic in \cite{Martelli:2006yb,Nekrasov:2021ked,Cassia:2022lfj}. Examples of calculations of characters are provided in the appendix \ref{app:character} as this is not the main theme of this paper.

The equivariant index for the twisted Dolbeault complex on $\Morb_{2m}$ is defined by  
\be \mathds{Z}(q_i, \Lambda_a) = \sum_{p=0}^m (-1)^p {\rm Tr} \{ q | H^{(0,p)}(\Morb_{2m}, O(\sum_a \Lambda_a D_a))\} \,\ee
where $q\in \mathbb{T}^m$ is an element of the torus  and  $\Lambda_a\in \mathbb{Z}$, $a=1,\ldots,\fan$ are integers specifying a choice of line bundle. The trace is taken on the induced action on the co-homology. The index can be computed with the Hirzebruch-Riemann-Roch theorem. We start with the smooth case that is considerably simpler and we
take all the labels to be one, the vectors $v_a$ to be primitive and  all the orders $d_A=1$. The fixed point formula for the equivariant index is
\be \label{a1} \mathds{Z}(\Lambda_a,q_i)=\sum_{A=(a_1,\ldots, a_m)} \frac{ \underline{q}^{ - \sum_{i=1}^m \Lambda_{a_i}\underline{u}^{a_i}_A} } 
{ \prod_{i=1}^m (1-\underline{q}^{ \underline{u}^{a_i}_A}) } 
 \, ,\ee
 where the symbol $\underline{q}^{\underline{n}}$ is an abbreviation for $q_1^{n_1}\ldots q_m^{n_m}$. The geometrical interpretation of the equivariant index is generically to count integer points
 \be \mathds{Z}(\Lambda_a,q_i) = \sum_{\underline{m}\in \Delta(\Lambda_a)} \underline{q}^{\underline{m}} \, ,\ee
 in the polytope 
 \be \label{pol} \Delta(\Lambda_a) =\{ m\cdot v^a \ge - \Lambda_a \} \, ,\qquad m\in \mathbb{Z}^m \, ,\ee
 as discussed, for example, in \cite{10.2307/41062}. The equivariant volume can be obtained by setting $q_i=\ex^{-\hbar \epsilon_i}$ and taking the limit $\hbar\rightarrow 0$. The limit of 
 \eqref{a1} is singular but exhibits the equivariant volume as the coefficient of the leading pole
 \be \label{Klimit} \mathds{Z}(\Lambda_a,q_i) \underset{\hbar\rightarrow 0}{=} \frac{\evol(\lambda_a,\epsilon_i)}{\hbar^m} + \ldots \, \ee
 where we scale $\Lambda_a=-\lambda_a/\hbar$.
 
 Analogously, the Molien-Weyl formula reads \cite{Nekrasov:2021ked,Cassia:2022lfj}
 \be \label{a2} \mathds{Z}_{MW}(T_A,\bar q_a)=  (-1)^{\fan-m} \int \prod_{A=1}^{\fan-m} \frac{dz_A}{2\pi i z_A} \frac{\prod_A z_A^{ -T_A}}{\prod_{a=1}^d (1- \prod_A z_A^{Q^A_a} \bar q_a )} \, ,\ee
 where the contour is defined again through the JK prescription. Setting $\bar q_a=\ex^{-\hbar \bar \epsilon_a}$, $T_A=t_a/\hbar \in \mathbb{Z}$ and changing variables to $z_a=\ex^{-\hbar \phi_A}$ we find
 \be \mathds{Z}_{MW}(T_A,\bar q_a) \underset{\hbar\rightarrow 0}{=} \frac{\evol_{MW}(t_A,\epsilon_i)}{\hbar^m} + \ldots \, .\ee
The Molien-Weyl formula \eqref{a2} can be interpreted as an average over the complexified group $U(1)^{\fan-m}$ acting on the coordinates of the ambient space $\mathbb{C}^{\fan}$.
The continuous average in \eqref{a2} should be supplemented by a discrete average if the group $K$ appearing in the symplectic reduction  $\Morb_{2m}=\mathbb{C}^\fan//K$ has a torsion part.

The  formulas \eqref{a1} and \eqref{a2} are  familiar in the context of non-compact conical Calabi-Yau singularities. For $\Lambda_a=T_A=0$ they  have been used to compute Hilbert series for the mesonic moduli space of the dual ${\cal N}=1$ superconformal theories \cite{Benvenuti:2006qr,Martelli:2006vh}. For $\Lambda_a\,  ,T_A\ne 0$ they compute Hilbert series for the baryonic moduli space \cite{Butti:2006au}. 
 
The relation between the two formulas is 
 
 \be\label{ccc} \mathds{Z}_{MW}\left ( T_A= \sum_a Q^A_a \Lambda_a, \bar q_a\right ) = \prod_{a=1}^d \bar q_a^{\Lambda_a} \mathds{Z}\left (\Lambda_a , q_i =\prod_a \bar q_a^{v^a_i} \right) \, .\ee  

The corresponding fixed point formulas for orbifolds are more complicated \cite{Vergne1996EQUIVARIANTIF,Silva}.  Each fixed point contribution is replaced by a discrete Molien formula implementing the quotient $\mathbb{Z}_{d_A}$
 \be \label{aa1} \mathds{Z}(\Lambda_a,q_i)=\sum_{A=(a_1,\ldots, a_m)} \frac{1}{d_A} \sum_{k=0}^{d_A-1} \frac{ \ex^{-2\pi i k \sum_{i=1}^m J_i \Lambda_{a_i}} \underline{q}^{ - \sum_{i=1}^m \Lambda_{a_i}\underline{u}^{a_i}_A/d_A} } 
{ \prod_{i=1}^m (1- \ex^{2\pi i k J_i}\underline{q}^{ \underline{u}^{a_i}_A/d_A}) } 
 \, ,\ee
 where 
 \be (\ex^{2\pi i J_1},\ldots ,\ex^{2\pi i J_m})\in \mathbb{T}^m\, ,\qquad \sum_{i=1}^m v^{a_i} J_i \in \mathbb{Z} \, ,\ee
 is a generator of the local orbifold singularity $\mathbb{Z}_{d_A}$. In the limit $\hbar\rightarrow 0$ only the term with $k=0$ contributes at the leading order and we recover the fixed point formula \eqref{fp} as the coefficient of the most singular term (see \eqref{Klimit}).

In the special case of non-trivial labels but no extra orbifold singularities  we can use the formulas in Appendix \ref{app:character} to resum \eqref{aa1} and we obtain 
\be \label{aaa1} \mathds{Z}(\Lambda_a,q_i)=\sum_A \frac{ \prod_{i=1}^m \underline{q}^{ -\underline{\hat u}^{a_i}_A \lfloor \frac{\Lambda_{a_i}}{ n_{a_i}}\rfloor} } 
{ \prod_{i=1}^m (1-\underline{q}^{ \underline{\hat u}^{a_i}_A}) } 
 \, ,\ee
 where the floor function $\lfloor x \rfloor$
denotes the integer part of $x$, $v^a=n_a \hat v^a$ and $u^a=n_a \hat u^a$ have been decomposed into primitive vectors $\hat v^a, \, \hat u^a$ and the label $n_a$ and we assume that $\hat d_{a,a+1}=\det (\hat v^a, \hat v^{a+1})=1$. One can check with elementary methods that this formula computes indeed the number of integer points in the polytope \eqref{pol}.

\section{Examples from geometry and holography}\label{sec:examples} 

To illustrate the previous discussion, in this section we provide explicit examples  of the equivariant volumes for some simple compact and non-compact orbifolds.
The specific examples are chosen among the orbifolds that appear as building blocks of various supergravity solutions with a holographic dual. We also derive some results  that will be used in sections \ref{anomal:sec} and \ref{sec:CY},  where we provide more physical applications.

\subsection{The spindle}\label{sec:spindle}

\label{spindle_example:sec}

The first example that we consider is the weighted projective line $\mathbb{WP}^1_{[n_+,n_-]}$, namely the spindle $\spindle$, which is at heart of the recent novel supergravity constructions.  It can be defined as the set of pairs of complex numbers
$(z_1,z_2)$ under the identification
\be\label{WP0} (z_1,z_2) \sim (\lambda^{n_+} z_1, \lambda^{n_-} z_2) \, ,\qquad \lambda\in \mathbb{C}^* \, .\ee

\begin{figure}[hhhh!]
\begin{center}
 
\begin{tikzpicture}[font = \footnotesize]

\node[draw=none, blue] at (-0.7,1) {$\CC/\ZZ_{n_+}$};
\draw[thick,dashed, -]  (0,1)--(4,1) node[right,blue] {$\CC/\ZZ_{n_-}$};
 \draw (2,1) ellipse (0.3 cm and 0.6cm);
  \draw (1.2,1) ellipse (0.2 cm and 0.4cm);
 \draw (2.8,1) ellipse (0.2 cm and 0.4cm);
 \draw (0.6,1) ellipse (0.1 cm and 0.2cm);
\draw (3.4,1) ellipse (0.1 cm and 0.2cm);

\filldraw (4,1) circle (0.5pt) ;
\filldraw (0,1) circle (0.5pt) ;

\end{tikzpicture}

\caption{The spindle $\spindle$ as a circle fibration over a segment.} \label{fig:spindle}
\end{center}
\end{figure}
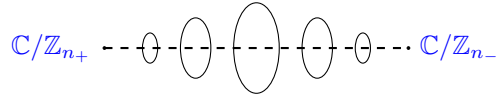

 This is the most general toric symplectic orbifold in two real dimensions. 
 It is  the simplest example of toric orbifold that is not a toric variety, as its singular points occur in complex co-dimension one. In particular, the orbifold points, that are fixed under the $U(1)$ action, are also the 
 divisors and therefore $d=n=2$.
 The image of $\spindle$ under the moment map is simply a segment $\mathcal{P}=I$, exactly as for $\mathbb{P}^1\simeq S^2$ and the 
 two facets 
 $\mathcal{F}_a$  are its
 end-points, defined by the   linear equations 
 \be
   l_a (y) = v^a y  - \lambda_a  =0 \, , \qquad a=1,2\, , 
 \ee 
 where the one-dimensional non-primitive vectors are $v^1 = n_1$, $v^2 \equiv -n_2$, where we will also denote $n_1\equiv n_+$ and $n_2\equiv n_-$.\footnote{We use this notation to facilitate comparison with the spindle literature in supergravity.} The two  labels are
 $n_+,n_-\in \mathbb{N}$. Thus $\spindle$ is a circle  fibration over the segment $I$, with the circle collapsing at the end-points, where are the two $\CC/\ZZ_{n_+}$, $\CC/\ZZ_{n_-}$ singularities, see figure \ref{fig:spindle}.  
 
 In the following
 it will be convenient to define two rescaled K\"ahler parameters $\lambda_1=n_+\hat\lambda_1$, $\lambda_2=n_-\hat\lambda_2$,  so that 
 \be
   l_1 (y) = n_+( y  - \hat\lambda_1 )   \, , \qquad  l_2 (y) = - n_-( y  + \hat\lambda_2 )   \, .
 \ee 
 We can then write explicitely the one-dimensional (canonical) metric (\ref{metric}) 
 \begin{equation}
 G_{11} = \frac{1}{2}\left( \frac{(v^1)^2}{ l_1} + \frac{(v^2)^2}{l_2} \right) = \frac{1}{2}\left(\frac{n_+}{y-\hat\lambda_1} - \frac{n_-}{y+\hat\lambda_2}  \right) \, , 
\end{equation}
 and the related moment maps 
 \begin{equation}
\label{spindlemomapsa}
\mu_1 =  \frac{1}{2\pi} \frac{y+\hat\lambda_2}{n_-(y-\hat\lambda_1) -n_+(y+\hat\lambda_2)}  \, , \qquad \mu_2 =   \frac{1}{2\pi} \frac{y-\hat\lambda_1}{n_-(y-\hat\lambda_1) -n_+(y+\hat\lambda_2)} \, . 
\end{equation}
 In the conventions of section \ref{sec:geo}, in which $l_a(y)\geq 0$,  we have that\footnote{Thus, in these conventions, the fibration exists for $\hat\lambda_1<-\hat\lambda_2$.} 
 $y\in [y_1,y_2]= [\hat\lambda_1,-\hat\lambda_2]$, with the extrema of the interval corresponding to the north and south poles of $\spindle$, respectively.

 We can then write all the relations discussed in section \ref{sec:geo} explicitly in terms of the moment maps (\ref{spindlemomapsa}). The explicit representatives of the first Chern classes $c_1(L_a)$ 
 associated to the line bundles $L_a$
defined by the extrema of the interval (the two divisors) are 
\be
c_1 (L_a) = \dd (\mu_a \dd \phi )\, , 
\ee
 where $\phi\sim \phi +2\pi$ is the azimutal coodinate, and denoting by $\epsilon$ the equivariant parameter,  their equivariant versions are 
 \be
 c_1^{\mathbb{T}}(L_a)  = \dd ( \mu_a \dd \phi) + 2\pi \epsilon  \mu_a\, .
\ee
 The  equivariant K\"ahler form is 
\be  \omega^{\mathbb{T}} = \omega + 2\pi H = \dd  y \wedge \dd \phi  + 2\pi \epsilon  y\, ,
\ee
where $H=\epsilon y$ is 
the Hamiltonian for the vector field $\xi= \epsilon  \partial_{\phi}$. With this information, we can check various relations explicitely, without appealing to the fixed point theorem. For example, it is immediate to compute
\begin{equation}
\label{defc1Laspindle}
\int_\spindle c_1({ L}_a)= 2\pi \int_{y_1}^{y_2} \dd \mu_a = 2\pi ( \mu_a (-\hat\lambda_2) -\mu_a (\hat\lambda_1 ) ) =  \frac{1}{n_a}\, .
\end{equation}
The co-homological relation  (\ref{kform})
can also be verified explicitly, noting that 
\begin{equation}
y =  -2\pi \sum_a \lambda_a \mu_a + \Phi(y)\, , 
\end{equation}
where 
\begin{equation}
\Phi (y) = \frac{(n_+-n_-) (y+\hat\lambda_2)(y-\hat\lambda_1 ) }{n_+ (\hat\lambda_2+y)+n_- (\hat\lambda_1-y)}\, ,
\end{equation}
is such that $\Phi(y_a)=0$, implying that 
$\Phi (y) \dd\phi$ is an exact one-form  on the spindle.

 Let us now turn to the fixed-point relations and the equivariant volume.
 From (\ref{spindlemomapsa}) it immediately follows that the restriction
 of the  moment maps  to the fixed points reads
 \begin{equation}
 \label{mamapfixspindle}
\left. \mu_a \right|_{y_b} =  - \frac{1}{2\pi}  \frac{u^a}{n_a} \delta_{ab} \, ,
\end{equation}
 where $u^1=1$, $u^2=-1$ and therefore
\be  
  c^{\mathbb{T}}_1(L_a) \Big |_{y_b} =    -   \frac{\epsilon u^a}{n_a} \delta_{ab}\,  .\ee
Using these and  the fact  that  the equivariant Euler class of the tangent bundle at a fixed point $y_a$  is
\be
e^{\mathbb{T}}|_{y_a} = -  \frac{\epsilon u^a }{n_a}  \, ,
\ee
one can  reproduce the integrals  (\ref{defc1Laspindle}) from the fixed point theorem (\ref{toric-equiv-integral0}).
The equivariant volume can be easily computed either directly
 \be
  \evol(\lambda_a,\epsilon) =  -\int_{\spindle} \ex^{- \epsilon y - \frac{\omega}{2\pi}}   = \int_{y_1}^{y_2}  \ex^{-\epsilon y } \dd y  
 =  \frac{1}{\epsilon}\left(  \ex^{-\tfrac{\epsilon \lambda_1}{n_+} } -\ex^{ \tfrac{\epsilon\lambda_2}{n_-} } \right) \, ,
 \ee
or using the fixed point formula 
\be \label{equivspindle}
 \evol(\lambda_a,\epsilon) = - \sum_{a=1}^2  \frac{ \ex^{- \frac{\omega^{\mathbb{T}}}{2\pi} |_{y_a}} } {n_a\,  e^{\mathbb{T}}|_{y_a}}  = 
  \frac{1}{\epsilon}\left(  \ex^{-\tfrac{\epsilon \lambda_1}{n_+} } -\ex^{ \tfrac{\epsilon\lambda_2}{n_-} } \right)\, , 
 \ee
which of course gives the same result. From this it is immediate to compute the non-zero equivariant intersection numbers (\ref{bbbbb}), 
\be
\label{intersectionsspindle}
\int_{\spindle} c_1^{\mathbb{T}}(L_{1})^s  = 
 \frac{(-\epsilon)^{s-1}}{n_+^s}\, ,\qquad \quad \int_{\spindle} c_1^{\mathbb{T}}(L_{2})^s  = 
  \frac{\epsilon^{s-1}}{n_-^s}\, ,
\ee
generalizing  (\ref{defc1Laspindle}) to $s>1$.

We end this subsection noting that it is straightforward to   check directly that the equivariant volume depends only on the co-homology class of $[\omega]$. Specifically, if we use a different representative of $[\omega]$, such as 
\be
\omega' = \dd y \wedge \dd \phi - \dd (\Phi (y)\dd \phi)\, , 
\ee
then we have
\begin{align} 
\evol(\lambda_a,\epsilon) &= - \int_{\spindle} \ex^{-\frac{\omega'^{\mathbb{T}}}{2\pi}} = \frac{1}{2\pi}\int_\spindle \ex^{-\epsilon y+ \epsilon \Phi (y)}  \omega'\nonumber\\
& = \int_{y_1}^{y_2}  \mathrm{e}^{- \epsilon y + \epsilon\Phi(y) }  \,  \dd (y-  \Phi (y) ) = \int_{y_1}^{y_2}  \mathrm{e}^{- \epsilon y }  \,  \dd y \, , 
\end{align}
where the last equality follows trivially from a change of variables and the fact that $\Phi (y_a)=0$.

\subsection{Four dimensional toric orbifolds}\label{sec:2d}

\begin{figure}[hhhh!]
\begin{center}
 
\begin{tikzpicture}[font = \footnotesize]
\draw[->] (0,0)--(2.5,0) node[below,blue] {$v^{a}$};
\draw[->] (0,0)--(1.7,1.7) node[above right,blue] {$v^{a+1}$};
\filldraw (1.2,0.5) circle (2pt) ;
\draw [-] (1.2,0.5) -- (1.2,-1) node[below right] {$p_{a-1}$};
\draw[<-] (1.4,0.2) --(1.4,-0.65) node[right,blue] {$u^2_{a}$};
\filldraw (1.2,-1) circle (2pt) ;
\draw [-] (1.2,0.5)--(0,1.7) node[above right] {$p_{a+1}$};
\draw[<-] (1.23,0.8) --(0.53,1.5) node[right,blue] {$u^1_{a}$};

\filldraw (0,1.7) circle (2pt) ;
\node[draw=none] at (1.5,0.5) {$p_a$}; 
\draw [-] (0,1.7) -- (-0.8,2);
\draw [-] (1.2,-1) --(0.8,-1.7);
\end{tikzpicture}

\caption{Fixed points are associated with vertices $p_a$ of the polytope. The vectors in the fan are orthogonal to the facets. For each vertex there is a corresponding cone $(v^a,v^{a+1})$ in the fan. The  two inwards normals $u_a^i$, $1=1,2$ to the cone, which lie along the edges of the polytope, enter in the fixed point formula through the quantities $\epsilon_i^a=\frac{\epsilon \cdot u_a^i}{d_{a,a+1}}$.}\label{fan2d}
\end{center}
\end{figure}
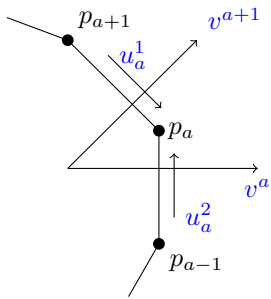

We now move to compact four-dimensional orbifolds ($m=2$), that have applications in holography \cite{Cheung:2022ilc,Faedo:2022rqx,Couzens:2022lvg}, 
as well as in geometry \cite{abreu2009,legendre2009,Apostolov:2013}. The image of $\Morb_{4}$ under the moment map
 is a rational simple convex  polygon  $\mathcal{P}$ in $\mathbb{R}^2$ and its facets $\mathcal{F}_a$, defined by the 
  linear equations $l_a(y)=0$, are the edges of the polygon, as shown in figure \ref{fan2d}.  Of course  each vertex $p_a$ of the polygon lies at the intersection of precisely $2$ edges, so that the simplicity condition is automatic.  
   Moreover,    the number of facets/edges  $\fan$ of $\mathcal{P}$ coincides with the number of vertices $\fix$, that are also the fixed points of the $\mathbb{T}^2$ action. 
   The condition that $\mathcal{P}$ be rational is equivalent to the fact  that the $\fan$ normals to the edges, $v^a$, have integer entries; however, they do not need to be primitive. 
The highest common factor  $n_a$ of the two entries of each of the $v^a$ is precisely the label  associated to  an edge/facet  $\mathcal{F}_a$.

Each facet defines a symplectic subspace of $\Morb_4$ of real dimension two, namely a divisor $D_a$ of complex dimension one, 
 with a local $\CC/\mathbb{Z}_{n_a}$ singularity. The vertices $p_a$ of  $\mathcal{P}$ correspond precisely to points where  two divisors $D_a$, $D_{a+1}$ intersect, 
 with local singularity $\CC^2/\mathbb{Z}_{d_{a,a+1}}$, where $d_{a,a+1}=|\det (v^{a},v^{a+1})|$. 
The linear dependence of the $\fan$ vectors $v^a$ in $\RR^2$ is expressed by the relations
\be 
 \sum_{a=1}^d Q_a^A v^a =0 \, , 
\ee
defining the GLSM charges $Q_a^A\in \mathbb{Z}$, with $A=1,\ldots \fan -2$,
 that are used to present $\Morb_4$ as the symplectic quotient 
$ \Morb_{4} = \mathbb{C}^d//U(1)^{d-2}$, up to torsion factors. Correspondingly there are
 two  linear relations among the divisors
\be \label{linreldivisors} \sum_{a=1}^\fan v^a_i D_a =0\, ,  \qquad i=1,2\, .\ee

 The equivariant volume and equivariant intersection numbers can be conveniently written in terms of the quantities
 \be \epsilon_i^a =\frac{\epsilon \cdot u_a^i}{d_{a,a+1}} \, ,\qquad i=1,2 \, ,\ee
 where $u_a^i$ are the inward normals to the  cones $(v^a,v^{a+1})$ as discussed in section \ref{sec:fixed} and pictured in figure \ref{fan2d}. We
 can rewrite these as 
 \begin{equation} 
  \epsilon^{a}_1 = -\frac{\det(v^{a+1}, \epsilon)}{\det( v^a , v^{a+1} )} \,,  \qquad
	\epsilon^{a}_2 = \frac{\det (v^a, \epsilon )}{\det (v^a ,v^{a+1} )}\,,
\end{equation}
 where $\epsilon \equiv (\epsilon_1,\epsilon_2)$.  In writing these formulas, we have assumed that the vectors $v^a$ lie on the plane in anticlockwise order and we have identified cyclically $v^{a+\fan}\equiv v^a$. In particular, the equivariant Euler class of the tangent bundle at a fixed point $y_a$  reads
\be
e^{\mathbb{T}}|_{y_a} =  \epsilon^{a}_1  \, \epsilon^{a}_2\, .
\ee
The restriction to the fixed points of the equivariant Chern classes $c_1^{\mathbb{T}}(L_a)$  (\ref{rest3})  can be written as   
\begin{equation}
 c_1^{\mathbb{T}}(L_a)|_{y_b}  =  -( \delta_{a,b} \epsilon_1^b +  \delta_{a,b+1}\epsilon_2^b )\, , 
 \end{equation}
so that the general formula for the equivariant intersection numbers reads 
\be \label{bbbbb4dorbs}\int_{\Morb_{4}} c_1^{\mathbb{T}}(L_{a_1}) \ldots c_1^{\mathbb{T}}(L_{a_p}) = \sum_a \frac{ c_1^{\mathbb{T}}(L_{a_1})|_{y_a} \ldots  c_1^{\mathbb{T}}(L_{a_p})|_{y_a}}{d_{a,a+1} \epsilon_1^a\epsilon_2^a}\, . 
\ee

 Below we report  explicit formulas for intersection numbers up to $p=4$. 
 First of all, for $p=0,1$ the integrals vanish automatically, giving rise to identities that may also be verified with elementary algebra. Specifically, we have 
\begin{eqnarray}
0 = \int_{\Morb_4} 1 &=& \sum_a \frac{1}{d_{a,a+1}\epsilon_1^a\epsilon_2^a}  \, , \\
0 = \int_{\Morb_4} c_1^{\mathbb{T}}(L_{a}) &=& -\frac{1}{d_{a,a+1}\epsilon_2^{a}} -\frac{1}{d_{a-1,a}\epsilon_1^{a-1}}  \, .
\end{eqnarray}

For the double intersection numbers we reproduce the intersections matrix of divisors (see \emph{e.g.} \cite{Faedo:2022rqx}), which is independent of the equivariant parameters $\epsilon_1,\epsilon_2$:
\begin{equation}
D_a\cdot D_b= D_{ab} = \int_{\Morb_{4}} c_1^{\mathbb{T}}(L_{a}) c_1^{\mathbb{T}}(L_{b}) =
\begin{cases}
\frac{1}{d_{a-1,a}} & \text{if $b=a- 1$}\, ,\\
\frac{1}{d_{a,a+1}} & \text{if $b=a +1$}\, ,\\
 -\frac{d_{a-1,a+1}}{d_{a-1,a}d_{a,a+1}} & \text{if $b=a$}\, ,\\
 0 & \text{otherwise}\, ,
\end{cases} \label{intersections2}
\end{equation}
where we used the identity 
\begin{equation} \frac{\epsilon_1^{a}}{d_{a,a+1} \epsilon_2^{a}} + \frac{\epsilon_2^{a-1}}{d_{a-1,a} \epsilon_1^{a-1}}=  -\frac{d_{a-1,a+1}}{d_{a-1,a}d_{a,a+1}} \end{equation}
to deal with the terms occurring  in the self-intersection $D_a\cdot D_a$.
Similarly, we can compute the
integral of three equivariant first Chern  classes, that read:
\begin{eqnarray}
D_{a_1a_2a_3}  &=& \int_{\Morb_{4}} c_1^{\mathbb{T}}(L_{a_1}) c_1^{\mathbb{T}}(L_{a_2}) c_1^{\mathbb{T}}(L_{a_3})\nonumber\\[1.5mm]
&=&-\begin{cases}
\frac{\epsilon_2^{a-1}}{d_{a-1,a}} & \text{if $a_i=a_j=a_k+1\equiv a$}\, ,\\
\frac{\epsilon_1^{a}}{d_{a,a+1}} & \text{if $a_i=a_j=a_k-1\equiv a$}\, ,\\
 \frac{(\epsilon_1^a)^2}{d_{a,a+1} \epsilon_2^{a}} + \frac{(\epsilon_2^{a-1})^2}{d_{a-1,a} \epsilon_1^{a-1}} & \text{if $a_1=a_2=a_3\equiv a $}\, ,\\
 0 & \text{otherwise}\, ,
\end{cases}  \label{intersections3}
\end{eqnarray}
where the diagonal term can be shown to be linear in $\epsilon_i$, using the identity 
\begin{equation} 
\frac{(\epsilon_1^a)^2}{d_{a,a+1} \epsilon_2^{a}} + \frac{(\epsilon_2^{a-1})^2}{d_{a-1,a} \epsilon_1^{a-1}} =   -\frac{d_{a-1,a+1}}{d_{a-1,a}^2d_{a,a+1}} (2 d_{a-1,a} \epsilon_1^{a} + d_{a-1,a+1} \epsilon_2^{a})  \, . \end{equation}
For the integrals of four equivariant first Chern  classes we have
\begin{eqnarray} 
D_{a_1a_2a_3a_4}  &= &  \int_{\Morb_{4}} c_1^{\mathbb{T}}(L_{a_1}) c_1^{\mathbb{T}}(L_{a_2}) c_1^{\mathbb{T}}(L_{a_3}) c_1^{\mathbb{T}}(L_{a_4})\nonumber\\[1.5mm]
& =&\begin{cases}
\frac{\epsilon_2^{a-1} \epsilon_1^{a-1}}{d_{a-1,a}} & \text{if $a_i=a_j=a_k+1=a_l+1\equiv a$}\, ,\\
\frac{\epsilon_1^{a} \epsilon_2^{a} }{d_{a,a+1}} & \text{if $a_i=a_j=a_k-1=a_l-1\equiv a$}\, ,\\
\frac{(\epsilon_2^{a-1})^2}{d_{a-1,a}} & \text{if $a_i=a_j=a_k=a_l+1\equiv a$}\, ,\\
\frac{(\epsilon_1^{a})^2}{d_{a,a+1}} & \text{if $a_i=a_j=a_k=a_l-1\equiv a$}\, ,\\
 \frac{(\epsilon_1^{a})^3}{d_{a,a+1} \epsilon_2^{a}} + \frac{(\epsilon_2^{a-1})^3}{d_{a-1,a} \epsilon_1^{a-1}}& \text{if $a_1=a_2=a_3=a_4\equiv a$}\, ,\\
 0 & \text{otherwise}\, ,
\end{cases}
\label{intersections4}
\end{eqnarray}
where the diagonal term can be shown to be quadratic in $\epsilon_i$, using the identity 
\begin{equation} \frac{(\epsilon_1^{a})^3}{d_{a,a+1} \epsilon_2^{a}} + \frac{(\epsilon_2^{a-1})^3}{d_{a-1,a} \epsilon_1^{a-1}}=  -\frac{d_{a-1,a+1}}{d_{a-1,a}^3d_{a,a+1}} (3 d_{a-1,a}^2 (\epsilon_1^{a})^2  +3 d_{a-1,a}d_{a-1,a+1} \epsilon_1^{a}\epsilon_2^{a}+ d_{a-1,a+1}^2 (\epsilon_2^{a})^2)\, . \end{equation}

The fixed point formula  (\ref{fp}) for the equivariant volume specializes to the expression\footnote{Notice that there is no summation on $a$ in the exponent.}
\begin{align}\label{eqvolume4d}
 \evol(\lambda_a,\epsilon_i ) &
=\sum_{a=1}^d \frac{1}{d_{a,a+1} \epsilon_1^{a}\epsilon_2^{a} }\ex^{ -\lambda_a \epsilon^a_1 -\lambda_{a+1} \epsilon^a_2}\, , 
 \end{align}
which is easily evaluated in examples. On the other hand, starting from the definition (\ref{evolH}), the equivariant volume can also be written as in integral over the polygon $\mathcal{P}$ as
\be 
\evol(\lambda_a,\epsilon_i ) = 
 \int_{\mathcal{P}} \mathrm{e}^{-y_i\epsilon_i} \dd y_1 \dd y_2\, , 
\ee
that can be evaluated explicitely using Stokes' theorem.  Notice that the dependence on the K\"ahler parameters $\lambda_a$ arises from the shape of ${\mathcal P}={\mathcal P}(\lambda_a)$. The details of this  calculation are given in  appendix \ref{fixptform4d_appendix}.

\subsubsection{The weighted projective space $\mathbb{WP}^2_{[N_1,N_2,N_3]}$}

As a first example of equivariant volume for four-dimensional toric orbifolds we consider the weighted projective space $\mathbb{W}\mathbb{P}^2_{[N_1,N_2,N_3]}$ defined as the set of triples of complex numbers
$(z_1,z_2,z_3)$ under the identification
\be\label{WP} (z_1,z_2,z_3) \sim (\lambda^{N_1} z_1, \lambda^{N_1} z_2, \lambda^{N_1} z_3) \, ,\qquad \lambda\in \mathbb{C}^* \, .\ee
There are various presentations in terms of labelled polytopes \cite{Abreu:2001to} and it is interesting to consider some of them
in order to understand better the role of the labels.

Consider first the fan
\be\label{nonminimal}  v^1=(n_3,n_3)\, ,\qquad v^2=(-n_1,0)\, ,\qquad v^3=(0,-n_2)\, ,\ee
where each of the three facets have non-trivial labels. We take the labels $n_a$ to be mutually coprime. The GLSM charges are given by
\be Q\equiv (N_1,N_2,N_3) =(n_1 n_2,n_2 n_3,n_1 n_3) \, ,\ee
and the symplectic quotient description $\mathbb{W}\mathbb{P}^2_{[N_1,N_2,N_3]}=\mathbb{C}^3//U(1)$, where $U(1)$ acts with charge $Q$, is just
the definition \eqref{WP} of the weighted projective space. Notice however that our choice of fan corresponds to ``non-minimal'' $Q_a=N_a$ that are products of integers.

The equivariant volume can be computed with the fixed point formula \eqref{eqvolume4d}
\be 
\evol(\lambda_a,\epsilon_i)=\frac{\ex^{-\epsilon_2\lambda_1/n_3-(\epsilon_2 -\epsilon_1)\lambda_2/n_1}}{\epsilon_2(\epsilon_2-\epsilon_1)}+\frac{\ex^{\epsilon_1\lambda_2/n_1+\epsilon_2 \lambda_3/n_2}}{\epsilon_1\epsilon_2} +\frac{\ex^{-\epsilon_1\lambda_1/n_3-(\epsilon_1 -\epsilon_2)\lambda_3/n_2}}{\epsilon_1(\epsilon_1-\epsilon_2)} \, .\ee
This expression is a rational function of $\epsilon_i$ but when expanded in power series in $\lambda_a$ as
\be\label{expandevol} \evol(\lambda_a,\epsilon_i)=\sum_{k=0}^\infty \evol^{(k)}(\lambda_a,\epsilon_i) \, ,\ee
where $\evol^{(k)}$ is the component of degree $k$ in $\lambda_a$, all singular terms cancel. The constant and linear terms vanish  and the other $\evol^{(k)}(\lambda_a,\epsilon_i)$ are homogeneous 
polynomials of degree $k-2$ in $\epsilon$
that encode the equivariant intersection numbers of the line bundles $L_a$. In particular, $\evol^{(2)} (\lambda_a,\epsilon_i)$ coincides with the non-equivariant limit
\be \evol(\lambda_a,0)= \frac12 \left ( \frac{n_1 n_2 \lambda_1 + n_2  n_3\lambda_2 +n_1 n_3\lambda_3}{n_1 n_2 n_3} \right)^2 = \frac12  \frac{ (N_1 \lambda_1 + N_2 \lambda_2 + N_3\lambda_3)^2}{N_1 N_2 N_3}  \, ,\ee
encoding the classical intersections numbers of the divisors $D_a$. Since only one divisor is geometrically independent because of \eqref{diveq}, this is a quadratic form of rank one.

We can compare the result with  the Molien-Weyl formula \eqref{molien}, 
\be \evol_{MW}(t,\bar\epsilon) = \int \frac{\dd\phi}{2\pi} \frac{\ex^{t \phi}}{(\bar \epsilon_1 + n_1 n_2 \phi)(\bar \epsilon_2 + n_3 n_2 \phi)(\bar \epsilon_3 + n_1 n_3 \phi)}\, ,\ee
where we use the charges $Q$. The JK prescription for a single integration is very simple.\footnote{See \cite{Benini:2013xpa} for a physics-oriented review.} It prescribes 
to take all residues associated with $\phi$ with positive charge for $t>0$, and minus the residues for $\phi$ with negative charge for $t<0$.
Taking $t>0$ and performing the residue computation, we obtain
\bea  \evol_{MW}(t,\bar\epsilon) = \frac{\ex^{-\bar\epsilon_1 t/(n_1 n_2)}}{(n_3 \bar\epsilon_1 -n_1 \bar\epsilon_2) (n_3 \bar\epsilon_1 -n_2 \bar\epsilon_3) } 
&+\frac{\ex^{-\bar\epsilon_2 t/(n_2 n_3)}}{(n_3 \bar\epsilon_1 -n_1 \bar\epsilon_2) (n_2 \bar\epsilon_3 -n_1 \bar\epsilon_2) } \\
&+\frac{\ex^{-\bar\epsilon_3 t/(n_1 n_3)}}{(n_3\bar\epsilon_1 -n_2 \bar\epsilon_3) (n_1 \bar\epsilon_2 -n_2 \bar\epsilon_3) } \, ,
\eea
with non-equivariant limit 
\be \frac 12  \frac{t^2}{(n_1 n_2 n_3)^2}= \frac 12  \frac{t^2}{N_1 N_2 N_3}\, .\ee

We see that each residue corresponds to a fixed point.  It is easy to see that the two master volumes coincides under the identification \eqref{conj0}
 \be\label{conj2WP} \evol_{MW}(t=  -\sum_{a=1}^3 N_a  \lambda_a, \bar\epsilon_a) = \ex^{\sum_{a=1}^3\lambda_a \bar\epsilon_a} \evol(\lambda_a, \epsilon_1 =
\bar \epsilon_1 n_3-\bar\epsilon_2 n_1,  \epsilon_2 =
\bar \epsilon_1 n_3-\bar\epsilon_3 n_2) \, .\ee

It is interesting to consider also the fan 
\be\label{minimal}  v^1=(-n_3,0)\, ,\qquad v^2=(0,-n_3)\, ,\qquad v^3=(n_1,n_2)\, ,\ee
where now the GLSM charges are the ``minimal'' ones
\be Q\equiv (N_1,N_2,N_3) =(n_1 ,n_2 , n_3) \, .\ee
However in this case there is also a torsion part in the symplectic quotient description. Indeed
 the symplectic action on $\mathbb{C}^3$ is given by 
\be (\ex^{2\pi i Q_1},\ex^{2\pi i Q_2},\ex^{2\pi i Q_3}) \, ,\ee
where 
\be \sum_a Q_a v^a_i\in \mathbb{Z} \, .\ee For the minimal fan \eqref{minimal} with  relatively prime $n_a$, the relations 
\be -n_3 Q_1 +n_1 Q_3\in \mathbb{Z} \, ,\qquad -n_3 Q_2 +n_2 Q_3\in \mathbb{Z} \, ,\ee
have solution 
\be\label{qminimal} Q_1 = n_1 \alpha + \frac{k_1}{n_3}\, ,\qquad Q_2 = n_2 \alpha + \frac{k_2}{n_3}\, ,\qquad Q_3 = n_3 \alpha \, ,\qquad k_i=0,\ldots ,n_3-1\, ,\,\,\, \alpha\in \mathbb{R} \, .\ee 
One $k_i$ can be further gauged away by choosing $\alpha={\rm integer}/n_3$ (assuming again that the $n_a$ are relatively prime). So we are left with a continuous $U(1)$ action generated by $\alpha$ and an extra discrete group\footnote{For the non-minimal case \eqref{nonminimal}, we have $-n_3 Q_1+n_1 Q_2, -n_3 Q_1 +n_2 Q_3\in \mathbb{Z}$ which implies $Q_1=n_1 n_2 \alpha + k_1/n_3$, $Q_2=n_3 n_2 \alpha + k_2/n_1$ and $Q_3=n_1 n_3 \alpha + k_3/n_2$ but all the $k_i$ can be absorbed by taking $\alpha={\rm integer}/n_i$ if the $n_i$ are coprime and we are left with the $U(1)$ action with charges $(n_1 n_2,n_2 n_3,n_1 n_3)$.}  
\be\label{qminimal2} \Gamma= \frac{\mathbb{Z}_{N_3}\times \mathbb{Z}_{N_3}}{\mathbb{Z}_{N_3}}\, ,\ee
so that
\be \Morb_4 = \mathbb{W}\mathbb{P}^2_{[N_1,N_2,N_3]} /\Gamma\, .\ee
As an exercise, the interested reader can  check   that the relation 
\be \evol_{MW}(t=  -\sum_a \lambda_a N_a, \bar\epsilon_a)\equiv \ex^{\sum_a\lambda_a \bar\epsilon_a} \evol( \lambda_a, \epsilon_i =\sum_a v_i^a \bar\epsilon_a) \, ,\ee
still holds.  The right hand side can be computed as before with the fixed point formula \eqref{eqvolume4d} while the left hand side with the  Molien-Weyl formula. However, since the Molien-Weyl integral is blind to the torsion part,  the formula  \eqref{molien} should be further divided by the order $N_3$ of the discrete group $\Gamma$. For example, 
the non-equivariant  volume is now 
\be   \frac12  \frac{ (N_1 \lambda_1 + N_2 \lambda_2 + N_3\lambda_3)^2}{N_1 N_2 N_3^2} = \frac 12  \frac{t^2}{N_1 N_2 N_3^2}  \, ,\ee
with an extra factor of $N_3$.

Finally, we can consider an example  taken from \cite{Abreu:2001to}\footnote{The reader should be aware that, in \cite{Abreu:2001to},  the weighted projective space is denoted with $\mathbb{W}\mathbb{P}^2_{(N_1,N_2,N_3)}$ and the example we are discussing with $\mathbb{W}\mathbb{P}^2_{[N_1,N_2,N_3]}$.}. The  fan is
\be\label{minimal2}  v^1=(-n_2 n_3,0)\, ,\qquad v^2=(0,-n_1 n_3)\, ,\qquad v^3=(n_1 n_2,n_1 n_2)\, ,\ee
and the GLSM charges are the ``minimal" ones
\be Q\equiv (N_1,N_2,N_3) =(n_1 ,n_2 , n_3) \, .\ee
However, the equations $\sum_a Q_a v^a_i\in \mathbb{Z}$ 
\be -n_3 n_2 Q_1 +n_1 n_2 Q_3\in \mathbb{Z} \, ,\qquad -n_3 n_1 Q_2 +n_1 n_2 Q_3\in \mathbb{Z} \, ,\ee
have solution 
\be Q_1 = n_1 \alpha + \frac{k_1}{n_2 n_3}\, ,\qquad Q_2 = n_2 \alpha + \frac{k_2}{n_1 n_3}\, ,\qquad Q_3 = n_3 \alpha+ \frac{k_3}{n_1 n_2} \, ,\,\,\, \alpha\in \mathbb{R} \, ,\ee 
and one $k_i$ can be further gauged away by choosing $\alpha={\rm integer}/(n_1 n_2 n_3)$. So we are left with a continuous $U(1)$ action generated by $\alpha$ and an extra discrete group  so that
\be \Morb_4 =\mathbb{W}\mathbb{P}^2_{[N_1,N_2,N_3]}/\Gamma\, ,\ee
where 
\be\label{residual} \Gamma= \frac{\mathbb{Z}_{N_2 N_3}\times \mathbb{Z}_{N_1 N_3} \times \mathbb{Z}_{N_2 N_3}}{\mathbb{Z}_{N_1 N_2 N_3}} \, ,\ee
is a discrete group of order  $N_1 N_2 N_3$. We still find 
\be \evol_{MW}(t=  -\sum_a \lambda_a N_a, \bar\epsilon_a)\equiv \ex^{\sum_a\lambda_a \bar\epsilon_a} \evol(\lambda_a, \epsilon_i =\sum_a v_i^a \bar\epsilon_a) \, ,\ee
where the Molien-Weyl integral should be further divided by $N_1 N_2 N_3$ .  
The non-equivariant  volume is
\be   \frac12  \frac{ (N_1 \lambda_1 + N_2 \lambda_2 + N_3\lambda_3)^2}{N_1^2 N_2^2 N_3^2} = \frac 12  \frac{t^2}{N_1^2 N_2^2 N_3^2}  \, .\ee

\subsubsection{Quadrilaterals}

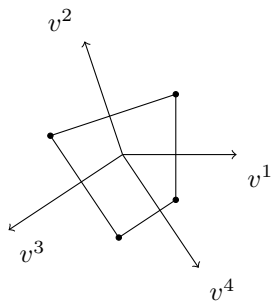
\begin{figure}[hhhh!]
\begin{center}
 
\begin{tikzpicture}[font = \footnotesize]
\draw[->] (0,0)--(1.5,0) node[below right] {$v^1$};

\draw[->] (0,0)--(-0.5,1.5) node[above left] {$v^2$};

\draw[->] (0,0)--(-1.5,-1) node[below right] {$v^3$};

\draw[->] (0,0)--(1,-1.5) node[below right] {$v^4$};

\draw[-] (-0.95,0.25)--(-0.05,-1.1)--(0.7,-0.6)--(0.7,0.8)--(-0.95,0.25);

\filldraw (-0.95,0.25) circle (1pt) ;
\filldraw (-0.05,-1.1) circle (1pt) ;

\filldraw (0.7,-0.6) circle (1pt) ;
\filldraw (0.7,0.8) circle (1pt) ;
\end{tikzpicture}

\caption{The fan and the polytope for a generic quarilateral.} \label{quad}
\end{center}
\end{figure}

Let us now discuss the most general four-dimensional toric orbifold with four fixed points, that we refer to as \emph{quadrilaterals}, following \cite{legendre2009}.  A sub-class of this family of orbifolds has been discussed in \cite{Faedo:2022rqx} and below we will discuss  how one can 
retrieve those results. 
It is convenient to parameterize the data of this orbifold in term of the six vector products
\be
d_{a,b}\equiv \det( v^a,v^b ) \,  \in\,   \ZZ\, , 
\ee
where $v^a$, $a=1,\dots,4$ is the set of toric data, as in figure \ref{quad}. Then a vector identity implies that these satisfy the following relation
\begin{equation} 
\label{quadconstraint}
	d_{1,2} \, d_{3,4} - d_{2,3}\,  d_{4,1} =  d_{1,3} \, d_{2,4}\,,
\end{equation}
showing that there are  five independent integers characterising a quadrilateral. One can show that the charges of the $U(1)^2$ action on $\CC^4$ can be written in 
 a particularly symmetric form using this parameterization, and read
\begin{align}
Q^1=\left(d_{2,4} , d_{4,1} ,0, d_{1,2}  \right)\, ,\qquad \quad Q^2=\left( d_{2,3} ,-  d_{1,3} , d_{1,2} ,0\right)\, .
\end{align}
Without loss of generality, one can use an SL$(2,\ZZ)$ transformation to set $v^1 = (n_-,0)$, where $n_-\in\ZZ$.  The remaining  vectors solving the constraint 
\be
\sum_{a=1}^4 Q_a^A v^a  =0 \, , 
\ee
are then easily determined and the full set reads  
\begin{align}
\label{quadrilfan}
&v^1 =  (n_-,0) \, ,  &v^2 = ( a_+ d_{2,4},  d_{1,2}/n_-)\, , ~~\,\nonumber\\
 &v^3 =    (a_-  d_{2,3} + a_+  d_{3,4}, d_{1,3}/n_-) \, ,   &v^4 = (a_-  d_{2,4}, -d_{4,1}/n_-) \,  , 
  \end{align}
 where $d_{1,2}, d_{1,3}, d_{4,1}$ are integer multiples of $n_-$ and $a_\pm \in \ZZ$ such that   
 \be
 \label{quadbezout}
 a_- d_{1,2} +  a_+ d_{4,1} =-n_- \, ,
 \ee 
 which always exist by Bezout's lemma.  
 
  The equivariant volume $\evol (\lambda_a,\epsilon_i)$ is easily computed using the general fixed point formula (\ref{eqvolume4d}), but we shall not write it explicitly as it does not have a compact expression.  Expanding it in Taylor series in 
  powers of $\lambda_a$ as in (\ref{expandevol}),
  one can check that $ \evol^{(0)} (\epsilon_i) = \evol^{(1)} (\lambda_a,\epsilon_i) = 0$, whereas the quadratic part is independent of $\epsilon_i$ and reads
  \be
  \evol^{(2)} (\lambda_a) = \frac{1}{2}\sum_{a,b} \lambda_a\lambda_b D_{ab}\, ,
  \ee
  where $D_{ab}$ is the intersection matrix of divisors (\ref{intersections2}), which reads
  \be
D_{ab} =  \left(
\begin{array}{cccc}
 \frac{ d_{2,4}}{ d_{1,2}  d_{4,1}} & \frac{1}{d_{1,2}} & 0 & \frac{1}{ d_{4,1}} \\
 \frac{1}{ d_{1,2}} & \frac{- d_{1,3}}{d_{1,2}d_{2,3}} & \frac{1}{d_{2,3}} & 0 \\
 0 & \frac{1}{ d_{2,3}} & -  \frac{ d_{2,4}}{d_{2,3} d_{3,4}} & \frac{1}{d_{3,4}} \\
 \frac{1}{ d_{4,1}} & 0 & \frac{1}{ d_{3,4}} & \frac{ d_{1,3}}{d_{3,4} d_{4,1}} \\
\end{array}
\right)\, . 
  \ee

It may be useful to compare with the orbifold geometry studied in \cite{Faedo:2022rqx} (see also \cite{Cheung:2022ilc}), which may be viewed as spindle fibered over another spindle, and interpreted as a natural orbifold generalization of Hirzebruch surfaces. This is obtained setting
\begin{align}
 d_{1,3} = 0 \, ,\qquad   d_{2,4} = - t \, ,  \qquad d_{1,2}= m_- n_-  \, , \qquad  \nonumber\\
 d_{2,3} =  m_- n_+ \, , \qquad  d_{3,4}= m_+ n_+  \, , \qquad  d_{4,1} =n_- m_+\, . 
\end{align}
 and $r_+ = - t a_+$, $r_- = - t a_-$. For these values the vectors of the fan reduce to those written in eq. (4.27) of \cite{Faedo:2022rqx} and the GLSM charges become  
 \begin{align}
Q^1=\left( -t  , m_+ n_- ,0, m_- n_-  \right)\, ,\qquad \quad Q^2=\left( n_+ , 0  , n_-  ,0\right)\, .
\end{align}
We can then perform the $U(1)^2$ quotient of $\CC^4$ in two stages. The first quotient using  weights $Q^1$ gives the line bundle $O(-t)\to \spindle_{m_+,m_-}$.
 The second quotient using weights $Q^2$ projectivize this bundle, giving a $\spindle_{n_+,n_-} \to \spindle_{m_+,m_-}$ bundle. Indeed for $n_+=n_-=m_+=m_-=1$ this reduces exactly to the Hirzebruch surface $\mathbb{F}_t$, which is the most general family of toric manifolds with four fixed points. 
 
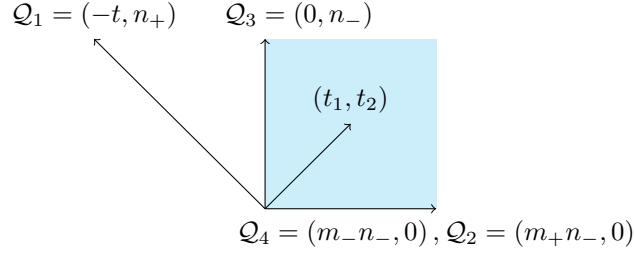
\begin{figure}\begin{center}
\begin{tikzpicture}[font = \footnotesize, scale =0.75]
\fill [cyan, opacity=0.2] (0,0)--(3,0)--(3,3)--(0,3)--(0,0);
\draw[->] (0,0)--(3,0) node[below] {$\mathcal{Q}_4=(m_- n_-,0) \, , \mathcal{Q}_2=(m_+ n_-,0)$};
\draw[->] (0,0)--(0,3) node[above ] {$\,\,\,\qquad \mathcal{Q}_3=(0,n_-)$};
\draw[->] (0,0)--(-3,3) node[above ] {$\mathcal{Q}_1=(-t,n_+)$};
\draw[->] (0,0)--(1.5,1.5) node[above] {$(t_1,t_2)$};
\end{tikzpicture}
\caption{The JK prescription for the $\spindle_{n_+,n_-} \to \spindle_{m_+,m_-}$ bundle.}\label{JKfib}
\end{center} \end{figure}

It is instructive to compare with the Molien-Weyl formula. We do it explicitly  for the case of the $\spindle_{n_+,n_-} \to \spindle_{m_+,m_-}$ bundle. The Molien-Weyl integral reads
\be \evol_{MW}=n_- \int \frac{\dd\phi_1}{2\pi} \frac{\dd\phi_2}{2\pi}  \frac{\ex^{t_1 \phi_1+ t_2 \phi_2}}{(\bar \epsilon_1 -t \phi_1 +n_+ \phi_2)(\bar \epsilon_2 + n_- m_+ \phi_1)(\bar \epsilon_3 + n_- \phi_2)(\bar \epsilon_4 + m_- n_- \phi_1)}\, .\ee
The multiplicative factor $n_-$ takes into account that the $U(1)^2$ action is not effective since
\be g_1 = \ex^{2\pi i \frac{n_+ }{n_-}Q_1} \, ,\qquad g_2 = \ex^{-2\pi i \frac{t }{n_-}Q_2} \, ,\ee
acts in the same way on all points in $\mathbb{C}^4$. 
The Molien formula mods out twice
by a discrete subgroup $\mathbb{Z}_{n_-}$  and we need to multiplicate by $n_-$ to obtain the right result. The poles in the integrand are associated with two-dimensional
vectors $\mathcal{Q}^a=(Q^1_a, Q^2_a)$,  $a=1,2,3,4$. The JK prescription instructs us to take the  (simultaneous) residues at the poles  $\mathcal{Q}_{a_1}$ and $\mathcal{Q}_{a_2}$ if the vector of K\"ahler parameters $(t_1,t_2)$ is contained in the cone $(\mathcal{Q}_{a_1},\mathcal{Q}_{a_2})$. When $(t_1,t_2)$ lies in the first quadrant, it is contained in four different cones  (see figure \ref{JKfib}).
In the  K\"ahler chamber $t_1>0$ and $t_2>0$ we then add four contributions. One can check  that
 \be\label{conjsp}  \evol_{MW}(t_A=  -\sum_a \lambda_a Q_a^A, \bar\epsilon_a) = \ex^{\sum_a\lambda_a \bar\epsilon_a} \evol(\lambda_a, \epsilon_i =\sum_a v_i^a \bar\epsilon_a) \, ,\ee
where the gauge invariant variables \eqref{gaugeinv} are
\bea & t_1 = t \lambda_1 - m_+ n_- \lambda_2 - m_- n_- \lambda_4\, \qquad  &t_2=-n_+ \lambda_1 - n_- \lambda_3 \\
&\epsilon_1= n_- \bar\epsilon_1+r_+ \bar\epsilon_2 -n_+ \bar\epsilon_3+r_- \bar\epsilon_4\, \qquad &\epsilon_2 =m_- \bar\epsilon_2 -m_+ \bar\epsilon_4 \, .\eea
The non-equivariant volume is
\bea 
\evol_{MW}(t_A, 0)  =\frac{ 2 n_+ t_1 t_2 + m_- r_- t_1^2+ m_+ r_+ t_2^2}{2 m_- m_+ n_-^2 n_+^2}   \, .\eea

\subsection{Non-compact examples}\label{sec:nnoncompact}
\label{sec:noncompactexamples}

We consider now some  non-compact examples where the polytope $\mathcal{P}$ is asymptotically a {polyhedral cone.
We compute the volume using the fixed point formula, assuming that there is a finite number of fixed points and no contributions from infinity. 
As anticipated in section  \ref{sec:non-comp}, the equivariant volume is a {\it rational} function of $\epsilon_i$. 
As we will see, two of the singular terms  for the Calabi-Yau examples can be identified with the Sasakian volume of \cite{Martelli:2005tp,Martelli:2006yb} and the  master volume introduced in \cite{Gauntlett:2018dpc}. This identification will be discussed in more detail in section \ref{sec:CY}.

\subsubsection{$\CC^2/\ZZ_p$}

Our first non-compact example is  ${\CC^2/\ZZ_p}$, which is a conical orbifold singularity. The most general toric action is obtained using as fan the vectors
\be 
   \label{nonsusyC2Zp}  v^1=(1,0)\,  ,\qquad v^2=(q,p) \,  .
   \ee
The corresponding $\mathbb{Z}_p$ action on $(z_1,z_2)\in \mathbb{C}^2$ is given by 
\be 
(z_1,z_2)\to (\mathrm{e}^{2\pi i Q_1}z_1,\mathrm{e}^{2\pi i Q_2}z_2) \, ,
\ee
 where  the GLSM charges are pure torsion, namely 
\be 
\sum_a Q_a v^a \in \mathbb{Z}^2 \, ,
\ee
whose solution is, for $p$ and $q$ coprime, given by 
\be
Q_1 =  k\frac{q}{p}\, , \qquad  Q_2 =  -\frac{k}{p}\, ,\qquad   \qquad k =0,\ldots ,p-1\,   .
\ee 
This is then the $\mathbb{C}^2/\mathbb{Z}_p$  quotient 
\be 
(z_1,z_2)\to (\omega_p^q z_1,\omega_p^{-1}z_2) \, ,
\ee
with $\omega_p=\mathrm{e}^{\frac {2\pi i}{p}}$, which for $|z_1|^2+|z_2|^2=1$ gives the Lens space $L(p,q)$.
 The equivariant volume is obtained from the general formula (\ref{eqvolume4d}), and is entirely encoded by the the contribution of the  single fixed point with $\ZZ_p$ singularity, namely
 \begin{equation}
\label{correctVeqC2Zp}
\evol_{\CC^2/\ZZ_p} (\lambda_a,\epsilon_i) =
p\frac{\mathrm{e}^{\frac{1}{p} (\lambda_1 (q\epsilon_2 - p\epsilon_1) -  \lambda_2  \epsilon_2)}}{ (p \epsilon_1 - q\epsilon_2) \epsilon_2}\, .
\end{equation}
Expanding (\ref{correctVeqC2Zp})  in powers of $\lambda_a$ we obtain
\begin{equation} 
\label{eqvolC2Zpexpanded}
 \evol_{\CC^2/\ZZ_p} (\lambda_a,\epsilon_i) =
\frac{p}{ \epsilon_2 (  p \epsilon_1 -q\epsilon_2) } -
\left( \frac{\lambda_1}  {\epsilon_2} +\frac{\lambda_2}{p\epsilon_1 -q\epsilon_2}     \right) + {\cal O} (\lambda_a^2) \, ,
\end{equation}
where the constant and linear terms are non zero, as a difference with the compact case. They are also singular functions of $\epsilon$. The constant piece is actually the Sasakian volume of the Lens space $L(p,q)$  and the linear term is the corresponding master volume introduced in  \cite{Gauntlett:2018dpc}. 
In particular, for $q=1$ that corresponds to the ``supersymmetric'' Lens space $L(p,1)$, 
the linear term coincides precisely with (B.6) of  \cite{Gauntlett:2019pqg}.

 Since there are no continuous GLSM charges, the Molien-Weyl formula is trivial
 \be \evol_{\CC^2/\ZZ_p}^{MW}(\bar\epsilon_a) =\frac{1}{p \bar \epsilon_1 \bar\epsilon_2} \, ,\ee
 where there are no K\"ahler parameters $t$ and we divide by $p$ for the order of the discrete group $\ZZ_p$. Nevertheless, it is still true that
  \be\label{conjsp}  \evol_{\CC^2/\ZZ_p}^{MW}( \bar\epsilon_a) = \ex^{\sum_a\lambda_a \bar\epsilon_a} \evol_{\CC^2/\ZZ_p}(\lambda_a, \epsilon_i =\sum_a v_i^a \bar\epsilon_a) \, ,\ee
  where the gauge invariant combinations are
  \be \epsilon_1 =\bar\epsilon_1+q \bar\epsilon_2  \, ,\qquad \epsilon_2 =p \bar\epsilon_2 \, .\ee

\subsubsection{${\cal O}(-p)\to \spindle$}

Let us now consider the asymptotically conical non-compact orbifold\footnote{The non-compact examples of this section are the total space of some vector bundles over a base orbifold, but we denote them using the more schematic notation ``fibre $\to$ base''.} 
${\cal O}(-p)\to \spindle$, where $\spindle= \spindle_{n_+,n_-}$ is a spindle with $\ZZ_{n_+}$, $\ZZ_{n_-}$ singularities at its poles. This can be thought of as a  ``blow-up''
of the  ${\CC^2/\ZZ_p}$ of the previous section, where the apex of the cone is replaced with the spindle $\spindle$. In general, this is a partial resolution, but in the special case that 
$n_+=n_-=1$, it is a complete resolution.

The   GLSM charges are given  by
 \be 
 Q= (n_+,-p,n_-) \, ,
 \ee
where notice that the space is a Calabi-Yau if and only if $p=n_++n_-$. In particular,  only for $p=2$ we have a complete resolution of the CY singularity ${\CC^2/\ZZ_2}$. 
The vectors of the fan solving the condition 
  \be
  \sum_{a=1}^3 Q_a v^a=0\, , 
  \ee
can be taken to be 
   \be 
    \label{secondbasisC2Zp}  
    v^1=(1,0)\,  , \qquad v^2=(t,n_-)\ ,\qquad v^3=(q,p) \, , 
     \ee
   with  $t,q\in \ZZ$ satisfying 
 \be
 \label{Opspindlebezout}
  t p -  q n_-  =  n_+\, ,
  \ee
  which always exist by Bezout's lemma.  Indeed, we see that the basis (\ref{secondbasisC2Zp}) is obtained from (\ref{nonsusyC2Zp})  by adding the vector $(t,n_-)$, which is normal to the edge 
  representing the spindle, as illustrated in figure \ref{fanblowup}. 
  
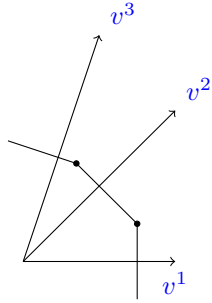
\begin{figure}[hhhh!]
\begin{center}
 
\begin{tikzpicture}[font = \footnotesize]
\draw[->] (0,0)--(2,0) node[below,blue] {$v^{1}$};
\draw[->] (0,0)--(1,3) node[above right,blue] {$v^{3}$};
\draw[->] (0,0)--(2,2) node[above right,blue] {$v^{2}$};
\draw[-] (1.5,-0.5)--(1.5,0.5);
\draw[-] (1.5,0.5)--(0.7,1.3);
\draw[-] (0.7,1.3)--(-0.2,1.6);
\filldraw (0.7,1.3) circle (1pt) ;
\filldraw (1.5,0.5) circle (1pt) ;
\end{tikzpicture}

\caption{The fan and the polytope for ${\cal O}(-p)\to \spindle$. The resolution introduces a compact facet in the polytope that is orthogonal to $v^2$. The corresponding  segment represents the toric polytope of a spindle
$\spindle$.} \label{fanblowup}
\end{center}
\end{figure}

  Alternatively, the basis  (\ref{secondbasisC2Zp}) can be obtained from (\ref{quadrilfan}) of the compact example, by removing $v^3$ and setting 
   $n_-=1$, $d_{4,1}=-p$,  $d_{1,2}=n_-$, $d_{2,4}=-n_+$ and
$a_+ d_{2,4}=t$, $a_-  d_{2,4}=q$. The condition (\ref{quadbezout}) obeyed by $a_+,a_-$ becoming the condition (\ref{Opspindlebezout}) for $t,q$.

The equivariant volume is  obtained from the general formula (\ref{eqvolume4d}), and now it receives contributions from the two orbifold fixed points, with  $\ZZ_{n_-}$, $\ZZ_{n_+}$ singularities, namely
\be
 \evol_{{{\cal O}(-p) \to \spindle} } (\lambda_a,\epsilon_i)  =n_-\frac{ \mathrm{e}^{\frac{1}{n_-}(\lambda_1 (t\epsilon_2 - n_-\epsilon_1 )-\epsilon_2 \lambda_2)}}{\epsilon_2 (n_-\epsilon_1 - t\epsilon_2)}
 +n_+\frac{ \mathrm{e}^{\frac{1}{n_+}(\lambda_2(q\epsilon_2-p\epsilon_1)+  \lambda_3(n_-\epsilon_1- t\epsilon_2))}}{(q\epsilon_2-p \epsilon_1 )(n_-\epsilon_1 - t \epsilon_2)}\, ,
\ee
and expanding this in $\lambda_a$ we get 
\begin{equation} 
\label{moreeqvolC2zp}
 \evol_{{{\cal O}(-p) \to \spindle} } (\lambda_a,\epsilon_i)  =
\frac{p}{ \epsilon_2 (  p \epsilon_1 -q\epsilon_2) } 
-\left( \frac{\lambda_1}  {\epsilon_2} + \frac{\lambda_3}{p\epsilon_1 -q\epsilon_2}     \right) + {\cal O} (\lambda_a^2) \, ,
\end{equation}
which coincides with (\ref{eqvolC2Zpexpanded}). Thus we see that up to linear order, the information about the (partial) resolution is washed out as there is no dependence on $n_-$ and $n_+$, nor on the K\"ahler parameter $\lambda_2$, corresponding to the compact divisor $\spindle$. Specifically, the leading and sub-leading terms reproduce precisely the Sasakian and GMS master volume of the Lens space $L(p,q)$.
Notice that this behaviour is independent of the Calabi-Yau condition  $p=n_++n_-$. Another notable  case is given by $n_+=n_-=p=1$, which corresponds to the smooth non-compact space   
${\cal O}(-1)\to \mathbb{P}^1$, namely the blow-up of $\CC^2$ at one point\footnote{In this case we have $t-q=1$ and therefore we can take $t=1$, $q=0$, reproducing the 
obvious fan $v^1=(1,0)  ,  v^2=(1,1), v^3=(0,1)$.}. This is not a Calabi-Yau and its link is the round $S^3$.

We can also compare with the Molien-Weyl formula
\be \evol^{MW}_{{\cal O}(-p) \to \spindle }( t, \bar\epsilon_a)=\int \frac{\dd \phi}{2\pi} \frac{\ex^{t \phi }}{(\bar\epsilon_1 +n_+ \phi)(\bar\epsilon_2 -p \phi)(\bar\epsilon_3 +n_- \phi)}\, ,\ee
where, according to the JK prescirption, for $t>0$ we take the residues for the terms with positive charge $Q_i$, $\phi=  -\bar\epsilon_1/n_+$ and $\phi =-\bar\epsilon_3/n_-$.
We find, as usual
\be\label{conjO(p)}  \evol_{{{\cal O}(-p) \to \spindle} }^{MW}( t =-\sum_a Q_a\lambda_a, \bar\epsilon_a) = \ex^{\sum_a \lambda_a \bar\epsilon_a} \evol_{{{\cal O}(-p) \to \spindle} }(\lambda_a, \epsilon_i =\sum_a v_i^a \bar\epsilon_a) \, ,\ee
  where the gauge invariant combinations are
  \be t = - n_+ \lambda_1 + p \lambda_2 -n_- \lambda_3\, ,\qquad \epsilon_1 =\bar\epsilon_1+t \bar\epsilon_2+q \bar\epsilon_3  \, ,\qquad \epsilon_2 =n_- \bar\epsilon_2  +p \bar\epsilon_3\, .\ee

\subsubsection{${\cal O}(-1) \oplus {\cal O}(-1) \to \mathbb{P}^1$}
\label{resconif}

We  now consider the  two small resolutions of the conifold singularity $\{ z_1 z_2 = z_3 z_4 | z_i \in \mathbb{C}^4    \}$, corresponding to the 
asymptotically conical non-compact manifold  ${\cal O}(-1) \oplus {\cal O}(-1) \to \mathbb{P}^1$, known in the physics literature as resolved conifold. 
The toric data consist in the fan
\be v^1=(1,0,0)\, ,\qquad v^2=(1,1,0)\, ,\qquad v^3=(1,1,1)\, ,\qquad v^4=(1,0,1) \, ,\ee
with GLSM charges 
\be Q=(1,-1,1,-1) \, .\ee
 The corresponding polytope is a conical non--compact polyhedron with four facets all intersecting
at the tip of the cone. 
Since the vectors $v^a$ lie on a plane, or, equivalently $\sum_a Q_a=0$,  this defines a conical Calabi-Yau three-fold.
In order to use the results of section \ref{sec:equivol} we need to resolve the conifold singularity. This can be done in a standard way 
by triangulating the fan as in figure \ref{conifold1}.
\begin{figure}[hhhh!]
\begin{center}

\begin{tikzpicture}[font = \footnotesize]
\draw (0,0)  node[below,blue] {$v^{1}$} --(3,0) node[below,blue] {$v^{2}$}--(3,3) node[above,blue] {$v^{3}$}--(0,3) node[above,blue] {$v^{4}$}--(0,0);
\draw (3,0)--(0,3);
\draw (1,1)--(2,2);
\draw (2,2)--(2,4);
\draw (2,2)--(4,2);
\draw (1,1)--(1,-1);
\draw (1,1)--(-1,1);
\filldraw (1,1) circle (2pt) ;
\filldraw (2,2) circle (2pt) ;

\end{tikzpicture}
\caption{One of the two small resolutions of the conifold singularity and the corresponding non-compact polytope projected on the plane where the vectors $v^a$ live.} \label{conifold1}
\end{center}
\end{figure}
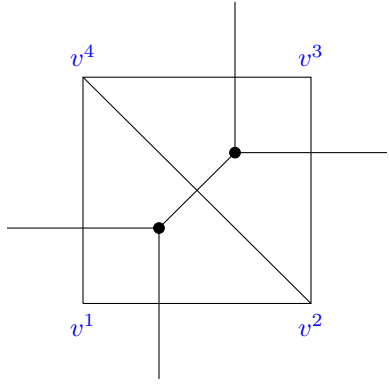
The fan is now the union of the two cones $(v^1,v^2,v^4)$ and $(v^2,v^3,v^4)$, each associated with a vertex of the polytope and therefore with a fixed point of the torus action.
The original conical singularity has been replaced by a compact two cycle, which can be visualised as a circle fibration over the segment connecting the two vertices of the polytope.
There are actually two small resolutions related by a flop. The second one is obtained by triangulating the fan in a different way, by adding the  line $v^1-v^3$ and considering the fan which is the union of the two cones $(v^1,v^2,v^3)$ and $(v^1,v^3,v^4)$.

The fixed point formula \eqref{fp} reads
\be \evol(\lambda_a,\epsilon_i)=\sum_{A=(a_1,a_2,a_3)} \frac{\ex^{-\epsilon \cdot (\lambda_{a_1}  u^{a_1}_A+ \lambda_{a_2}  u^{a_2}_A+\lambda_{a_3}  u^{a_3}_A)/d_{a_1,a_2,a_3}}}{d_{a_1,a_2,a_3}(\frac{\epsilon \cdot u^{a_1}_A}{d_{a_1,a_2,a_3}}) (\frac{\epsilon \cdot u^{a_2}_A}{d_{a_1,a_2,a_3}})(\frac{\epsilon \cdot u^{a_3}_A}{d_{a_1,a_2,a_3}}) } \, ,\ee
where $A$ runs over the triangular cones of the resolution, $d_{a_1,a_2,a_3}=|\det (v^{a_1},v^{a_2},v^{a_3})|$ is the order of the orbifold singularity at the fixed point and $u^{a}_A$ are the inward normals to the faces of the cones as in figure \ref{fp3}.
 \begin{figure}[hhh!]
 \begin{center}
\begin{tikzpicture}[font = \footnotesize]
\draw (0,0)  node[below,blue] {$v^{a_1}$} --(3,0) node[below,blue] {$v^{a_2}$}--(0,3) node[above,blue] {$v^{a_3}$}--(0,0);
\filldraw (1,1) circle (2pt) node[below right] {$y$};
\draw[<-] (1.1,1.1)--(2,2) node[above] {$u^{a_1}$};
\draw[<-] (1,0.9)--(1,-1) node[left] {$u^{a_3}$};
\draw[<-] (0.9,1)--(-1,1) node[below] {$u^{a_2}$};
\node[draw=none,red] at (5.1,-0.3) {$c_1(L_{a_2}) \big |_{y}= \frac{ \epsilon\cdot   u^{a_2}}{d_{a_1,a_2,a_3}}$};
\node[draw=none,red] at (2,3.2) {$c_1(L_{a_3}) \big |_{y}= \frac{ \epsilon\cdot   u^{a_3}}{d_{a_1,a_2,a_3}}$};
\node[draw=none,red] at (-2,-0.3) {$c_1(L_{a_1}) \big |_{y}= \frac{ \epsilon\cdot   u^{a_1}}{d_{a_1,a_2,a_3}}$};
\node[draw=none] at (7,1.5) {$d_{a_1,a_2,a_3}=| \det (v^{a_1}, v^{a_2}, v^{a_3})|$};
\end{tikzpicture}
\caption{The contribution of a single vertex of the polytope to the fixed point formula for $m=3$.}\label{fp3}
\end{center}
\end{figure}
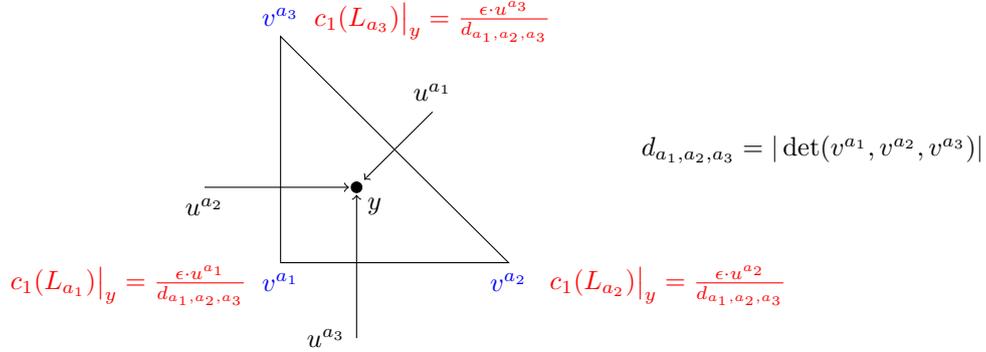

Consider first the resolution in figure \ref{conifold1}. The necessary data are
\bea (a_1,a_2,a_3)&=(1,2,4)\, , \qquad u^{a_1}_A=(1,-1,-1)\, ,\qquad u^{a_2}_A=(0,1,0)\, ,\qquad u^{a_3}_A=(0,0,1)\, , \\
(a_1,a_2,a_3)&=(2,3,4)\, , \qquad u^{a_1}_A=(1,0,-1)\, ,\qquad \,\,\, \,\, u^{a_2}_A=(-1,1,1)\, ,\,\,\,\,\,\,\,  u^{a_3}_A=(1,-1,0)\, , \nonumber \eea
and the equivariant volume is
\be\label{V1} \evol(\lambda_a,\epsilon_i)=\frac{\ex^{-(\epsilon_1-\epsilon_2-\epsilon_3) \lambda_1-\epsilon_2 \lambda_2-\epsilon_3 \lambda_4}}{\epsilon_2\epsilon_3(\epsilon_1-\epsilon_2-\epsilon_3)} +
\frac{\ex^{-(\epsilon_1-\epsilon_3) \lambda_2-(-\epsilon_1+\epsilon_2+\epsilon_3) \lambda_3-(\epsilon_1-\epsilon_2) \lambda_4}}{(\epsilon_1-\epsilon_2)(\epsilon_1-\epsilon_3)(-\epsilon_1+\epsilon_2+\epsilon_3)} \, .\ee
At $\lambda_a=0$ we recover the Sasakian volume of the 5-dimensional base  of the singular cone (cfr (7.30) in \cite{Martelli:2006yb}) 
\be\label{I} \evol^{(0)}(\epsilon_i) = \evol(\lambda_a=0,\epsilon_i)=\frac{\epsilon_1}{\epsilon_2\epsilon_3(\epsilon_1-\epsilon_2)(\epsilon_1-\epsilon_3)} \, ,\ee
    and the quadratic part is, up to a normalization,  the master volume introduced in  \cite{Gauntlett:2018dpc}\footnote{In this formula we identify cyclically $v^5=v^1$.}
\be\label{II}  2 \evol^{(2)}(\epsilon_i)=  \sum_{a=1}^4 \frac{\lambda_a \left ( \lambda_{a-1} \det (v^a,v^{a+1},\epsilon) -\lambda_{a} \det (v^{a-1},v^{a+1},\epsilon)+\lambda_{a+1} \det (v^{a-1},v^{a},\epsilon)\right)}{\det (v^{a-1},v^{a},\epsilon)\det (v^a,v^{a+1},\epsilon)}\ .\ee

Things work similarly for the second resolution 
\bea (a_1,a_2,a_3)&=(1,2,3)\, , \qquad u^{a_1}_A=(1,-1,0)\, ,\qquad u^{a_2}_A=(0,1,-1)\, ,\qquad u^{a_3}_A=(0,0,1)\, , \\
(a_1,a_2,a_3)&=(1,3,4)\, , \qquad u^{a_1}_A=(1,0,-1)\, ,\qquad u^{a_2}_A=(0,1,0)\, ,\qquad u^{a_3}_A=(0,-1,1)\, . \nonumber \eea
The equivariant volume now reads
\be\label{V2} \evol(\lambda_a,\epsilon_i)=
\frac{\ex^{-(\epsilon_1-\epsilon_2) \lambda_1-(\epsilon_2-\epsilon_3) \lambda_2-\epsilon_3 \lambda_3}}{\epsilon_3(\epsilon_1-\epsilon_2)(\epsilon_2-\epsilon_3) }+ \frac{\ex^{-(\epsilon_1-\epsilon_3) \lambda_1-\epsilon_2 \lambda_3-(\epsilon_3-\epsilon_2) \lambda_4}}{\epsilon_2(\epsilon_1-\epsilon_3)(\epsilon_3-\epsilon_2)} \, ,\ee
which is different from \eqref{V1}. The difference between the equivariant volumes for the two resolutions starts at order three in the K\"ahler parameters
\be\label{diff}  
\evol \Big |_{{\rm res} 2}-\evol \Big |_{{\rm res} 1}  = \frac{(\lambda_1-\lambda_2+\lambda_3-\lambda_4)^3}{6} +O(\lambda_a^4) \, .\ee
In particular, equations \eqref{I} and \eqref{II} are  valid for both resolutions. We see that  the first terms in the $\lambda_a$ expansion of $\evol$, which  are singular in $\epsilon_i$,  capture the geometry of the
asymptotic singular cone  and are therefore independent of the resolution.  The difference \eqref{diff}  is instead a power series in $\lambda_a$ with coefficients which are regular polynomials in $\epsilon_i$ and this reflects the fact that the different flops differ by a compact cycle.

It is also interesting to compare with the result of the Molien-Weyl formula \eqref{molien} which, using  $Q=(1,-1,1,-1)$, reads
\be \evol_{MW}(t,\bar\epsilon)=\int \frac{\dd\phi}{2 \pi i} \frac{\ex^{t \phi}}{(\bar \epsilon_1 +\phi)(-\bar \epsilon_2 +\phi)(\bar \epsilon_3 +\phi)(-\bar \epsilon_4 +\phi)} \, .\ee
We need to specify a prescription for the contour and the residues to take, which depends on the sign of $t$ and the choice of resolution. The JK  prescription requires to take the residues where the charge $Q$ is positive for $t > 0$ (first resolution) and minus the residues where the charge $Q$ is negative for $t < 0$ (second resolution). 
For example, for $t>0$ we take the two residues $\phi=-\bar \epsilon_1$ and $\phi=-\bar \epsilon_3$, obtaining
\be \evol_{MW}(t,\bar\epsilon) = \frac{\ex^{-t \bar\epsilon_3}}{(\bar\epsilon_1-\bar\epsilon_3)(\bar\epsilon_2+\bar\epsilon_3)(\bar\epsilon_3+\bar\epsilon_4)}-
\frac{\ex^{-t \bar\epsilon_1}}{(\bar\epsilon_1-\bar\epsilon_3)(\bar\epsilon_1+\bar\epsilon_2)(\bar\epsilon_1+\bar\epsilon_4)} \, , \ee
while for $t<0$ we take (minus) the residues associated with negative charge,  $\phi=\bar \epsilon_2$ and $\phi=\bar \epsilon_4$. 
It is easy to check that in both cases
 \be\label{conj22con} \evol_{MW}(t=  -\sum_a \lambda_a Q_a, \bar\epsilon_a) \Big |_{\substack{t>0\\t<0}} = \ex^{\sum_a\lambda_a \bar\epsilon_a} \evol(\lambda_a, \epsilon_i =\sum_a v_i^a \bar\epsilon_a) \Big |_{\substack{{\rm res} 1\\{\rm res 2}}}\, ,\ee
where  the gauge invariant variables explicitly read 
\be t=-\lambda_1+\lambda_2-\lambda_3+\lambda_4\, ,\qquad \epsilon_1=\bar \epsilon_1+\bar \epsilon_2+\bar \epsilon_3+\bar \epsilon_4\, ,\qquad
\epsilon_2=\bar \epsilon_2+\bar \epsilon_3\, ,\qquad \epsilon_3=\bar \epsilon_3+\bar \epsilon_4\, .\ee

\section{Integrating the anomaly polynomial on orbifolds}

\label{anomal:sec}

In this section we will consider 
field theories  associated to D3 and M5 branes and we will show how
the formalism of equivariant integration can be employed to determine the anomaly polynomials of lower dimensional 
 theories,  obtained compactifying the original theories on orbifolds. For  theories compactified on the spindle these results were obtained in \cite{Ferrero:2020laf,Boido:2021szx,Hosseini:2021fge,Ferrero:2021wvk}, but the equivariant formalism 
 allows for uniform derivations for various  branes wrapped on different orbifolds.  
 In the case of  theories compactified on manifolds the method of integration of the anomaly polynomial 
  is  well-known and
  is reviewed in 
  \cite{Hosseini:2020vgl}, to which we refer for more details. In this reference is also discussed  the extension in which 
  background gauge fields for the isometries of the compactification manifolds are turned on, and 
     the relation to equivariant integration is spelled out. 
  In a SCFT in (even) dimension ${\rm d}$, the anomaly polynomial can be thought of  as a formal 
$({\rm d}+2)$-form on an auxiliary $({\rm d}+2)$-dimensional space $Z_{{\rm d}+2}$ that is 
the total space of a $\Morb_{\rm p}$ fibration over a $({\rm d}+2-{\rm p})$-dimensional base space $B_{{\rm d}+2-{\rm p}}$, 
\begin{equation}
 \Morb_{\rm p}  \hookrightarrow  Z_{{\rm d}+2} \to B_{{\rm d}+2-{\rm p}}\, , 
\end{equation}
and integrating it on $\Morb_{\rm p}$ gives a $({\rm d}+2-{\rm p})$-form, that is the anomaly polynomial of a $({\rm d}-{\rm p})$-dimensional SCFT. In \cite{Ferrero:2020laf} 
an extension of  this construction to the  case that 
$\Morb_{\rm 2}=\spindle $ is a spindle has been proposed and succesfully matched to the dual supergravity solution. 
More generally, here we will assume that $\Morb_{\rm p}$ (and hence, generically,  also $Z_{{\rm d}+2}$ )  is an orbifold.

The anomaly polynomial is  a linear combination of characteristic classes of vector 
bundles on $Z_{{\rm d}+2}$ which (using the splitting principle)  can always be decomposed in terms of first Chern classes of holomorphic line (orbi-)bundles on $Z_{{\rm d}+2}$.
Specifically, for any   $U(1)$ symmetry acting on  $\Morb_{\rm p}$, the latter can be fibered over $B_{{\rm d}+2-{\rm p}}$ by  gauging this $U(1)$ with  
a connection on a line bundle  $\mathcal{J}$ over   $B_{{\rm d}+2-{\rm p}}$, denoted $A_\mathcal{J}$. When $\Morb_{2m}$ is a toric orbifold this  consists in replacing  the occurrences of each 
term $\dd \phi_i$ 
with  $\dd \phi_i +A_{\mathcal{J}_i}$, where ${\cal J}_i$
are auxiliary line bundles with  $c_1({\cal J}_i) = \left[F_{\mathcal{J}_i} \right]/2\pi \in H^2(B_{{\rm d}+2-{\rm p}},\ZZ)$ and $F_{\mathcal{J}_i} =\dd A_{\mathcal{J}_i}$. This recipe is  formally equivalent to consider \emph{equivariant first Chern classes}, as follows. Starting from the line bundles $L_a$ 
with equivariant first Chern classes $c_1^{\mathbb{T}}(L_a)$ 
discussed in section \ref{sec:geo}, one can  make the replacement
\begin{equation}
\label{anomprocedure}
c_1^{\mathbb{T}}(L_a) = c_1(L_a) + 2\pi\mu_a^i  \epsilon_i \quad  \mapsto \quad
 c_1({{\cal L}}_a) = 
  c_1({L}_a) + 2\pi \mu_{a}^i c_1({\cal J}_i)\, .
\end{equation}
One can then use these  $c_1({{\cal L}}_a)$ as a basis to parameterise the characteristic 
 classes appearing in the anomaly polynomials.

\subsection{D3 branes on the spindle}

For 4-dimensional SCFTs the anomaly polynomial is a $6$-form  defined on an six-dimensional orbifold $Z_6$ that we take to be
the total space of a $\spindle$ fibration over an auxiliary space $B_4$,
\begin{equation}
\spindle \hookrightarrow Z_6 \to B_4\, . 
\end{equation}
Neglecting terms that are sub-leading in the 
large $N$ limit
the six-form anomaly polynomial is given by
\begin{equation}
\label{intanompolspindle}
{\cal A}_{4{\rm d}} = \frac{1}{6} \sum_{I,J,K} c_{IJK} c_1(F_I) c_1(F_J)  c_1(F_K) \, , 
\end{equation}
where $c_{IJK}$ are the cubic 't Hooft anomaly coefficients  $c_{IJK}=$Tr$(F_IF_JF_K)$ and $F_I$ denote the generators of  global $U(1)$ symmetries. The $c_1(F_I)$ are formal first Chern classes associated 
to these $U(1)_I$ symmetries that can be  decomposed as\footnote{To avoid clumsiness in the formulas, we sistematically use the Einstein's notation for the indices $a,b,\dots$ in this section.}   
\begin{equation}
\label{c1FIspindle}
c_1(F_I) = \Delta_I c_1(F_R^{2{\rm d}}) -  {\flp}_I^{a} c_1 ( {\cal L}_a)\, , \qquad c_1({\cal L}_a) = c_1({L}_a) + 2\pi \mu_{a} c_1({\cal J})\, , 
\end{equation}
where  ${\flp}_I^{a}\in \ZZ$ are  ``fluxes stuck at the fixed points'',  $L_a$ are the line bundles discussed in section \ref{spindle_example:sec}
 and $c_1(F_R^{2{\rm d}})\in H^2(B_4,\ZZ)$ is the first Chern class of the $2{\rm d}$ R-symmetry line bundle.

The anomaly polynomial of the two-dimensional theory is obtained integrating  the anomaly polynomial of the four-dimensional theory  on the spindle,
 \begin{equation}
{\cal A}_{2{\rm d}} =  \int_{\spindle}{\cal A}_{4{\rm d}}\, ,
\end{equation}
and is a four-form on $B_4$. Substituting (\ref{c1FIspindle}) in (\ref{intanompolspindle})  it is immediate to see that this is exactly equivalent  to the  equivariant integral on the spindle
\begin{equation}
\label{2danomeqintegralspindle}
F (\Delta_I,\epsilon) = \frac{1}{6} \sum_{I,J,K} c_{IJK} \int_\spindle (\Delta_I - \flp_I^a  c_1^{\mathbb{T}}(L_a)  )  (\Delta_J  - \flp_J^a  c_1^{\mathbb{T}}(L_a) )  (\Delta_K  - \flp_K^a  c_1^{\mathbb{T}}(L_a)) \, ,
\end{equation}
 where we have expressed the resulting four-form as a function of the variables $\Delta_I$ and the equivariant paremeter $\epsilon$. 
This can also be understood as allowing the trial 2d R-symmetry to mix with the $U(1)$ symmetry of the spindle, formally setting  $c_1({\cal J}) = \epsilon c_1(F_R^{2{\rm d}})$
(thus ``undoing'' the replacement (\ref{anomprocedure})), and extracting the coefficient of the four-form
 ${\cal A}_{2d}  = F (\Delta_I,\epsilon) c_1(F_R^{2{\rm d}})^2$.  Expanding the integrand we obtain\footnote{We use that $c_{IJK}$ are totally symmetric to write (\ref{2danol}).}
\begin{align}
\label{2danol}
F  (\Delta_I,\epsilon)   = \frac{1}{6} \sum_{I,J,K} c_{IJK}  \flp_I^a      \left[ -3 \Delta_J \Delta_K D_a  +3   \flp_J^b \Delta_K   D_{ab}   -\flp_J^b \flp_K^c D_{abc}  \right]  \,, 
\end{align}
with the  equivariant intersection numbers
\begin{equation}\label{intspindle}
D_a = \int_\spindle c_1^{\mathbb{T}}(L_a)   \, , \quad D_{ab} = \int_\spindle c_1^{\mathbb{T}}(L_a)c_1^{\mathbb{T}}(L_b) \, , \quad D_{abc} = \int_\spindle c_1^{\mathbb{T}}(L_a)c_1^{\mathbb{T}}(L_b)c_1^{\mathbb{T}}(L_c)\, , 
\end{equation}
whose non-zero values are given in (\ref{intersectionsspindle}). Alternatively, the same result can be obtained by evaluating the equivariant integral (\ref{2danomeqintegralspindle}) using the fixed point theorem (\ref{toric-equiv-integral0}), explaining the 
observation made in \cite{Hosseini:2021fge}. Specifically, we get
\be
\label{2danomfixedpoints}
F (\Delta_I,\epsilon) = F^1  (\Delta_I,\epsilon) +  F^2 (\Delta_I,\epsilon)\, , 
\ee
where, 
in terms of the trial 4d central charge 
\be
a_{4{\rm d}} (\Delta_I) = \frac{9}{32} \sum_{I,J,K} c_{IJK}\Delta_I \Delta_J \Delta_K \, , 
\ee
we have
\be
F^a (\Delta_I,\epsilon)  = (-1)^a \frac{16}{27\epsilon}a_{4{\rm d}} (\Delta_I^{a})\, ,\qquad \qquad  \Delta_I^{a} \equiv \Delta_I - (-1)^a \flp_I^a \frac{\epsilon}{n_a} \, . 
\ee
Notice that the terms singular in $\epsilon$   (cubic in the $\Delta_I$) cancels out, as expected.

The expression above depends on the $\flp_I^a$, but it can be rewritten in terms of ``physical fluxes'' $\mathfrak{n}_I$,  defined as the integrals of 
the flavour line bundles  \cite{Hosseini:2020vgl} 
\begin{equation}
\label{degflavlinbundles}
c_1 (E_I) \equiv   -   {\flp}_I^{a} c_1 ( L_a)\, , 
\end{equation}
namely 
\begin{equation}
\mathfrak{n}_I \equiv - \int_\spindle c_1(E_I) 
=\frac{{\flp}_I^{1}}{n_1} +   \frac{{\flp}_I^{2}}{n_2} \, .
\end{equation}
Recall that in general supersymmetry can be preserved on the spindle by coupling with a background R-symmetry gauge field   with first Chern class
\be
c_1(E_R) = -\sigma^1c_1(L_1) -\sigma^2c_1(L_2)\, ,
\ee
where $\sigma^a=\pm 1$, corresponding to 
either twist or  anti-twist \cite{Ferrero:2021etw}.
Since the supercharge should couple to $2c_1(F_R^{2{\rm d}}) + c_1(E_R)$ we must have
\be
\sum_I \left( \Delta_I c_1(F_R^{2{\rm d}}) +  c_1 (E_I)  \right)  = 2c_1(F_R^{2{\rm d}}) + c_1(E_R)\, , 
\ee
implying  the constraints
\be 
\label{susyonspindles}
\sum_I \Delta_I = 2 \, , \qquad \quad \sum_I  {\flp}_I^a= \sigma^a  \, . 
\ee
As a consequence, the physical fluxes obey 
\be
\label{physfluxesspindle}
\sum_I \mathfrak{n}_I =\frac{\sigma^1}{n_1}+\frac{\sigma^2}{n_2} \, 
\ee
and we can introduce  new variables 
\begin{equation} 
\label{nicevarphivariables}
  \varphi_I \equiv   \Delta_I + \frac{1}{2}  (  \frac{{\flp}_I^{1}}{n_1}-  \frac{{\flp}_I^{2}}{n_2}   )\epsilon \,, \qquad \sum_I \varphi_I - \frac{1}{2}\left(\frac{\sigma^1}{n_1}-\frac{\sigma^2}{n_2}\right) \epsilon=2\, .
\end{equation}
In terms of these variables  the gravitational blocks depend only on  the physical fluxes, namely
\be
F^a (\varphi_I,\epsilon)  = (-1)^a \frac{16}{27\epsilon}a_{4{\rm d}} (\varphi_I - (-1)^a \tfrac{\mathfrak{n}_I }{2} \epsilon) \, ,
\ee
yielding 
\begin{equation}
\label{simplestA2d}
F  (\varphi_I,\epsilon) = -\frac{1}{24} \sum_{I,J,K} c_{IJK}\mathfrak{n}_I  \left[  2\varphi_J \varphi_K + \mathfrak{n}_J \mathfrak{n}_K\epsilon^2\right] \, , 
\end{equation}
in agreement with  \cite{Faedo:2021nub}.

Notice that (\ref{simplestA2d})  is much simpler than the analogous expression in eq. (5.9) of \cite{Hosseini:2021fge} and is therefore the most convenient form to be used in the 
extremization. The reason is that in 
the variables $\varphi_I$ the expression is manifestly independent of the redundant parameters $r_0^I$ used in the construction in 
 \cite{Hosseini:2021fge}, as we now explain.
  In this reference one starts from the background gauge fields 
 $A_I = \rho_I (y)\dd \phi$ inherited from the supergravity solution, where $y,\phi$ are coordinates on the spindle. 
 This is lifted to $Z_6$  by the gauging procedure $\dd \varphi \mapsto \dd \varphi + A_\mathcal{J}$, which leads to the connection one-forms ${\mathscr{A}}_I = \rho_I(y) (\dd \phi + A_\mathcal{J})$, where 
 $\dd A_\mathcal{J} = 2\pi c_1(\mathcal{J})$, so that 
 \be
 \dd {\mathscr{A}}_I = \rho_I '(y) (\dd \phi + A_\mathcal{J}) + 2 \pi \rho_I (y) c_1(\mathcal{J})\, . 
 \ee
 Comparing with (\ref{c1FIspindle}) leads us to identify $\rho_I (y)$ precisely with our moment maps, namely 
 \be
 \rho_I (y) =  - 2\pi  {\flp}_I^{a} \mu_a \, . 
 \ee
 In our 
 notation\footnote{We identify their $ {\flp}^i $ with our  $-\mathfrak{n}_I$. }, the functions $\rho_I(y)$ satisfy 
  \cite{Hosseini:2021fge}
 \begin{equation}
\rho_I (y_1) =  \frac{1}{2} \mathfrak{n}_I - \frac{1}{4} r^I_0 \left( \frac{1}{n_1}+\frac{1}{n_2}\right) \, , \qquad  \rho_I (y_2) =  -  \frac{1}{2}\mathfrak{n}_I - \frac{1}{4}r^I_0 \left( \frac{1}{n_1}+\frac{1}{n_2}\right) \, , 
\label{rhoIs}
\end{equation}
where $\sum_I r_0^I=2$, while  using (\ref{mamapfixspindle})
 we have 
 \be
 2\pi   {\flp}_I^{a} \mu_a  |_{y_b} = (-1)^b \frac{{\flp}_I^{b} }{n_b}\, ,
 \ee
implying that we must identify
\be
\label{r0Irational}
 \frac{r^I_0}{2}  \left( \frac{1}{n_1}+\frac{1}{n_2}\right) = -\frac{{\flp}_I^{1}}{n_1} + \frac{{\flp}_I^{2}}{n_2} \, . 
\ee
Summing over $I$ we then get 
\be
-\frac{\sigma^1}{n_1} +\frac{\sigma^2}{n_2}  = \sum_I \left(-\frac{{\flp}_I^{1}}{n_1} +   \frac{{\flp}_I^{2}}{n_2}\right)  =   \frac{1}{n_1}+\frac{1}{n_2} \, ,
\ee
consistently with 
$\sigma^1= -1$, $\sigma^2=1$, corresponding to the anti-twist   \cite{Hosseini:2021fge}. This shows that the constants $r_0^I$ 
 parameterise the redundancy in the relation
between the physical fluxes $\mathfrak{n}_I$ and the ${{\flp}_I^{a}}$. However, notice that (\ref{r0Irational}) implies that $r^I_0 \in \mathbb{Q}$.

\subsection{M5 branes on the spindle}

The anomaly polynomial of M5 brane 6-dimensional SCFTs 
 is an eight-form  defined on an eight-dimensional orbifold $Z_8$ that we take to be
the total space of a $\spindle$ fibration over an auxiliary space $B_6$,
\begin{equation}
\spindle \hookrightarrow Z_8 \to B_6\, .
\end{equation}
Neglecting terms that are sub-leading in the 
large $N$ limit, the eight-form anomaly polynomial is given by
\begin{equation}
\label{intanompolspindleM5}
{\cal A}_{6{\rm d}} = \frac{N^3}{24}p_2(R) = \frac{N^3}{24}c_1(F_1)^2c_1(F_2)^2  \, , 
\end{equation}
where $p_2(R)$ is the second Pontryagin class of the $SO(5)_R$ normal bundle to the M5-brane in the eleven-dimensional spacetime.
 The $F_I$, $I=1,2$ are the generators 
of $U(1)_1\times U(1)_2\subset SO(5)_R$ global symmetries that are preserved when the M5-brane is compactified on the spindle \cite{Ferrero:2021wvk}.
 The   $c_1(F_I)$ are the first Chern classes of the  line bundles on  $Z_8$ associated to these $U(1)_I$ symmetries and are decomposed as
 \begin{equation}
 \label{decomposeFIM5onspindle}
c_1(F_I) = \Delta_I c_1(F_R^{4{\rm d}}) -  \flp_I^a   c_1({\cal L}_a)\, .
\end{equation}

The anomaly polynomial of the four-dimensional theory is obtained integrating the anomaly polynomial the six-dimensional theory on the spindle, 
 \begin{equation}
{\cal A}_{4{\rm d}} =  \int_{\spindle}{\cal A}_{6{\rm d}}\, ,
\end{equation}
and is a six-form on $B_6$. 
We now substitute (\ref{decomposeFIM5onspindle}) in (\ref{intanompolspindleM5}), where $c_1({\cal L}_a)$ is as in eq. 
(\ref{c1FIspindle})
and $c_1(F_R^{4{\rm d}})\in H^2(B_6,\ZZ)$ is the first Chern class of the $4{\rm d}$ R-symmetry line bundle.
Formally setting  $c_1({\cal J}) = \epsilon c_1(F_R^{4{\rm d}})$
and extracting the coefficient of the six-form
 ${\cal A}_{4d}  = F (\Delta_I,\epsilon) c_1(F_R^{4{\rm d}})^3$ 
  leads 
to the  equivariant integral 
\begin{equation}
\label{2danomeqintegral}
F (\Delta_I,\epsilon) = \frac{N^3}{24}  \int_\spindle (\Delta_1 - \flp_1^a  c_1^{\mathbb{T}}(L_a)  )^2  (\Delta_2  - \flp_2^a  c_1^{\mathbb{T}}(L_a) )^2 \, ,
\end{equation}
which may be expanded as
\begin{align}
\label{anomM5spindleexpanded}
F  (\Delta_I,\epsilon)   & = \frac{N^3}{24} \big[ -2\Delta_1\Delta_2 ( \Delta_1\flp_2^a + \Delta_2\flp_1^a )D_a  + (\Delta_1^2 \flp_2^a \flp_2^b  + \Delta_2^2 \flp_1^a \flp_1^b + 4\Delta_1\Delta_2    \flp_1^a \flp_2^b )D_{ab}  \nonumber\\[1.5mm]
&\quad \qquad ~- 2 \flp_1^a(\Delta_1  \flp_2^b   + \Delta_2   \flp_1^b ) \flp_2^cD_{abc}+ \flp_1^a \flp_1^b \flp_2^c \flp_2^dD_{abcd}\big]\, ,
\end{align}
where the non-zero equivariant intersection numbers  are given in   (\ref{intersectionsspindle}). 
Alternatively, the same result can be obtained evaluating the equivariant integral (\ref{2danomeqintegral}) using the fixed point theorem (\ref{toric-equiv-integral0}). Specifically, we get
\be
\label{2danomfixedpoints2}
F (\Delta_I,\epsilon) = F^1  (\Delta_I,\epsilon) +  F^2 (\Delta_I,\epsilon)\, , 
\ee
where,
in terms of the ``trial 6d central charge"\footnote{There is no notion of trial central charge in a 6d SCFT, however we adopt this conventional definition, in analogy with the 4d case. }  
\be
a_{6{\rm d}} (\Delta_I) \equiv  \frac{N^3}{24} \Delta_1^2 \Delta_2^2 \, , 
\ee
we have
\be
F^a (\Delta_I,\epsilon)  = (-1)^a \frac{1}{\epsilon}a_{6{\rm d}} (\Delta_I^{a})\, ,\qquad \qquad  \Delta_I^{a} \equiv \Delta_I - (-1)^a \flp_I^a \frac{\epsilon}{n_a} \, . 
\ee
Notice that the terms singular in $\epsilon$  (quartic in the $\Delta_I$) cancel out, as expected.  The rest of the discussion proceeds exactly as for the D3 branes in the previous subsection.
In particular, supersymmetry implies that the constraints  (\ref{susyonspindles}) hold
and the physical fluxes $\mathfrak{n}_I$ obey
(\ref{physfluxesspindle}).
In terms of the variables $\varphi_I$ (\ref{nicevarphivariables}) the fixed point contributions read
\be
F^a (\varphi_I,\epsilon)  = (-1)^a \frac{1}{\epsilon}a_{6{\rm d}} (\varphi_I - (-1)^a \tfrac{\mathfrak{n}_I }{2} \epsilon) \, ,
\ee
which give the very simple expression
\begin{equation}
\label{veryniceFM5}
F  (\varphi_I,\epsilon) = -\frac{N^3}{48}  (\varphi_1 \mathfrak{n}_2 +\varphi_2 \mathfrak{n}_1 ) (4 \varphi_1 \varphi_2 +\mathfrak{n}_1 \mathfrak{n}_2\epsilon^2) \, , 
\end{equation}
in agreement with  \cite{Faedo:2021nub}.  

Let us briefly compare the results above with the corresponding  calculation in  \cite{Ferrero:2021wvk}. 
Again, the functions $\rho_I(y)$ used  in  \cite{Ferrero:2021wvk} should be identified with our moment maps as 
$ \rho_I (y) =  - 2\pi  {\flp}_I^{a} \mu_a $, and the resulting  ${\flp}_I^{a}$ are easily obtained from the expressions in eq. (A.2) of  \cite{Ferrero:2021wvk}.
However, due to the supergravity coordinates and the specific gauge used in  \cite{Ferrero:2021wvk}, these  are quadratic irrational functions of the spindle parameters and the physical fluxes and do not depend on any free parameter. One can check that the constraint  (\ref{susyonspindles}) on the ${\flp}_I^{a}$ is respected, with $\sigma^1=-1$ and $\sigma^2=-1$, consistently with the fact that 
the discussion in  \cite{Ferrero:2021wvk} concerns the twist 
case\footnote{We identify their $P_i $ with our  $-\mathfrak{n}_I$. }}. Notice that the form of (\ref{veryniceFM5}) is much simpler than the corresponding function computed  \cite{Ferrero:2021wvk}. In particular, in (\ref{veryniceFM5}) there are no
linear and cubic terms in $\epsilon$.

\subsection{M5 branes on 4d orbifolds}\label{sec:M5orb}

We now consider  M5-branes compactified on a four-dimensional toric 
orbifold\footnote{The M5 brane anomaly polynomial in an example of this type of compactification was studied in
 \cite{Cheung:2022ilc}.} as discussed in section \ref{sec:2d}.
The eight-form anomaly polynomial 
(\ref{intanompolspindleM5})
is defined on an 
 eight-dimensional orbifold $Z_8$ that we take to be
the total space of a $\Morb_4$ fibration over an auxiliary space $B_4$,
\begin{equation}
\Morb_4 \hookrightarrow Z_8 \to B_4\, , 
\end{equation}
and integrating it on $\Morb_4$ gives a four-form, that is the anomaly polynomial of a 2d SCFT:
 \begin{equation}
{\cal A}_{2{\rm d}} =  \int_{\Morb_4}{\cal A}_{6{\rm d}}\, .
\end{equation}
The $c_1(F_I)$ now decompose as 
\begin{equation}
\label{c1FImorb4}
c_1(F_I) = \Delta_I c_1(F_R^{2{\rm d}}) -  \flp_I^a   c_1({\cal L}_a)\, , \qquad c_1({\cal L}_a) = c_1({L}_a) + 2\pi \mu_{a}^i c_1({\cal J}_i)\, , 
\end{equation}
where  ${\cal J}_1, {\cal J}_2$ are the auxiliary line bundles with  $c_1({\cal J}_i)\in H^2(Z_4,\ZZ)$. 
The  $L_a$, $a=1,\dots, \fan$ are the  line bundles discussed in section \ref{sec:2d} and $c_1(F_R^{2{\rm d}})\in H^2(Z_4,\ZZ)$ is the first Chern class of the $2{\rm d}$ R-symmetry line bundle.

Substituting (\ref{c1FImorb4}) in (\ref{intanompolspindleM5}) 
and setting  $c_1({\cal J}_i) = \epsilon_i c_1(F_R^{2{\rm d}})$,
leads 
to the  equivariant integral 
\begin{equation}
\label{2danomeqintegralmorb4}
F (\Delta_I,\epsilon_i) = \frac{N^3}{24}  \int_{\Morb_4} (\Delta_1 - \flp_1^a  c_1^{\mathbb{T}}(L_a)  )^2  (\Delta_2  - \flp_2^a  c_1^{\mathbb{T}}(L_a) )^2 \, ,
\end{equation}
wihch has exactly the same form of (\ref{2danomeqintegral})! Indeed, expanding it, this gives  again  (\ref{anomM5spindleexpanded}), where now $D_a=0$ 
and the non-zero equivariant intersection numbers can be read off from (\ref{intersections2}),  (\ref{intersections3}), and  (\ref{intersections4}), respectively.
On the other hand, employing the fixed point theorem, we  can write (\ref{2danomeqintegralmorb4}) as a sum of contributions over the fixed points, 
namely\footnote{Notice that there is no summation on $a$ in the expressions inside the parenthesis.}
\begin{equation}
\label{matchFFMexp}
F (\Delta_I,\epsilon_i) = \frac{N^3}{24} \sum_{a=1}^n \frac{1}{d_{a,a+1}\epsilon^{a}_1  \epsilon^{a}_2} \left( \Delta_1  - \flp_1^a  \epsilon_1^a  - \flp_1^{a+1}\epsilon_2^a  \right)^2\left( \Delta_2  - \flp_2^a  \epsilon_1^a  - \flp_2^{a+1}\epsilon_2^a  \right)^2\, ,
\end{equation}
in agreement with the formula conjectured in  \cite{Faedo:2022rqx}.
In terms of the gravitational blocks we  have
\be
F (\Delta_I,\epsilon_i) = \sum_{a=1}^n  F^a  (\Delta_I,\epsilon_i) \, , \qquad \quad F^a   (\Delta_I,\epsilon_i)=  \frac{1}{d_{a,a+1} \epsilon^{a}_1  \epsilon^{a}_2}  a_{6{\rm d}} (\Delta_I^a)\, ,
\ee
with 
\be
\Delta_I^a = \Delta_I  - \flp_I^a  \epsilon_1^a  - \flp_I^{a+1}\epsilon_2^a\, . 
\ee
The physical fluxes are now defined by integrating the first Chern classes of the flavour line bundles, $c_1(E_I)=- \flp_I^a c_1(L_a)$, over the various divisors, namely
\be
\mathfrak{q}_I^a\equiv -  \int_{D_a}  c_1(E_I) = \flp_I^bD_{ab}\, .
\ee
However, recall that this relation cannot be inverted and therefore it is not manifest that 
(\ref{matchFFMexp}) depends only on the physical fluxes.
That this is true was proved in  \cite{Faedo:2022rqx}, as we now  recall, translating the arguments to  our current notation\footnote{Recall that in this paper we denote the non-primitive, ``long'', vectors of the fan by $v^a$, the corresponding divisors as $D_a$, and their intersection matrix as $D_{ab}$. While in  \cite{Faedo:2022rqx} these were denoted by $\hat v^a$,  $\hat D_a$, and $\hat D_{a,b}$, respectively. Similarly,  $(\flp_I^a)_\mathrm{here} =  (m_a \flp_I^a)_\mathrm{there}$ and  $(\mathfrak{q}_I^a)_\mathrm{here} =(\mathfrak{q}_I^a/m_a)_\mathrm{there}$.}. From (\ref{linreldivisors}) it immediately follows that 
\be 
 \sum_{a=1}^\fan v^a D_{ab} =0\qquad \Rightarrow \qquad\sum_{a=1}^\fan v^a   \mathfrak{q}_I^a  =0 \,,
  \ee
therefore, only $\fan - 2$
physical fluxes are linearly independent. In particular, the  ``gauge'' transformation 
\be
\label{gaugetrans}
\flp_I^a \, \to \, \tilde \flp_I^a= \flp_I^a + \det (\lambda_I,v^a) \, , 
\ee
where $\lambda_I\in \RR^2$ are arbitrary two-dimensional constant vectors\footnote{Notice that these $\lambda_I$ have nothing to do with the K\"ahler parameters $\lambda_a$ appearing elsewhere.}, leaves the physical fluxes  $\mathfrak{q}_I^a$ invariant. 
In order to discuss 
how this  transformation
affects (\ref{matchFFMexp}) we will assume\footnote{This holds in all known examples of supergravity solution corresponding to M5 branes wrapped on four-dimensional toric orbifolds
 \cite{Cheung:2022ilc,Faedo:2022rqx,Couzens:2022lvg,FFM2}. It should be possible to prove this, along the lines of the analysis in \cite{Ferrero:2021etw}.}, following \cite{Faedo:2022rqx}, that supersymmetry requires  the  background R-symmetry gauge field to have   first Chern class
\be\label{susysigma}
c_1(E_R) =-  \sum_{a=1}^\fan\sigma^a c_1(L_a) \, ,
\ee
where $\sigma^a=\pm 1$, generalizing the twist and anti-twist for the spindle \cite{Ferrero:2021etw}. This corresponds to 
 the following constraints 
 \be
\Delta_1 + \Delta_2 = 2  +\det (W,\epsilon)\, , \qquad \flp_1^a + \flp_2^a = \sigma^a + \det (W,v^a)   \, , 
\ee
 where $\epsilon = (\epsilon_1,\epsilon_2)$ and $W\in \RR^2$ is a two-dimensional constant vector
transforming as 
\be
W \, \to\,  \tilde W = W +\lambda_1 + \lambda_2\, 
\ee
under  (\ref{gaugetrans}). 
Performing this transformation in  (\ref{matchFFMexp}) 
implies that the variables $\Delta_I^a $ change as
\be
\Delta_I^a \, \to   \Delta_I  - \tilde \flp_I^a  \epsilon_1^a  - \tilde \flp_I^{a+1}\epsilon_2^a =   \Delta_I  -  \flp_I^a  \epsilon_1^a  -  \flp_I^{a+1}\epsilon_2^a - \det (\lambda_I ,\epsilon)\, , 
\ee
thus 
\begin{equation}
F (\Delta_I,\epsilon_i) \to F (\tilde \Delta_I,\epsilon_i)\, , \qquad \tilde \Delta_I \equiv \Delta_I   - \det (\lambda_I ,\epsilon)\, , 
\end{equation}
where $\tilde \Delta_I$ obey the same constraint as the $\Delta_I$, namely 
\be
\tilde \Delta_1 + \tilde \Delta_2 = 2  +\det (\tilde W,\epsilon)-  \det (\lambda_1 ,\epsilon)- \det (\lambda_2 ,\epsilon) = 2+\det (W,\epsilon)\, . 
\ee
This is  exactly the same function as the initial one, completing the proof.

\section{Non-compact Calabi-Yau singularities}\label{sec:CY}

The geometry of many type II and M theory solutions arising from branes can be modelled on singular Calabi-Yau cones.
In this section we consider the equivariant volume of partial resolutions of non-compact (asymptotically conical) Calabi-Yau singularities and provide 
many applications to holography.

Consider a non-compact toric  Calabi-Yau  $X$ of complex dimension $m$   defined by a fan with primitive vectors $v^a$, $a=1,\ldots, \fan$. The Calabi-Yau condition requires the vectors $v^a$ to lie on a plane. We choose an $\mathrm{SL}(m;\mathbb{Z})$ basis  where the first component of all the vectors $v^a$ is one
\be\label{CYplane} v^a = (1, w^a) \, \qquad w^a\in \mathbb{Z}^{m-1}\, . \ee
Notice that \eqref{chargesGLSM}  implies that the GLSM charges $Q^A_a$ are traceless 
\be \sum_{a=1}^\fan Q^A_a =0 \, .\ee

We assume that $X$ is the (partial) resolution of a Calabi-Yau conical singularity, as in some of the examples in section \ref{sec:noncompactexamples}.
This means that the convex rational polyhedron 
\be \mathcal{P} =\{ l_a(y)= v^a_i y_i -\lambda_a \ge 0 \, \qquad a=1,\ldots, \fan \} \, ,\ee
is asymptotically a cone. More precisely, we assume that  the large $y_i$ approximation of the polyhedron
\be \mathcal{P}^\prime =\{ l_{a_k}(y)= v^{a_k}_i y_i \ge 0 \, \qquad k=1,\ldots, \fan^\prime \} \, ,\ee
is  a non-compact cone with a single vertex in $y=0$ and $\fan^\prime$ facets.
Notice that only the non-compact facets of  $\mathcal{P}$ are relevant for the asymptotic behaviour and, in general $\fan^\prime \le \fan$.
The vectors $v^{a_k}$,  $k=1,\ldots, \fan^\prime$, define the fan of the singular cone $X_{sing}$ of which $X$ is a resolution. 
The fan of $X_{sing}$ consists of a single cone and  it corresponds to a singularity  which is, in general, not of orbifold type. 

The resolution  replaces the singularity of $\mathcal{P}^\prime$ with compact cycles of real dimension $2 m-2$, corresponding  to the bounded facets of $\mathcal{P}$, as well as sub-cycles of smaller dimensions.\footnote{The example of the conifold discussed in section \ref{resconif} is special in that there are no compact four-cycles. This is due to the fact we just subdivide the fan without introducing new vectors.} We assume that $\mathcal{P}$ has only orbifold singularities. From the dual point of view, the fan of $X$ is the union of $m$-dimensional cones, each corresponding to a vertex of $\mathcal{P}$ and therefore to a fixed point. Each  cone is specified by a choice of $m$ adjacent vectors $(v^{a_1},\ldots, v^{a_m})$. We use the notation $A=(a_1,\ldots, a_m)$ to identify the set of such cones and we assume that there are $\fix$ of them.

Using the results of section \ref{sec:fixed},  the equivariant volume is given by
\be \label{fpm} \evol(\lambda_a,\epsilon_i)=\sum_{A=(a_i,\ldots,a_m)} \frac{\ex^{-\epsilon \cdot (\lambda_{a_1}  u^{a_1}_A+\ldots+\lambda_{a_m}  u^{a_m}_A)/d_{a_1,\ldots,a_m}}}{d_{a_1,\ldots ,a_m}(\frac{\epsilon \cdot u^{a_1}_A}{d_{a_1,\ldots,a_m}}) \ldots(\frac{\epsilon \cdot u^{a_m}_A}{d_{a_1,\ldots,a_m}}) } \, ,\ee
where $A$ runs over the $m$-dimensional cones of the resolution, $d_{a_1,\ldots,a_m}= | \det (v^{a_1},,\ldots, v^{a_m})|$, and $u^a_A$ are the inward normal to the facets of $A$ defined in section \ref{sec:fixed}.

\subsection{GK geometry and the GMS master volume}

In this section we explore some general properties of the equivariant volume for non-compact Calabi-Yau singularities and its relation with other {\it volumes} appearing
in the literature in similar contexts, like the the Sasakian volume of \cite{Martelli:2005tp,Martelli:2006yb} and the  master volume introduced in \cite{Gauntlett:2018dpc}. 

Consider the formal expansion
\be \label{expV} \evol(\lambda_a,\epsilon_i)=\sum_{k=0}^\infty \evol^{(k)}(\lambda_a,\epsilon_i) \, ,\ee
where $\evol^{(k)}(\lambda_a,\epsilon_i)$ is homogeneous of degree $k$ in $\lambda_a$. We start by observing some general properties of the homogeneous quantities $\evol^{(k)}(\lambda_a,\epsilon_i)$. From \eqref{1der}, by matching degrees in $\lambda_a$, we have
\be \label{1derh} \sum_{a=1}^\fan v^a_i \frac{\partial \evol^{(k)}}{\partial \lambda_a}   = -\epsilon_i \evol^{(k-1)} \, .\ee
By considering the case $i=1$ and the fact that, for Calabi-Yaus, $v_1^a=1$, we also have
\be \label{1derh1} \sum_{a=1}^\fan  \frac{\partial \evol^{(k)}}{\partial \lambda_a}   = -\epsilon_1 \evol^{(k-1)} \, .\ee
We can combine \eqref{1derh} and  \eqref{1derh1} into
\be \label{1derMV} \sum_{a=1}^\fan \frac{\partial \evol^{(k)}}{\partial \lambda_a} (\epsilon_1 v^a_i  -\epsilon_i) =0 \, .\ee
 It follows from this equations that $\evol^{(k)}$ are invariant
 under the gauge transformation
 \be\label{gfreed} \lambda_a \rightarrow \lambda_a + \sum_{i=1}^m \gamma_i (\epsilon_1 v^a_i -\epsilon_i  ) \, ,\ee
 which allows to eliminate  $m-1$\footnote{$i=1$ is trivial.} unphysical $\lambda_a$ that do not correspond to non-trivial co-homology classes.

In the compact case, the terms with $k<m$ in the expansion  \eqref{expV} are identically zero. Indeed they can be written as in \eqref{bbbbb} and they are integrals of forms of degree less than $2 m$. As we already discussed  in section \ref{sec:non-comp}, in the non-compact case this is not true and
also the terms with $k<m$ are not vanishing. They are rational functions of $\epsilon$  encoding some interesting geometrical information that we will now elucidate.

It is important to observe that $\evol^{(k)}(\lambda_a,\epsilon_i)$ with $k<m$ only depends on the  $\fan^\prime$ K\"ahler parameters $\lambda_a$ associated with the singular fan and the non-compact directions. The K\"ahler parameter $\lambda_a$ associated with the compact directions start contributing at order $m$ in the expansion. This can be   understood as follows.
The compact K\"ahler parameters  are associated with the bounded facets of the original polytope. We can always modify the polytope by adding new facets and making it compact. We can also assume that the bounded facets of the original polytope are not modified by this operation.   The compact K\"ahler parameter enters in the fixed point formula only in the terms associated with the vertices of the bounded facets and, therefore, their contribution to $\evol$ for the old and new  polytope is the same. Since the new polytope is compact, this contribution must start at order $m$.

The term of zero degree is simply
\be\label{evolMSY} \evol(0,\epsilon_i) = \frac{1}{(2 \pi)^m}\int_{X} \ex^{-H}  \frac{\omega^m}{m!} =\int_{\mathcal{P}^\prime} \ex^{-\epsilon_i y_i} \dd y_1\ldots \dd y_m \, \ee
and it computes the equivariant volume of the singular Calabi-Yau cone $X_{sing}$, or equivalently, the regularized volume of the polyhedron $\mathcal{P}^\prime$. At $\lambda_a=0$, the
metric of $X_{sing}$ becomes conical
\be \dd s^2(X) = \dd r^2 + r^2 \dd s^2(Y)\, .\ee
A choice of K\"ahler metric exhibits $X_{sing}$ as a cone over a Sasakian manifold $Y$ of real dimension $2 m -1$. As shows in \cite{Martelli:2005tp,Martelli:2006yb} there is family
of Sasakian metrics parameterized by the Reeb vector $\xi =\epsilon_i \partial_{\phi_i}$. They correspond to different choices of the
radial coordinate $\frac{r^2}{2}=H$, where $H=\epsilon_i y_i$ is the Hamiltonian for $\xi$. 
Expression \eqref{evolMSY} then reduces, up to a numerical factor, to the Sasakian volume of Y \cite{Martelli:2005tp,Martelli:2006yb}
\be \boxed{{\rm vol}[Y](\epsilon_i) = \frac{2\pi^m}{(m-1)!}\evol(0,\epsilon_i) }\, .\ee
The specialization of \eqref{fpm} to $\lambda_a=0$ in the smooth case
\be\label{MSY} \evol(0,\epsilon_i)=\sum_{A=(a_i,\ldots,a_m)} \frac{1}{d_{a_1,\ldots ,a_m}(\frac{\epsilon \cdot u^{a_1}_A}{d_{a_1,\ldots,a_m}}) \ldots(\frac{\epsilon \cdot u^{a_m}_A}{d_{a_1,\ldots,a_m}}) } \, ,\ee
was derived indeed in \cite{Martelli:2006yb}.

The term of degree $m-1$ is also interesting. It coincides, up to a numerical factor, with the master volume introduced in \cite{Gauntlett:2018dpc}. To define the master volume, one
foliates the $2m-1$ base $Y$  as
\be \dd s^2(Y) = \eta^2 +  \dd s^2_{2m-2}\, ,\ee
where $\eta$ is the dual one form to $\xi$ ($i_\xi \eta =1$) and the metric $\dd s^2_{2m-2}$ is conformally K\"ahler with K\"ahler form $\omega_{B}$. We turn on the $\fan^\prime$ K\"ahler parameters $\lambda_a$  associated with the non-compact facets of the polyhedron by letting $\omega_{B}$ vary \cite{Gauntlett:2018dpc}
\begin{equation}
\label{kform2}
\frac{[\omega_B]}{2\pi} = -\sum_a \lambda_a c_a \, ,
\end{equation}
where $c_a$ are the co-homology classes that uplift to $c_1(L_a)$ on the Calabi-Yau $X_{sing}$.  Notice that  $[\dd \eta] = 2\pi \sum_a  c_a$ \cite{Gauntlett:2018dpc}.
The master volume is then defined as
\be\label{GMS} \mathcal{V}(\lambda_a,\epsilon_i) = \int_Y \eta \wedge \frac{\omega_B^{m-1}}{(m-1)!} = \frac{(-2\pi)^{m-1}} {(m-1)!} \sum_{a_1,\ldots, a_{m-1}} \lambda_{a_1} \cdots \lambda_{a_{m-1}} \int_Y \eta \wedge c_{a_1} \wedge \cdots \wedge c_{a_{m-1}}\, .\ee
For example, for $m=3$, the master volume reads \cite{Gauntlett:2018dpc}
\be\label{masvol}  \mathcal{V} (\lambda_a,\epsilon_i) = \frac{(2\pi)^3}{2!} \sum_{a=1}^{\fan^\prime} \frac{\lambda_a \left ( \lambda_{a-1} \det (v^a,v^{a+1},\epsilon) -\lambda_{a} \det (v^{a-1},v^{a+1},\epsilon)+\lambda_{a+1} \det (v^{a-1},v^{a},\epsilon)\right)}{\det (v^{a-1},v^{a},\epsilon)\det (v^a,v^{a+1},\epsilon)}\ ,\ee
where the vectors $v^a$ with $a=1, \ldots, \fan^\prime$ runs over the fan of the singular cone $X_{sing}$. 

Notice that, in the original approach of \cite{Gauntlett:2018dpc}, the master volume is defined by considering conical non-K\"ahler metrics on $X$.  
In our approach instead, we give up the conical condition on $X$ and we use a K\"ahler metric. Despite the difference of approaches, we
can recover the master volume as the term of degree $m-1$ in the equivariant volume 
 \be\label{GKv}  \boxed{\mathcal{V} (\lambda_a,\epsilon_i)  =  (2\pi)^m  \evol^{(m-1)}(\lambda_a,\epsilon_i)} \, ,\ee
 where we stress that 
 \be \evol^{(m-1)} =\frac{1}{(m-1)!}\sum_{a_1,\ldots,a_{m-1}=1}^{\fan^\prime}  \frac{\partial^{m-1} \evol}{\partial \lambda_{a_1} \cdots \partial \lambda_{a_{m-1}}} \Big |_{\lambda=0} \lambda_{a_1} \cdots \lambda_{a_{m-1}}\, ,\ee is a function only of the   K\"ahler parameters $\lambda_a$, $a=1, \ldots, \fan^\prime$,  associated with the fan of the singular cone.
 The identity \eqref{GKv} can be checked by direct computation. The case $m=3$ is explicitly worked out in section \ref{masterform4d_appendix}.
By differentiating $m-1$ times  equation \eqref{1der} for $i=1$
\be \label{1derh22} \sum_{a_1=1}^\fan  \frac{\partial \evol}{\partial \lambda_{a_1}}   = -\epsilon_1 \evol\, ,\ee
 setting $\lambda_a=0$ and multiplying by $\lambda_{a_2} \ldots \lambda_{a_m}$, we can also rewrite \eqref{GKv} as
 \be\label{GKv2}  \mathcal{V} (\lambda_a,\epsilon_i) =-\frac{(2\pi)^m  }{\epsilon_1 (m-1)!}\sum_{a_1,\cdots,a_{m}=1}^{\fan^\prime}  \frac{\partial^m \evol}{\partial \lambda_{a_1} \ldots \partial \lambda_{a_{m}}} \Big |_{\lambda=0} \lambda_{a_1} \cdots \lambda_{a_{m-1}} \, ,\ee
where the analogy with \eqref{GMS} is manifest.

 As observed in \cite{Gauntlett:2018dpc}, the master volume reduces to the Sasakian volume when all the K\"ahler parameters are equal. We can understand this statement
 in our formalism as follows
 \be  \evol^{(0)} (\epsilon_i) = \frac{(-1)^{m-1}}{\epsilon_1^{m-1}} \sum_{a_1,\dots, a_{m-1}=1}^{\fan^\prime}  \frac{\partial^n {\evol}}{\partial \lambda_{a_1} \cdots \partial \lambda_{a_{m-1}}} \Big |_{\lambda =0} =   \frac{(m-1)!}{(-\epsilon_1)^{m-1} (2\pi)^{m}}  \mathcal{V} (\lambda_a=1,\epsilon_i)\, , \ee
 where we used \eqref{nderid} with all $i_k=1$ and $\lambda_a=0$ and equation \eqref{GKv}.
 
 The master volume plays an important role in supergravity solutions based on GK geometry \cite{Couzens:2018wnk,Gauntlett:2018dpc}. In this context the so-called {\it supersymmetric action}  
\be \label{GKsusy} S_{\rm SUSY} =-\sum_{a=1}^d \frac{\partial \mathcal{V}}{\partial \lambda_a} \, , \ee
plays an even more important role. It is the object that needs to be extremized in order to find a solution of the equation of motions.
Using \eqref{1derh1} we find
\be  \label{GKsusy3} \boxed{S_{\rm SUSY} = \epsilon_1 (2\pi)^m  \evol^{(m-2)}(\lambda_a,\epsilon_i)} \, .\ee

\subsection{Gravitational blocks from the equivariant volume}

In this section we use the equivariant volume to study type II and M theory branes compactified on a spindle.
Supersymmetry is preserved either with a topological twist or an antitwist \cite{Ferrero:2021etw}. From the field theory point of view, 
we consider the case where a superconformal field theory compactified on a spindle flows in the IR to a
superconformal quantum mechanics or to a two-dimensional superconformal field theory. The corresponding
supergravity solutions describing the IR limit have a geometry AdS$_2\times Z$ or AdS$_3\times Z$, respectively,
and can be interpreted as  the near-horizon geometry of black holes or black strings.
 
The local geometry of such brane systems   can be modelled in terms of  CY $m$-folds. We can describe
many D-branes and M-branes configurations in terms of a formal fibration
\be\label{fibre} {\rm CY}_m \hookrightarrow {\rm CY}_{m+1} \to \spindle \, ,\ee
where the CY$_m$ encode the geometry and the information about the original higher-dimensional CFT.

 It is often useful to define an off-shell free energy $F(\Delta_I,\epsilon)$, or \emph{extremal function}\footnote{Often also called entropy functions, even if the related physical observable is not the entropy of a black hole.}, depending on chemical potentials $\Delta_I$ for the continuous global 
 symmetries of the higher-dimensional CFT and an equivariant parameter $\epsilon$ for the rotation along the spindle, whose extremization
gives the entropy of the black hole  or the central charge of the two-dimensional CFT.  All these extremal functions should arise by evaluating the supergravity action on supersymmetric solutions that obey 
a subset of  the equations of motion. 
The extremization with respect to $\Delta_I$ and $\epsilon$
is then equivalent to imposing the remaining equations of motion. Explicit examples of this construction are given in  \cite{Martelli:2005tp,Gauntlett:2018dpc}.   Extremal functions of known black holes and black strings
can be expressed in terms  of  gravitational blocks \cite{Hosseini:2019iad}. The  general form of the extremal functions for branes compactified on a spindle in this context was proposed in \cite{Faedo:2021nub}, covering also cases where the explicit
computation of the supergravity  action from first principles is still missing. The characteristic form of the off-shell free energy $F$ is given by a gluing 

\be\label{glue} \boxed{F (\Delta_I,\epsilon) = \frac{1}{\epsilon} \left ( {\cal F} (\Delta^+_I)\pm {\cal F} (\Delta^-_I)\right )} \ee
where the block ${\cal F}$ encodes some universal properties of the higher-dimensional SCFT and it is related to the geometry of CY$_m$, and the gluing  depends
on the details of the fibration \eqref{fibre}. For D3 and M5 branes, ${\cal F}$ is related to the central charge of the higher-dimensional CFT. For other types of branes, ${\cal F}$ is the sphere-free energy of the higher-dimensional CFT at large $N$ \cite{Hosseini:2019iad,Faedo:2021nub}.

In this section we focus on the gravitational interpretation of \eqref{glue}.  In the case of M2 and D3 branes, the off-shell free energy can be expressed in terms of the supersymmetric action in the formalism of  GK geometry and the decomposition \eqref{glue}
was explicitly proved in \cite{Boido:2022mbe} from the gravitational point of view. We will show how to recover and reinterpret this results in terms of the equivariant volume.
In the case of D2, D4 and M5 brane systems, we cannot apply the GK formalism but we will propose a possible and intriguing extension.

\subsubsection{The geometry of CY$_m$ fibred over the spindle}
 
 As a preliminary, in this section we derive a general expression for the equivariant volume of the fibration \eqref{fibre}.
 Consider  a CY cone $m$-fold defined by $m$-dimensional  vectors $v^a$, with $v^a_1=1$, and the $(m+1)$-dimensional toric geometry specified by the fan \cite{Boido:2022mbe}
\be  V^a=(0,v^a)\, ,\qquad V^+ = (n_+,w_+) \, ,\qquad V^- = (-\sigma n_-,w_-) \, ,\ee
where 
$\sigma =\pm 1$. This is a fibration over a spindle $\mathbb{WP}^1_{[n_+,n_-]}$ specified by the  vectors $w_\pm$. Notice that the $(m+1)$-dimensional geometry is still a Calabi-Yau if the first component of the vectors $w_\pm$ is one, $w_{\pm 1}=1$, which we will assume. As shown in \cite{Boido:2022mbe}, this geometry
explicitly appears in the gravity solutions of M2 and D3-branes compactified on a spindle with
\be w_+=(1,-a_+ \vec p)\, ,\qquad w_-=(1,- \sigma a_- \vec p)\, ,\ee
where $a_- n_+ + a_+ n_- =1$ and $\sigma =\pm 1$ and $\vec p$ is a ($m-1$)-dimensional vector. In this context, supersymmetry is preserved with a twist ($\sigma=1)$ or an anti-twist  ($\sigma=-1)$.  Notice also that, in the anti-twist case, the toric diagram is not convex and it does not strictly define a toric geometry. We will nevertheless proceed also in this case, considering it as an extrapolation from the twist case. 

The  fixed point formula is
\be \evol_{CY_{m+1}}(\lambda_a,\epsilon_i)=\sum_{(a_1,\ldots ,a_{m+1})\in A} \frac{\ex^{-\epsilon_{(m+1)} \cdot (\sum_{i=1}^{m+1} \lambda_{a_i}  U^{a_i})/d_{a_1,\ldots,a_{m+1}} } }{d_{a_1,\ldots,a_{m+1}}\prod_{i=1}^{m+1} (\frac{\epsilon_{(m+1)} \cdot U^{a_i}}{d_{a_1,\ldots,a_{m+1}}}) } \, ,\ee
where $A$ runs over the polyhedral cones of a resolution and  $\epsilon_{(m+1)}=(\epsilon_0,\epsilon_1,\ldots,\epsilon_m)$. For ease of notation, we drop the label $A$ from the normal vectors $U^a$. We can choose a resolution for the CY$_m$ by subdividing the $m$-dimensional fan. We then obtain a resolution of the CY$_{m+1}$ by considering polyhedra where $V^+$ and $V^-$ are added to the $m$-dimensional cones $(V^{a_1}, \ldots, V^{a_m})$. Let's assume also $d_{a_1,\ldots,a_{m}}=1$ for the CY$_m$. The inward normals to the tetrahedra are 
\bea &(V^{a_1}, \ldots, V^{a_m},V^+) \, \qquad \rightarrow \qquad &U^{a_i}=(- u^{a_i}\cdot w_+, n_+ u^{a_i}) \, ,\qquad\,\,\,\,\,\, &U^{+} =(1,0,\ldots,0)\, , \\
&(V^{a_1}, \ldots, V^{a_m},V^-) \, \qquad \rightarrow  \qquad &U^{a_i}=-( -u^{a_i}\cdot w_-, -\sigma n_- u^{a_i}) \, ,\qquad &U^{-} =(-1,0,\ldots,0)\, , \eea
where $u^{a_i}$ are the inward normals to the CY$_m$ cones, we identified $a_{m+1}=\pm$ and $\sigma=-1$ is obtained by analytic continuation.

The contribution of the tetrahedra with vertex $V^+$ is
 \bea 
&\sum_{(a_1,\ldots,a_3)\in A} \frac{\ex^{-( \epsilon_{(m)} -\epsilon_0 w_+/n_+) \cdot (\sum_{i=1}^m \lambda_{a_i}  u^{a_i})  -\epsilon_0 \lambda_{+}/n_+}}{\epsilon_0\prod_{i=1}^m ( \epsilon_{(m)} -\frac{\epsilon_0 w_+}{n_+}) \cdot u^{a_i}    } \\
&= \frac{1}{\epsilon_0}\evol_{CY_{m}} \left(\lambda_a +\frac{\epsilon_0}{n_+ \epsilon_1  -\epsilon_0} \lambda_+ \, , \epsilon_{(m)} -\frac{\epsilon_0 w_+}{n_+} \right)\, , 
 \eea
 where  $\epsilon_{(m)}=(\epsilon_1,\ldots,\epsilon_m)$,  
 and we used $d_{a_1,\ldots,a_{m+1}}=n_+$ and
 the identity among normals $\sum_i u^{a_i}=(1,0,\ldots,0)$.\footnote{Consider for simplicity $m=3$. Assuming an order such that $\det (v^1,v^2,v^3)=1$, $\sum_i u^{a_i} = v^2\wedge v^3+v^3\wedge v^1+v^1\wedge v^2$. Let $e_i$ be the canonical basis in $\mathbb{R}^3$. Then $\sum_i u^{a_i}\cdot e_{2,3}=0$. For example $\sum_i u^{a_i}\cdot e_{2}=\det (e_2,v^2,v^3) +\det (e_2,v^3,v^1)+\det (e_2,v^1,v^2)=v^3_3-v^2_3+v^1_3-v^3_3 +v^2_3-v^1_3=0$ where we used $v^a_1=1$. On the other hand, $\sum_i u^{a_i}\cdot e_{1}=\det (e_1,v^2,v^3) +\det (e_1,v^3,v^1)+\det (e_1,v^1,v^2) =1$ since it is the sum of the areas of three triangular cones obtained by triangulating $(v^1,v^2,v^3)$ with the insertion of $e_1$ (it lies in the same plane). The sum of the three areas is the area of the original cone $d_{1,2,3}=1$.} 
  
 The contribution of the tetrahedra of $V^-$ is obtained by replacing $w_+$ with $w_-$, $n_+$ with $-\sigma n_-$ and $\lambda_4\equiv \lambda_+$ with a new variable $\lambda_-$. The signs in $U^{a_i}$ are compensated by $d_{a_1,a_2,a_3,a_4}=-(-\sigma n_-)$ (we work for positive $\sigma$ and analitically continue the result), which
 also brings an overall extra sign. 

  The final result is 
 \be
 \label{FIN} 
\boxed{ \evol_{CY_{m+1}} = \frac{1}{\epsilon_0}\evol_{CY_{m}} \left(\lambda^+_a \, , \epsilon^+_i  \right) -\frac{1}{\epsilon_0}\evol_{CY_{m}} \left( \lambda^-_a \, , \epsilon^-_i  \right) }\, \ee
where
\bea &\epsilon^+= \epsilon_{(m)} -\frac{\epsilon_0 w_+}{n_+}\, , \qquad &\lambda_a^+=\lambda_a +\frac{\epsilon_0}{n_+ \epsilon_1  -\epsilon_0} \lambda_+ \,, \\
&\epsilon^-=\epsilon_{(m)} +\frac{\epsilon_0 w_-}{\sigma n_-}\, , \qquad\,\,\,\,\,\,\,\,\,\,\,\, &\lambda_a^-=\lambda_a -\frac{\epsilon_0}{\sigma n_- \epsilon_1  +\epsilon_0} \lambda_- \, ,\eea
and we see that $\evol_{CY_{m+1}} $ can be obtained by gluing two copies of $\evol_{CY_{m}}$.

\subsubsection{The case of D3 branes}

The case of D3 and M2 branes can be described in terms of GK geometry \cite{Gauntlett:2018dpc}. We consider a system of branes
 sitting at the tip of a conical toric Calabi-Yau three-fold singularity (CY$_{m}$) with Sasaki-Einstein base $Y_{2m-1}$  and further compactified on a spindle. The dual supergravity solution has  an AdS$_2 \times Z_9$ and   AdS$_3 \times Z_7$ near horizon geometry, for M2 and D3 branes respectively. The internal manifolds $Z_{2m+1}$  are obtained by fibering the Sasaki-Einstein base $Y_{2m-1}$ over the spindle.
Supersymmetry requires that the  cone $C(Z_{2m+1})$ is topologically a CY$_{m+1}$, although the supergravity metric is not Ricci-flat. 

The supergravity solution can be described in terms of  a family of backgrounds that depends on the equivariant parameters $\epsilon_i$ and the K\"ahler parameters
 $\lambda_\alpha$ of the CY$_{m+1}$ \cite{Gauntlett:2018dpc}.  To avoid confusion, we will use Greek
 letters to label the vectors in the $m+1$-dimensional  fan and the associated K\"ahler parameters.  The conditions for supersymmetry can be compactly written as
 \be\label{cGK}
\sum_{\alpha} \frac{\partial S_{\rm SUSY}}{\partial\lambda_\alpha} =0 \, , \qquad \nu_m \, M_\alpha = - \frac{\partial S_{\rm SUSY}}{\partial\lambda_\alpha} \, ,\ee
where $S_{\rm SUSY}$ is the supersymmetric action \eqref{GKsusy} of the CY$_{m+1}$ 
\be \label{GKsusy2} 
S_{\rm SUSY} =\epsilon_1 (2\pi)^{m+1} \evol_{CY_{m+1}}^{(m-1)}(\lambda_\alpha,\epsilon_i) \, , 
\ee
and $M_\alpha$ are integer fluxes, encoding the flux quantization conditions of the M-theory four-form or the type IIB RR five-form. $\nu_m$ is a normalization constant that depends on the 
dimension. The fluxes $M_\alpha$ contain the information about the number of branes $N$ and the topological details of the Sasaki-Einstein fibration over the spindle.
Notice that for consistency of \eqref{cGK} with \eqref{1derMV}  we must have
\be \label{cflux}
\sum_{\alpha} V^\alpha M_\alpha =0 \, ,
\ee
where $V^\alpha$ are the vectors in the fan of  the CY$_{m+1}$.\footnote{In our examples, the index $\alpha$ will split into an index $a$ associated 
with  the CY$_{m}$ fan and the two indices $\pm$ associated with the spindle.} 
Given \eqref{1derMV} and \eqref{cflux}, only $\fan-m+1$ equations in \eqref{cGK}  are actually independent. Combining the equations \eqref{cGK}  with the $m-1$ gauge invariances \eqref{gfreed}, we can eliminate all the $\lambda_\alpha$. We are left with a functional of the equivariant parameters $\epsilon_i$ that needs to be extremized in order to find the 
 solution of the equations of motion. 
 
 Let us now specialize to the case of D3-branes where $m=3$. The solution AdS$_3 \times Z_7$ is dual to a two-dimensional CFT. In this context,
 the Killing vector $\xi =\sum_{i=1}^m \epsilon_i \partial_{\phi_i}$ is interpreted as the R-symmetry of the dual CFT and the extremization of the supersymmetric action is the geometrical dual of $c$-extremization \cite{Benini:2013cda} in two-dimensional CFTs  \cite{Couzens:2018wnk,Gauntlett:2018dpc}.
 
 More explicitly, the on-shell value of the supersymmetric action, up to a normalization coefficient,  is the exact central charge of the two-dimensional CFT, while the off-shell value of $S_{\rm SUSY}$ as a function of $\epsilon_i$ after imposing \eqref{cGK} equals the  charge $c$ as a function of a trial R-symmetry. This has been proved explicitly in \cite{Hosseini:2019use} for the case of a compactification on $S^2$, or on a Riemann surface,  and in \cite{Boido:2022mbe} for a spindle. More precisely, $S_{\rm SUSY}$ as a function of $\epsilon_i$ coincides with the trial $c$-function after  the baryonic directions in the trial R-symmetry have been extremized.\footnote{This is similar to what happens for the Sasakian volume and the trial $a$-charge for D3 branes sitting at the tip of a conical toric Calabi-Yau three-fold singularity \cite{Martelli:2005tp,Butti:2005vn}.} 
 It is also shown in \cite{Boido:2022mbe} how to write the  supersymmetric action as the gluing of two gravitational blocks \cite{Hosseini:2019iad}. We now recover this result in our formalism.
 
We can extract the supersymmetric action  from the gluing  formula \eqref{FIN}
 
 \be\label{FIN0} \evol_{CY_4} = \frac{1}{\epsilon_0}\evol_{CY_3} \left( \lambda^+_a \, ,\epsilon^+_i   \right) -\frac{1}{\epsilon_0}\evol_{CY_3} \left( \lambda^-_a \, ,\epsilon^-_i  \right) \, ,\ee
where
\bea &\epsilon^+= \epsilon_{(3)} -\frac{\epsilon_0 w_+}{n_+}\, , \qquad &\lambda_a^+=\lambda_a +\frac{\epsilon_0}{n_+ \epsilon_1  -\epsilon_0} \lambda_+ \, , \\
&\epsilon^-=\epsilon_{(3)} +\frac{\epsilon_0 w_-}{\sigma n_-}\, , \qquad\,\,\,\,\,\,\,\,\,\,\,\, &\lambda_a^-=\lambda_a -\frac{\epsilon_0}{\sigma n_- \epsilon_1  +\epsilon_0} \lambda_- \, .\eea

Notice that the redefinitions are homogeneous in $\lambda_\alpha$. This means that the previous identity can be easily truncated at a given order in $\lambda_\alpha$. Taking the quadratic piece 
and using \eqref{GKv} and \eqref{GKsusy3} we obtain
 \be\label{SS} S_{\rm SUSY}|_{CY_4} =2\pi \frac{\epsilon_1}{\epsilon_0} \left (\mathcal{V}_{CY_3}( \lambda^+_a \, ,\epsilon^+_i )  -\mathcal{V}_{CY_3}(\lambda^-_a \, ,\epsilon^-_i) \right )  \, , \ee
 since the supersymmetric action is the quadratic part of $\evol_{CY_4}$ and the 3d master volume is the quadratic part of $\evol_{CY_3}$.
 We thus recover the factorization in gravitational blocks derived in \cite{Boido:2022mbe} (see for example (7.27) and (7.31) in that paper).
 
We can be more explicit in  the special case of $S^5$ and the dual $\mathcal{N}=4$ SYM. This example is discussed in \cite{Boido:2022mbe} in a different gauge. The vectors are
\be v^1=(1,0,0)\, ,\qquad v^2=(1,1,0)\, ,\qquad v^3 =(1,0,1) \, ,\ee
and the equivariant and master volume read
\be \label{evC3} \evol_{\mathbb{C}^3}=\frac{\ex^{-(\epsilon_1-\epsilon_2-\epsilon_3) \lambda_1-\epsilon_2 \lambda_2-\epsilon_3 \lambda_3}}{(\epsilon_1-\epsilon_2-\epsilon_3)\epsilon_2\epsilon_3}\, ,\qquad \mathcal{V}_{\mathbb{C}^3}= \frac{(2\pi)^3((\epsilon_1-\epsilon_2-\epsilon_3) \lambda_1+\epsilon_2 \lambda_2+\epsilon_3 \lambda_3)^2}{2(\epsilon_1-\epsilon_2-\epsilon_3)\epsilon_2\epsilon_3}\, .\ee
We parameterize the fluxes $M_\alpha$ as 
\be\label{fluxN4} (M_1,M_2,M_3,M_+,M_-)\equiv  N (  \mathfrak{n}_1, \mathfrak{n}_2, \mathfrak{n}_3, \tfrac{1}{n_+},\tfrac{1}{\sigma n_-})\, ,\ee
where the constraint \eqref{cflux} further requires  
\be \sum_{I=1}^3 \mathfrak{n}_I = -\frac{1}{n_+}-\frac{1}{\sigma n_-}\, ,\,\,\,\quad  \mathfrak{n}_2=-\frac{w_{+2}}{n_+}-\frac{w_{-2}}{\sigma n_-}
\, ,\,\,\, \quad \mathfrak{n}_3=-\frac{w_{+3}}{n_+}-\frac{w_{-3}}{\sigma n_-}
\, .\ee
We will be cavalier about the normalization of the fluxes in the following, but it is clear than $N$ should be proportional to the number of colors of the dual theory.
We can use the gauge freedom \eqref{gfreed} to set $\lambda_1=\lambda_2=\lambda_3=0$. The master volume evaluated at the two poles in this gauge reads
\be 
\mathcal{V}_{\mathbb{C}^3}(\lambda^\pm \, ,\epsilon^\pm_i ) = \frac{(2\pi)^3 \epsilon_0^2 \lambda_\pm^2}{2 n_\pm^2(\epsilon_1^\pm-\epsilon_2^\pm-\epsilon_3^\pm)\epsilon_2^\pm\epsilon_3^\pm} \, ,\ee
 and we can find $\lambda_{\pm}$ by solving the $\alpha=\pm$ components of the second equation in \eqref{cGK} (recall that only two of them are independent)
 \be -\nu_3 N= \frac{(2\pi)^4 \epsilon_1 \epsilon_0 \lambda_+}{ n_+(\epsilon_1^+-\epsilon_2^+-\epsilon_3^+)\epsilon_2^+\epsilon_3^+} \, ,\qquad -\nu_3 \sigma N= -\frac{(2\pi)^4 \epsilon_1 \epsilon_0 \lambda_-}{ n_-(\epsilon_1^--\epsilon_2^--\epsilon_3^-)\epsilon_2^-\epsilon_3^-} \, .\ee
 Defining
\be \label{epsN4} \epsilon_0=\epsilon \, ,\quad \epsilon_1=\Delta_1+\Delta_2+\Delta_3\, ,\quad \epsilon_2=\Delta_2\, ,\quad \epsilon_3=\Delta_3\, ,\ee
the extremization in \cite{Gauntlett:2018dpc} must be done under the condition $\epsilon_1=2/(m-3)=2$ which correponds to  
\be\label{constrN4} \Delta_1+\Delta_2+\Delta_3=2\, .\ee 
We then see that $\Delta_I$, with $I=1,2,3$,  parameterize the R-charges of the three chiral fields  of $\mathcal{N}=4$ SYM.
The supersymmetric action \eqref{SS} is then obtained by gluing blocks
\be S_{\rm SUSY}|_{CY_4}  = \frac{ {\cal F}(\Delta_1^+,\Delta_2^+,\Delta_3^+)}{\epsilon} -\frac{{\cal F}( \Delta_1^-,\Delta_2^-,\Delta_3^- )}{\epsilon} \, ,\ee 
with the appropriate function   for ${\cal N}=4$ SYM  \cite{Hosseini:2019iad}
\be {\cal F}(\Delta_1,\Delta_2,\Delta_3) = \frac{\nu_3^2 }{64 \pi^4} N^2 \Delta_1 \Delta_2 \Delta_3 \, ,\ee
and
\bea\label{N4spindle} & \Delta_1^+=\Delta_1 -\frac{\epsilon}{n_+} (1- \sum_{I=2}^3 w_{+I})\, ,\qquad \Delta_2^+=\Delta_2 -\frac{\epsilon}{n_+} w_{+2} \, , \qquad \Delta_3^+=\Delta_3 -\frac{\epsilon}{n_+} w_{+3} \,,  \\
&\Delta_1^-=\Delta_1 +\frac{\epsilon}{\sigma n_-} (1- \sum_{I=2}^3 w_{-I}) \, ,\,\,\,\,\, \Delta_2^-=\Delta_2 +\frac{\epsilon}{\sigma n_-} w_{-2}\, ,\,\,\,\,\, \Delta_3^-=\Delta_3 +\frac{\epsilon}{\sigma n_-} w_{-3}  \, .\eea

We can see that this result reproduces the anomaly polynomial computed in section \ref{anomal:sec}. In doing so, we need to be careful that there is some ambiguity in identifying the chemical potentials associated with the R-symmetry in \eqref{epsN4}. Any redefinition $\Delta_I\rightarrow \Delta_I+\delta_I\epsilon$ with $\sum_{I=1}^3 \delta_I=0$ would respect the constraint \eqref{constrN4} and would lead to a potentially good choice of R-symmetry chemical potential. To avoid this ambiguity, we can compare the objects that are invariant under
this redefinition
\bea \sum_{I=1}^3 \Delta_I^+= 2  -\frac{\epsilon}{n_+}\, ,\qquad \sum_{I=1}^3 \Delta_I^-= 2  +\frac{\epsilon}{\sigma n_-} \, ,\qquad \Delta_I^+ - \Delta_I^- =  \epsilon\,  \mathfrak{n}_I\, .\eea
It is then easy to see that the same relations hold for the anomaly as written in  \eqref{2danomfixedpoints} with $\sigma^1=-1$ and $\sigma^2=-\sigma$.

Notice that the supersymmetric action scales like $N^{2}$ as expected for $\mathcal{N}=4$ SYM. This can be understood from the fact that $S_{\rm SUSY}|_{CY_4}$ is quadratic 
in $\lambda_\alpha$ and, from \eqref{cGK}, $\lambda_\alpha$ scales linearly with $N$.

\subsubsection{The case of M2 branes}

We consider now the case of M2 sitting at the tip of a conical toric CY$_4$ with Sasaki-Einstein base $Y_{7}$  and further compactified on a spindle. The dual M theory solutions correspond to 4d black holes  and have an AdS$_2 \times Z_9$ near horizon geometry.  We can again describe the system in terms of GK geometry. The supersymmetric action provides an entropy functional for the black hole.  The construction is the gravitational dual of
$\cal{I}$-extremization \cite{Benini:2015eyy} and has been applied  to the case of a compactification on $S^2$  in \cite{Hosseini:2019ddy,Gauntlett:2019roi,Kim:2019umc}\footnote{Or, more generally, in the case of a compactification on a Riemann surface.} and in the case of a compactification on the spindle in \cite{Boido:2022iye,Boido:2022mbe}.

From the gluing formula \eqref{FIN} we find
\be\label{FIN00} \evol_{CY_5} = \frac{1}{\epsilon_0}\evol_{CY_4} \left(\lambda^+ _a \, ,\epsilon^+_i  \right) -\frac{1}{\epsilon_0}\evol_{CY_4} \left(\lambda^-_a \, ,\epsilon^-_i \right) \, ,\ee
where
\bea &\epsilon^+= \epsilon_{(4)} -\frac{\epsilon_0 w_+}{n_+}\, , \qquad &\lambda_a^+=\lambda_a +\frac{\epsilon_0}{n_+ \epsilon_1  -\epsilon_0} \lambda_+ \, , \\
&\epsilon^-=\epsilon_{(4)} +\frac{\epsilon_0 w_-}{\sigma n_-}\, , \qquad\,\,\,\,\,\,\,\,\,\,\,\, &\lambda_a^-=\lambda_a -\frac{\epsilon_0}{\sigma n_- \epsilon_1  +\epsilon_0} \lambda_- \, .\eea
 Taking the piece cubic in $\lambda$ we recover the result  in \cite{Boido:2022mbe}
\be\label{SS0} S_{\rm SUSY}|_{CY_5} =2\pi \frac{\epsilon_1}{\epsilon_0} \left (\mathcal{V}_{CY_4}(\lambda^+_a\, ,\epsilon^+_i )  -\mathcal{V}_{CY_4}( \lambda^- _a\, ,\epsilon^-_i) \right )  \, . \ee 
 
To simplify the discussion we restrict to the case $Z_7=S^7$
which is dual to the ABJM theory with $k=1$ compactified on the spindle. The vectors are
\be v^1=(1,0,0,0)\, ,\qquad v^2=(1,1,0,0)\, ,\qquad v^3 =(1,0,1,0) \, , \qquad v^4=(1,0,0,1) \, ,\ee
and the equivariant  volume is
\be \evol_{\mathbb{C}^4}=\frac{\ex^{-(\epsilon_1-\epsilon_2-\epsilon_3-\epsilon_4) \lambda_1-\epsilon_2 \lambda_2-\epsilon_3 \lambda_3-\epsilon_4 \lambda_4}}{(\epsilon_1-\epsilon_2-\epsilon_3-\epsilon_4)\epsilon_2\epsilon_3\epsilon_4}\, 
.\ee
We parameterize the fluxes $M_\alpha$ as  
\be (M_1,M_2,M_3,M_4,M_+,M_-)\equiv N ( \mathfrak{n}_1, \mathfrak{n}_2, \mathfrak{n}_3, \mathfrak{n}_4, \tfrac{1}{n_+},\tfrac{1}{\sigma n_-})\, ,\ee
where the constraint \eqref{cflux} further requires 
\be \sum_{I=1}^4 \mathfrak{n}_I = -\frac{1}{n_+}-\frac{1}{\sigma n_-}\, ,\,\,\,\qquad  \mathfrak{n}_{I}=-\frac{w_{+I}}{n_+}-\frac{w_{-I}}{\sigma n_-}
\, ,\,\,\, I=2,3,4 \, ,\ee
and  $N$ will be related to the number of colors of the dual theory.
We use again the gauge freedom \eqref{gfreed} to set $\lambda_1=\lambda_2=\lambda_3=\lambda_4=0$ and we determine $\lambda_{\pm}$  solving \eqref{cGK}. 
This time, the master volume evaluated at the two poles in this gauge reads
\bea 
\mathcal{V}_{\mathbb{C}^4}(\lambda^+\, ,\epsilon^+_i ) &= \frac{(2\pi)^4 \epsilon_0^3 (- \lambda_+)^3}{6 n_+^3(\epsilon_1^+-\epsilon_2^+-\epsilon_3^+-\epsilon_4^+)\epsilon_2^+\epsilon_3^+ \epsilon_4^+} \, ,\\ 
\mathcal{V}_{\mathbb{C}^4}(\lambda^-\, ,\epsilon^-_i ) &= \frac{(2\pi)^4 \epsilon_0^3 ( \lambda_-)^3}{6 (\sigma n_-)^3(\epsilon_1^--\epsilon_2^--\epsilon_3^--\epsilon_4^-)\epsilon_2^-\epsilon_3^-\epsilon_4^-} \, , \eea
 and we can find $\lambda_{\pm}$ by solving the $\alpha=\pm$ components of equation \eqref{cGK} (recall that only two of them are independent)
 \be\label{fluxesGKM2} \nu_4 N= \frac{(2\pi)^5 \epsilon_1\epsilon_0^2 \lambda_+^2}{ 2n_+^2(\epsilon_1^+-\epsilon_2^+-\epsilon_3^+-\epsilon_4^+)\epsilon_2^+\epsilon_3^+\epsilon_4^+} \, ,\,\, \nu_4  N= \frac{(2\pi)^5 \epsilon_1\epsilon_0^2 \lambda_-^2}{ 2 n_-^2(\epsilon_1^--\epsilon_2^--\epsilon_3^--\epsilon_4^-)\epsilon_2^-\epsilon_3^-\epsilon_4^-} \, .\ee
There is a sign ambiguity in solving \eqref{fluxesGKM2} that we fix to match the gravitational entropy function result and the explicit geometric analysis in \cite{Boido:2022mbe}.\footnote{ We see from \eqref{fluxesGKM2} that  $(\epsilon_1^\pm-\epsilon_2^\pm-\epsilon_3^\pm-\epsilon_4^\pm)\epsilon_2^\pm\epsilon_3^\pm\epsilon_4^\pm$ should be positive. We then choose the solution $\lambda_+<0,\lambda_->0$ which 
is consistent with  \cite{Boido:2022mbe} -- for example, one can compare with (7.16) and (7.26) in that reference.}
Define
\be \epsilon_0=\frac{\epsilon}{2} \, ,\quad \epsilon_1=\frac{\Delta_1+\Delta_2+\Delta_3+\Delta_4}{2}\, ,\quad \epsilon_2=\frac{\Delta_2}{2}\, ,\quad \epsilon_3=\frac{\Delta_3}{2}\, ,\quad \epsilon_4=\frac{\Delta_4}{2} \, .\ee
The extremization in \cite{Gauntlett:2018dpc} must be done under the condition $\epsilon_1=2/(m-3)=1$ which corresponds to  
\be \Delta_1+\Delta_2+\Delta_3+\Delta_4=2 \, .\ee 
We then see that $\Delta_I$, with $I=1,2,3,4$, parameterize the R-charges of the four chiral fields  of ABJM. 
The supersymmetric action \eqref{SS} is then given by gluing blocks
\be\label{CY5gluing}  S_{\rm SUSY}|_{CY_5}  = \frac{ {\cal F}(\Delta_1^+,\Delta_2^+,\Delta_3^+,\Delta_4^+)}{\epsilon} -\sigma \frac{{\cal F}( \Delta_1^-,\Delta_2^-,\Delta_3^- ,\Delta_4^-)}{\epsilon} \, ,\ee 
with the appropriate funnction for the ABJM theory \cite{Hosseini:2019iad}
\be {\cal F}(\Delta_1,\Delta_2,\Delta_3,\Delta_4) = \frac{\nu_4^{3/2}}{24 \pi^{5/2}} N^{3/2} \sqrt{\Delta_1\Delta_2\Delta_3 \Delta_4} \, ,\ee
and

\bea\label{ABJMspindle} 
& \Delta_1^+=\Delta_1 -\frac{\epsilon}{n_+} (1- \sum_{I=2}^4 w_{+I}) \, ,\,\,\, \qquad \Delta_{I}^+=\Delta_{I} -\frac{\epsilon}{n_+} w_{+I} \, ,\,\,\,\,\,\, I=2,3,4\, , \\
& \Delta_1^-=\Delta_1 +\frac{\epsilon}{\sigma n_-} (1- \sum_{I=2}^4 w_{+I}) \, , \,\,\,\qquad  \Delta_{I}^-=\Delta_{I} +\frac{\epsilon}{\sigma n_-} w_{-I} \, , \,\,\, \,\,\,I=2,3,4
 \, .\eea
Notice that
\bea \sum_{I=1}^4 \Delta_I^+= 2  -\frac{\epsilon}{n_+}\, ,\,\,\,\, \sum_{I=1}^4 \Delta_I^-= 2  +\frac{\epsilon}{\sigma n_-} \, ,\,\,\,\, \Delta_I^+ - \Delta_I^- =  \epsilon\,  \mathfrak{n}_I\,, \qquad I=1,2,3,4 \, .\eea
This formula matches the entropy function found in \cite{Faedo:2021nub,Boido:2022mbe}.\footnote{ To compare with (5.22) in \cite{Faedo:2021nub} one needs to identify 
 $\mathfrak{n}_I^{\text{here}}=-\mathfrak{n}_i^{\text{there}}$, $n_\pm^{\text{here}}=n_\mp^{\text{there}}$  and $2\epsilon^{\text{there}}=-\epsilon$. The chemical potentials are related by the redefinition $\varphi_I=\Delta_I -\frac{\epsilon}{n_+} w_{+I} -\frac{\epsilon}{2} \mathfrak{n}_I$ for $I=2,3,4$ and $\varphi_1=\Delta_1 -\frac{\epsilon}{n_+} (1-w_{+2}-w_{+3}-w_{+4}) -\frac{\epsilon}{2} \mathfrak{n}_1$. }

Notice that the entropy function scales like $N^{3/2}$ as expected for the ABJM theory. The reason is that $S_{\rm SUSY}|_{CY_5}$ is cubic in $\lambda_\alpha$ and, from 
\eqref{cGK}, $\lambda_\alpha$ scales as $\sqrt{N}$. Notice also that the gluing \eqref{CY5gluing} involves a different relative sign for the twist and anti-twist. This is typical of M2 and D4 branes and it is not present for D3 and M5 branes \cite{Faedo:2021nub}.

\subsubsection{The case of M5 branes}\label{sec:M5}

The case of M5 branes compactified on a spindle has no known description in terms of GK geometry. However, a stack of M5 branes in flat space in M theory probes a transverse geometry
that is $\mathbb{C}^2 \times \mathbb{R}$ suggesting that the relevant CY$_2$ is simply $\mathbb{C}^2$. In the associated supergravity solution AdS$_7\times S^4$, the base $S^3$ of the cone $\mathbb{C}^2$ is fibred over a real direction to give the $S^4$. Notice that there is a supersymmetric generalization where the M5 branes probes a  $\mathbb{C}^2/\mathbb{Z}_p \times \mathbb{R}$ and the CY$_2$ could be replaced by $\mathbb{C}^2/\mathbb{Z}_p$.

 We now consider the CY$_3$ consisting of $\mathbb{C}^2$  fibered over the 
 spindle with fan
\bea V^1=(0,1,0)\, ,\qquad  V^2=(0,1,1)\, ,\qquad &V^+ = (n_+,w_+) \, ,\qquad V^- = (-\sigma n_-,w_-) \, ,
\eea
with 
\bea
w_+=(1,p_+)\, ,\qquad \quad w_-=(1,p_-)\, .\eea
From the gluing formula \eqref{FIN} we have
 \be\label{FIN2} \evol_{CY_3} = \frac{1}{\epsilon_0}\evol_{\mathbb{C}^2} \left(\lambda^+_a\, ,\epsilon^+_i   \right) -\frac{1}{\epsilon_0}\evol_{\mathbb{C}^2} \left(\lambda^-\, ,\epsilon^-_i  \right) \, ,\ee
 where
\bea &\epsilon^+= \epsilon_{(2)} -\frac{\epsilon_0 w_+}{n_+}\, , \qquad &\lambda_a^+=\lambda_a +\frac{\epsilon_0}{n_+ \epsilon_1  -\epsilon_0} \lambda_+ \, ,\\
&\epsilon^-=\epsilon_{(2)} +\frac{\epsilon_0 w_-}{\sigma n_-}\, , \qquad\,\,\,\,\,\,\,\,\,\,\,\, &\lambda_a^-=\lambda_a -\frac{\epsilon_0}{\sigma n_- \epsilon_1  +\epsilon_0} \lambda_- \, ,\eea
 and  the equivariant volume of $\mathbb{C}^2$ can be read from  \eqref{correctVeqC2Zp} with $p=q=1$
\be\label{evC2} \evol_{\mathbb{C}^2} =  \frac{\ex^{- \lambda_1(\epsilon_1-\epsilon_2)-\lambda_2 \epsilon_2}}{\epsilon_2(\epsilon_1-\epsilon_2)} \, .\ee

In analogy with $S_{\rm SUSY}$ in the context of GK geometry, we expect to be able to extract an off-shell free energy $\Fext$,  
for the spindly M5 branes AdS$_5$ solutions found in \cite{Ferrero:2021wvk} from the equivariant volume of $\mathbb{C}^2$ fibred over the spindle. We now show that 
this is the case by an appropriate generalization of the  equations \eqref{cGK}. A scaling argument for $N$ suggests that the correct generalization is
\be \nu_{M5} M_\alpha=-\frac{\partial \evol_{CY_3}^{(2)}}{\partial \lambda_\alpha} \, ,\qquad  \Fext= \evol_{CY_3}^{(3)} \, , \qquad \sum_\alpha M_\alpha V^\alpha=0 \, ,\ee
where the flux equation uses the quadratic piece but the extremal function
is the cubic piece of the equivariant volume. In this way, $\lambda_\alpha$ is linear in $N$ and the extremal function
$\Fext$ cubic in $N$, as expected for M5 branes. Here and in the following sub-sections we will not be concerned with precise normalizations.

Following the same logic as in previous sections, we solve the constraint on the fluxes
 \be\label{fluxC2} M_\alpha =N (\mathfrak{n}_1,\mathfrak{n}_2, \tfrac{1}{n_+}, \tfrac{1}{\sigma n_-}) \, , \ee
  with $\mathfrak{n}_1+\mathfrak{n}_2=- \frac{1}{n_+}- \frac{1}{\sigma n_-}$ and define
\be\label{epsC2}  \epsilon_0=\frac{\epsilon}{2} \, ,\quad \epsilon_1=\frac{\Delta_1+\Delta_2}{2}\, ,\quad \epsilon_2=\frac{\Delta_2}{2} ,\ee
with $\Delta_1+\Delta_2=2$. The chemical potentials $\Delta_1$ and $\Delta_2$ can be associated with the Cartan subgroup of the $SO(5)$ R-symmetry of the $(2,0)$ theory.
 We find that, after solving for $\lambda_\pm$ in the gauge $\lambda_1=\lambda_2=0$, the extremal function is given by gluing two blocks
\be \Fext  = \frac{ {\cal F}(\Delta_1^+,\Delta_2^+)}{\epsilon} -\frac{{\cal F}( \Delta_1^-,\Delta_2^-)}{\epsilon}\, ,\ee 
with the appropriate function for the $(2,0)$ theory \cite{Hosseini:2019iad}
\be {\cal F}(\Delta_1,\Delta_2) = \frac{\nu_{M5}^3}{48} N^3  (\Delta_1 \Delta_2)^2 \, ,\ee
and
\bea\label{20spindle} & \Delta_1^+=\Delta_1 -\frac{\epsilon}{n_+} (1- p_+)\, ,\qquad \Delta_2^+=\Delta_2 -\frac{\epsilon}{n_+} p_+ \, , \\
& \Delta_1^-=\Delta_1 +\frac{\epsilon}{\sigma n_-} (1- p_-) \, ,\qquad \Delta_2^-=\Delta_2 +\frac{\epsilon}{\sigma n_-} p_-\, .\eea
Notice that
\bea \sum_{I=1}^2 \Delta_I^+= 2  -\frac{\epsilon}{n_+}\, ,\qquad \sum_{I=1}^2 \Delta_I^-= 2  +\frac{\epsilon}{\sigma n_-} \, ,\qquad \Delta_I^+ - \Delta_I^- =  \epsilon\,  \mathfrak{n}_I\, .\eea
We see that we have  reproduced the $(2,0)$ anomaly integrated on the spindle \eqref{2danomfixedpoints2} with $\sigma^1=-1$ and $\sigma^2=-\sigma$.

\subsubsection{The case of D4 branes}\label{sec:D4}

We now turn to the case of D4 branes compactified on a spinde.  The massive type IIA  supergravity solution associated with a system of D4 and D8 branes is topologically AdS$_6$ times an hemisphere $S^4$ \cite{Brandhuber:1999np} suggesting that the relevant CY$_2$ is again $\mathbb{C}^2$. Supergravity solutions with an AdS$_4$ factor corresponding to compactifications of the dual 5d SCFT on a spindle  have been found in \cite{Faedo:2021nub,Giri:2021xta}.

It si intriguing to observe that we can again extract an off-shell free energy for these spindly  solutions from a generalization of the  equations \eqref{cGK} valid for GK geometry.
We consider as before the equivariant volume \eqref{FIN2} of $\mathbb{C}^2$ fibred over the spindle. 
This time we take
\be \nu_{D4}  M_\alpha=-\frac{\partial \evol_{CY_3}^{(3)}}{\partial \lambda_\alpha} \, ,\qquad  \Fext= \evol_{CY_3}^{(5)} \, , \qquad \sum_\alpha M_\alpha V^\alpha =0 \, ,\ee
so that $\lambda_\alpha \sim N^{1/2}$ and $\Fext \sim N^{5/2}$ as expected for D4 branes.

We use the same definitions \eqref{fluxC2} and \eqref{epsC2} of the previous section. After solving for $\lambda_\pm$ in the gauge $\lambda_1=\lambda_2=0$, the extremal function is given by\footnote{We made a choice of determination for the fractional power that is similar to the M2 brane case and correctly reproduce the supergravity result.}
\be \Fext  = \frac{ {\cal F}(\Delta_1^+,\Delta_2^+)}{\epsilon} -\sigma \frac{{\cal F}( \Delta_1^-,\Delta_2^-)}{\epsilon}\, ,\ee 
where  for the 5d SCFT \cite{Hosseini:2019iad}
\be {\cal F}(\Delta_1,\Delta_2) =\frac{\nu_{D4}^{5/2}}{ 60\sqrt{2}} N^{5/2}  (\Delta_1 \Delta_2)^{3/2} \, ,\ee
and
\bea\label{20spindle0} & \Delta_1^+=\Delta_1 -\frac{\epsilon}{n_+} (1- p_+)\, ,\qquad \Delta_2^+=\Delta_2 -\frac{\epsilon}{n_+} p_+ \, , \\
& \Delta_1^-=\Delta_1 +\frac{\epsilon}{\sigma n_-} (1- p_-) \, ,\qquad \Delta_2^-=\Delta_2 +\frac{\epsilon}{\sigma n_-} p_-\, .\eea
Notice that
\bea \sum_{I=1}^2 \Delta_I^+= 2  -\frac{\epsilon}{n_+}\, ,\qquad \sum_{I=1}^2 \Delta_I^-= 2  +\frac{\epsilon}{\sigma n_-} \, ,\qquad \Delta_I^+ - \Delta_I^- =  \epsilon\,  \mathfrak{n}_I\, .\eea
We have reproduced the entropy function in \cite{Faedo:2021nub,Giri:2021xta}.\footnote{  To compare with (5.33) in \cite{Faedo:2021nub} one needs to identify  $\mathfrak{n}_I^{\text{here}}=-\mathfrak{n}_i^{\text{there}}$, $n_\pm^{\text{here}}=n_\mp^{\text{there}}$  and $2\epsilon^{\text{there}}=-\epsilon$. The chemical potentials are related by the redefinition  $\varphi_1=\Delta_1 -\frac{\epsilon}{n_+} (1-p_{+}) -\frac{\epsilon}{2} \mathfrak{n}_1$ and $\varphi_2=\Delta_2 -\frac{\epsilon}{n_+} p_{+} -\frac{\epsilon}{2} \mathfrak{n}_2$.}

\subsubsection{The case of D2 branes}

We finally consider the case of D2 branes compactified on a spinde.  The massive type IIA  supergravity solution associated with a system of D2-branes  is topologically AdS$_4\times X_6$, where the internal manifold is a Sasaki-Einstein five-manifold foliated over a segment \cite{Guarino:2015jca,Fluder:2015eoa}.
 Supergravity solutions with an AdS$_2$ factor corresponding to compactifications of the dual 3d SCFT on a spindle  have been found in \cite{Couzens:2022yiv}. 
 
 We expect that the corresponding extremal function is related to the equivariant volume of a CY$_3$ fibred over the spindle. A scaling argument suggests to use
\be\label{cGKD2} \nu_{D2} M_\alpha=-\frac{\partial \evol_{CY_4}^{(4)}}{\partial \lambda_\alpha} \, ,\qquad  \Fext = \evol_{CY_4}^{(5)} \, , \qquad \sum_\alpha M_\alpha V^\alpha=0 \, ,\ee
so that $\lambda_\alpha \sim N^{1/3}$ and $\Fext\sim N^{5/3}$ as expected for D2 branes.

We can verify that this is the case for the solution  associated with $\mathbb{C}^3$ where the three-dimensional SCFT is a pure ${\cal N}=2$ Chern-Simons theory with the same matter content and superpotential of the maximal supersymmetric Yang-Mills theory \cite{Guarino:2015jca} that we indicate as D2$_k$. We have
\be\label{FIN000} \evol_{CY_4} = \frac{1}{\epsilon_0}\evol_{\mathbb{C}^3} \left( \lambda^+_a\, ,\epsilon^+_i   \right) -\frac{1}{\epsilon_0}\evol_{\mathbb{C}^3} \left(\lambda^-_a  \, ,\epsilon^-_i\right) \, ,\ee
where $\evol_{\mathbb{C}^3}$ is given  in \eqref{evC3}. The fluxes can be written as in \eqref{fluxN4}. Since the matter content of the theory is the same as for ${\cal N}=4$ SYM we can still use the parameterization
\be \label{epsD2} \epsilon_0=\epsilon \, ,\quad \epsilon_1=\Delta_1+\Delta_2+\Delta_3\, ,\quad \epsilon_2=\Delta_2\, ,\quad \epsilon_3=\Delta_3\, ,\ee
with $\Delta_1+\Delta_2+\Delta_3=2$, and interpret the  $\Delta_I$, with $I=1,2,3$, as the R-charges of the three chiral fields  of the theory.

We can use the gauge freedom \eqref{gfreed} to set $\lambda_1=\lambda_2=\lambda_3=0$ and we can find $\lambda_{\pm}$ by solving the first equation in \eqref{cGKD2}.
The extremal function \eqref{cGKD2} is then obtained by gluing two blocks
\be \Fext= \frac{ {\cal F}(\Delta_1^+,\Delta_2^+,\Delta_3^+)}{\epsilon} -\frac{{\cal F}( \Delta_1^-,\Delta_2^-,\Delta_3^- )}{\epsilon}\, ,\ee 
where  for the D2$_k$ theory 
\be {\cal F}(\Delta_1,\Delta_2,\Delta_3) \propto N^{5/3} ( \Delta_1 \Delta_2 \Delta_3)^{2/3} \, ,\ee  
and
\bea\label{D2spindle} & \Delta_1^+=\Delta_1 -\frac{\epsilon}{n_+} (1- \sum_{I=2}^3 w_{+I}) \, ,\qquad \Delta_2^+=\Delta_2 -\frac{\epsilon}{n_+} w_{+2} \, ,\qquad \Delta_3^+=\Delta_3 -\frac{\epsilon}{n_+} w_{+3}  \, , \\
& \Delta_1^-=\Delta_1 +\frac{\epsilon}{\sigma n_-} (1- \sum_{I=2}^3 w_{-I}) \, ,\,\,\,\,\, \Delta_2^-=\Delta_2 +\frac{\epsilon}{\sigma n_-} w_{-2}\, ,\,\,\,\,\, \Delta_3^-=\Delta_3 +\frac{\epsilon}{\sigma n_-} w_{-3}  \, .\eea
Notice that
\bea \sum_{I=1}^3 \Delta_I^+= 2  -\frac{\epsilon}{n_+}\, ,\qquad \sum_{I=1}^3 \Delta_I^-= 2  +\frac{\epsilon}{\sigma n_-} \, ,\qquad \Delta_I^+ - \Delta_I^- =  \epsilon\,  \mathfrak{n}_I\, .\eea
This is the natural expectation for the extremal function for a D2 brane theory. It would be interesting to have a more complete analysis in supergravity to compare with.

\section{Discussion}
\label{sec:discuss}

In this paper we provided several  applications of equivariant localization to quantum field theory and holography. 
In particular, we have shown that the equivariant volume, a basic and well-studied geometrical object in symplectic geometry, is at the heart of many constructions characterising supersymmetric geometries. It generalises at once   the Sasakian volume of \cite{Martelli:2005tp,Martelli:2006yb} and the  master volume introduced in \cite{Gauntlett:2018dpc} that have been proven very useful in the study of superconformal field theories dual to branes sitting at Calabi-Yau singularities, and generalizations thereof. The corresponding quantum (or K-theoretical)  version, the equivariant index-character, is analogously crucial for studying the Hilbert series of the moduli spaces of the corresponding  SCFTs and we expect that it still has many surprises to unveil.

More specifically, we proposed  that the equivariant volume should be  the key object for all the extremal problems characterising supersymmetric geometries with a holographic interpretation, in different supergravity theories. We showed in this paper that all the extremization problems associated with compactifications on the spindle of M2 and D3 brane at Calabi-Yau singularities as well as  M5, D4 and D2 brane configurations in flat space,  can be re-expressed in terms of the equivariant volume. In the case of M2 and D3 branes it was known before that the extremization problem can be expressed in terms of the master volume  \cite{Gauntlett:2018dpc} and we showed how this object is 
encoded in the equivariant volume of the associated fibered Calabi-Yaus. 
In the M5, D4 and D2 branes there is not yet an analogous of the construction in \cite{Gauntlett:2018dpc}, but we showed that the relevant extremal functions can be written in terms of
the equivariant volume. We leave for future work the analysis of  more complicated situations, like M5 and D4 branes compactified on a four-dimensional orbifolds $\Morb_4$ or D2 branes
in massive type IIA associated with a generic Sasaki-Einstein five-manifold foliated over a segment \cite{Fluder:2015eoa}. These examples are more complicated but we are confident that the corresponding extremal functions and extremization problems can be written and formulated solely in terms of the equivariant volume.

In this paper we discussed the factorization \eqref{glue} of the extremal functions associated to
spindly black objects in terms of gravitational blocks both from the gravitational and the field theory point of views. On the geometry side, in \emph{all} cases  the 
factorization is an immediate corollary of the fixed-point localization formula for the equivariant volume, applied to the relevant non-compact geometry -- see eq.  (\ref{FIN}). 
For D3 and M5 branes, $F$ and ${\cal F}_m$ are related to the central charge of the lower and higher-dimensional SCFT, respectively,  and the gluing \eqref{glue} is nothing else that equivariant localization applied to the computation of the higher-dimensional  anomaly polynomial  that we discussed in section \ref{anomal:sec}. For other types of branes, ${\cal F}_m$ is the sphere-free energy of the higher-dimensional SCFT at large $N$ \cite{Hosseini:2019iad,Faedo:2021nub}, and \eqref{glue} is typically related to the large $N$ factorization properties of SCFT partition functions of theories  with a holographic dual \cite{Choi:2019dfu,Hosseini:2021mnn,Hosseini:2022vho}. In this context, the factorization becomes visible after taking the large $N$ limit of the SCFT free energy, which is usually expressed in terms of a matrix model. The partition function of three-dimensional ${\cal N}=2$  SQFTs compactified on a spindle times a circle was derived
in \cite{Inglese:2023wky}, generalizing at once the superconformal index and the topologically twisted index. The large $N$ limit of this spindle index is currently under investigation \cite{workinp} and we expect to reproduce the factorization \eqref{glue}, thus closing the circle.

The derivation of partition functions of SQFTs compactified on orbifolds is another arena where the results of this paper could be useful. Indeed, quantum field theory localization relies on the use of the equivariant index theorem \cite{Pestun:2016jze} and the building blocks of this construction will involve generalizations of the index-character discussed in section \ref{sec:eqindex} and appendix \ref{app:character}. More precisely, the index-character enters as a basic building block for partition functions on $\Morb\times S^1$ where supersymmetry is preserved with a topological twist on $\Morb$, while it needs to be generalized in the case of general $\sigma^a$ as in \eqref{susysigma}. In particular, it would be interesting to generalize the five-dimensional indices of \cite{Hosseini:2018uzp,Crichigno:2018adf,Hosseini:2021mnn} to the case of a five-dimensional SCFT defined on $\Morb_4\times S^1$ and apply the result to reproduce the entropy function for the supergravity solutions associated with D4 branes compactified on a four-dimensional orbifold \cite{Faedo:2022rqx}, as well as recover the
anomaly polynomial results for M5 branes compactified on a four-dimensional orbifold discussed in section  \ref{sec:M5orb}.\footnote{This can be done by considering the five-dimensional ${\cal N}=2$ SYM theory that decompactifies to the $(2,0)$ theory in the UV.}

\section*{Acknowledgements}

DM is supported in part by the INFN. AZ is partially supported by the INFN and the MIUR-PRIN contract 2017CC72MK003. We thank F. Faedo for comments on a draft of this paper. 
We gratefully acknowledge support from the Simons Center for Geometry and Physics, Stony Brook University, at which some of the research for this paper was performed.

\appendix

\section{Direct proof of the fixed point formula for $\Morb_4$}  
\label{fixptform4d_appendix}

In this section we derive the fixed-point formula  \eqref{eqvolume4d} for four-dimensional toric orbifolds. We use the notations introduced in section \ref{sec:2d}.
We start by writing the equivariant volume as an integral over the polytope, as in \eqref{polyvol}
\begin{equation}
\label{intermediatemaster}
\evol (\lambda_a,\epsilon_i) =  \frac{1}{(2\pi)^2} \int_{\Morb_4} \mathrm{e}^{-H} \frac{\omega^2}{2} =  \int_{\mathcal{P}} \mathrm{e}^{-y_i\epsilon_i} \dd y_1 \dd y_2\, , 
\end{equation}
which can be evaluated by elementary methods.
Defining the one-form
\begin{equation}
\nu \equiv  \frac{ \mathrm{e}^{-y_i\epsilon_i}}{\epsilon\cdot \epsilon}   (\epsilon_2 \dd y_1 - \epsilon_1 \dd y_2)
\end{equation}
and,  using Stoke's theorem, we have
\be
\evol (\lambda_a,\epsilon_i)  =\int_{\mathcal{P}} \dd \nu = \int_{\partial \mathcal{P}} \nu =  \sum_a \int_{{\cal F}_a} \nu\, , 
\ee
where a facet ${\cal F}_a$ of the polytope  is defined by the linear equation
 \be {\cal F}_a = \{l_a\equiv v_i^a y_i -\lambda_a = 0 \}\, . \ee
We can introduce a coordinate\footnote{With a slight abuse of notation we do not denote this with an index $s_a$.} $s\in [s_a^\mathrm{min},s_a^\mathrm{max}]$ on each facet ${\cal F}_a$, writing 
\begin{equation}
\label{parayi}
y_i |_{{\cal F}_a} = \tilde v_i^a s + \frac{\lambda_a}  {v^a\cdot v^a }v_i^a\, , 
 \end{equation}
where $\tilde v^a_i\equiv  \varepsilon_{ij} v^a_j$. The extrema of the interval can be determined by intersecting ${\cal F}_a$ with ${\cal F}_{a-1}$ and
${\cal F}_{a+1}$, respectively, and read 
 \begin{align}
 \label{sminmax}
 s_a^\mathrm{min}& = \frac{1}{\langle v^{a-1},v^a \rangle }\left( \lambda_{a-1} - \frac{v^{a-1}\cdot v^a}{v^{a}\cdot v^a} \lambda_a\right) \, , \nonumber\\
  s_a^\mathrm{max} & = \frac{1}{\langle v^{a+1},v^a \rangle }\left( \lambda_{a+1} - \frac{v^{a+1}\cdot v^a}{v^{a}\cdot v^a} \lambda_a\right)\, , 
  \end{align} 
  where $ \langle v, w\rangle \equiv \mathrm{det}(v, w)$.
Plugging (\ref{parayi}) into $\nu$ and integrating we have
 \begin{align}
 \int_{{\cal F}_a} \nu &=  \frac{v^a \cdot  \epsilon}{\epsilon\cdot \epsilon\,  \langle v^a, \epsilon\rangle}   \, \mathrm{exp}\left[-\frac{ \epsilon \cdot v^a } {v^a\cdot v^a}\lambda_a\right]
\left( \mathrm{exp} \left[  \langle v^a, \epsilon\rangle s_a^\mathrm{max}\right] - \mathrm{exp} \left[  \langle v^a, \epsilon\rangle s_a^\mathrm{min}\right]\right) \, .
  \end{align} 
Recalling that 
  \begin{equation} 
  \epsilon^{a}_1 = -\frac{\langle v^{a+1}, \epsilon\rangle}{\langle v^a , v^{a+1} \rangle} \,,  \qquad
	\epsilon^{a}_2 = \frac{\langle v^a, \epsilon\rangle}{\langle v^a , v^{a+1} \rangle}\, ,
\end{equation}
and  using the vector identities
 \begin{align}
& \langle  v^{a+1}, v^{a} \rangle \,\epsilon \cdot v^a + \langle  v^a , \epsilon \rangle  \, v^a \cdot v^{a+1}   = -\langle  \epsilon, v^{a+1} \rangle \, v^a  \cdot  v^a  \, , \nonumber \\
 &
\langle  v^{a-1}, v^{a} \rangle \, \epsilon \cdot  v^a + \langle  v^a , \epsilon \rangle \,  v^{a-1} \cdot v^{a}   = - \langle  \epsilon , v^{a-1} \rangle\,  v^a  \cdot  v^a  \,, 
 \end{align} 
 we obtain the compact expression
  \begin{align}
 \int_{{\cal F}_a} \nu &=  \frac{v^a \cdot  \epsilon}{\epsilon\cdot \epsilon\,  \langle v^a, \epsilon\rangle}   \left( \mathrm{e}^{-\left(  \epsilon_1^a  \lambda_a + \epsilon_2^a \lambda_{a+1}\right)}-
 \mathrm{e}^{-\left(  \epsilon_1^{a-1}  \lambda_{a-1} + \epsilon_2^{a-1} \lambda_{a}\right)}\right) \, . 
 \end{align}

Now, for a four-dimensional orbifold
\begin{equation}
 c_1^{\mathbb{T}}(L_a)|_{y_b}  = -\left (\delta_{b,a} \epsilon_1^b +  \delta_{b,a-1}\epsilon_2^{b} \right )\, , 
 \end{equation}
 so that
 \be   \int_{{\cal F}_a} \nu =  \frac{ v^a \cdot \epsilon}{\epsilon\cdot\epsilon \langle v^a, \epsilon\rangle}\left (\ex^{\sum_b \lambda_{b} c^{\mathbb{T}}_1(L_b) |_{y_a}}-\ex^{\sum_b \lambda_{b} c^{\mathbb{T}}_1(L_b) |_{y_{a-1}}}\right   ) \, . \ee
Rearranging  the contributions of the vertices in the sum 
\begin{align}
\evol (\lambda_a,\epsilon_i) & =  \sum_a \int_{{\cal F}_a} \nu = 
\sum_a \left(\frac{ v^a \cdot \epsilon}{\epsilon\cdot\epsilon \langle v^a, \epsilon\rangle}-\frac{ v^{a+1} \cdot \epsilon}{\epsilon\cdot\epsilon \langle v^{a+1}, \epsilon\rangle}\right ) \ex^{\sum_b \lambda_{b} c_1(L_b)  |_{y_a}} \nonumber \\
&=\sum_a \frac{1}{d_{a,a+1} \epsilon_1^{a}\epsilon_2^{a} }\ex^{\sum_b \lambda_{b} c_1(L_b)  |_{a}}\, , \end{align}
where we used another other standard vector identity
\be \langle  v^{a+1}, \epsilon \rangle \,\epsilon \cdot v^a - \langle  v^a , \epsilon \rangle  \, v^{a+1} \cdot \epsilon   = -\langle  v^a, v^{a+1} \rangle \, \epsilon  \cdot  \epsilon  \, ,\ee
we obtain the localization formula \eqref{eqvolume4d}. 

 \section{Direct proof of the master volume formula for CY$_3$}  \label{masterform4d_appendix}

We consider a singular Calabi-Yau cone with fan specified by the vectors $v^a=(1,w^a)$, with $a=1,\ldots , \fan^\prime$ and $w^a\in \mathbb{Z}^2$. The projection on the plane with first coordinate equal to one is a convex polytope called toric diagram in the physics literature. We label the vectors $v^a$ along the toric diagram in anticlockwise order and we identify vectors cyclically, $v^{\fan^\prime +1}\equiv v^1$. In order to apply the fixed point formula we need to resolve the singularity. This can be done by triangulating the toric diagram. We add new vectors $v^\alpha=(1, \bar w^\alpha)$, $\alpha=\fan^\prime +1,\ldots ,\fan$ with $\bar w^\alpha$ lying inside the toric diagram until it becomes the union of triangles. 

\begin{figure}[hhh!]
 \begin{center}
\begin{tikzpicture}[font = \footnotesize]
\draw (-4,0)--(-2,1.5)  node[above] {$v^{a+1}$} --(0,2) node[above] {$v^{a}$}--(2,1) node[above right] {$v^{a-1}$}--(4,-1);
\draw (-1,-1) node[below] {$\bar v$}--(0,2);
\draw (-1,-1) node[below] {$\bar v$}--(-2,1.5);
\draw (-1,-1) node[below] {$\bar v$}--(2,1);
\filldraw (0,2) circle (2pt);
\filldraw (-2,1.5)  circle (2pt);
\filldraw (2,1)  circle (2pt);
\filldraw (-1,-1) circle (2pt);
\node[draw=none] at (0.2, 0.8) {I}; 
\node[draw=none] at (-1, 0.8) {II}; 
\end{tikzpicture}
\caption{The triangles in the toric diagram contributing a $\lambda_a$ dependence in the fixed point formula for $\evol$.}\label{fig:CY3}
\end{center}
\end{figure}
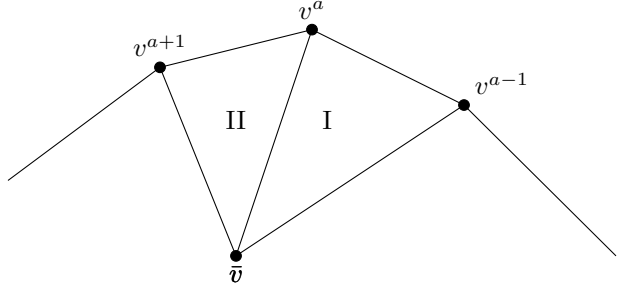

The master volume $\mathcal{V}$ is a function of the $\lambda^a$ associated with the external vectors $v^a=(1,w^a)$ only. The equivariant volume $\evol$ is a function of all $\lambda$ and can be obtained from the fixed point formula \eqref{fpm}, where the sum is extended to all triangles in which the toric diagram has been partitioned. The computation is fortunately local.  Every particular external $\lambda_a$ enters in just two fixed point contributions, associated with the two triangles in figure \ref{fig:CY3}, where $\bar v$ is an internal point.
Denoting $ \langle v, w, u \rangle \equiv \mathrm{det}(v, w, u)$, and taking into account the orientations of the inward normals $u^a_A$, we can write
\bea
\evol = \frac{d_I^2 \ex^{-(\bar \lambda \langle v^{a-1}, v^a, \epsilon \rangle +\lambda_{a-1}\langle v^{a}, \bar v, \epsilon \rangle +\lambda_a \langle \bar v, v^{a-1}, \epsilon \rangle)/d_I}}{\langle v^{a-1}, v^a, \epsilon \rangle \langle v^{a}, \bar v, \epsilon \rangle \langle \bar v, v^{a-1}, \epsilon \rangle} + \frac{d_{II}^2\ex^{-(\lambda_a \langle v^{a+1}, \bar v, \epsilon \rangle +\bar \lambda \langle v^{a}, v^{a+1}, \epsilon \rangle +\lambda_{a+1}\langle \bar v, v^{a}, \epsilon \rangle)/d_{II}}}{\langle v^{a+1}, \bar v, \epsilon \rangle \langle v^{a}, v^{a+1}, \epsilon \rangle \langle \bar v, v^{a}, \epsilon \rangle} +\ldots
\eea
where $\bar \lambda$ is the K\"alher parameter associated with the internal point $\bar v$, $d_I=| \langle v^{a-1}, v^a, \bar v\rangle |$  and $d_{II}=| \langle v^{a+1}, v^a, \bar v\rangle |$ are the orders of the local singularities and the dots refer to terms independent of $\lambda_a$. We can then compute
\bea
\frac{\partial^2\evol}{\partial \lambda_a^2 } \Big |_{\lambda=0} &=\frac{\langle \bar v, v^{a-1},\epsilon\rangle }{\langle v^{a-1}, v^a, \epsilon\rangle \langle v^a, \bar v, \epsilon \rangle}+  
\frac{\langle  v^{a+1}, \bar v ,\epsilon\rangle }{\langle v^{a}, v^{a+1}, \epsilon\rangle \langle \bar v, v^a,  \epsilon \rangle} = -\frac{\langle  v^{a-1}, v^{a+1} ,\epsilon\rangle }{\langle v^{a-1}, v^{a}, \epsilon\rangle \langle v^a, v^{a+1},  \epsilon \rangle}\\
\frac{\partial^2 \evol}{\partial \lambda_a \partial \lambda_{a-1}} \Big |_{\lambda=0} &=\frac{1}{\langle v^{a-1}, v^a, \epsilon\rangle} \\
\frac{\partial^2\evol}{\partial \lambda_a \partial \lambda_{a+1}} \Big |_{\lambda=0} &=\frac{1}{\langle v^{a}, v^{a+1}, \epsilon\rangle} \, ,
\eea
thus recovering the master volume expression \eqref{masvol}.

\section{The equivariant index }\label{app:character}

In this appendix we present some simple examples of character indices. It is an interesting topic, with many applications for example to localization computations \cite{Inglese:2023wky}, but, since it is not  the main theme of this paper we will be brief.

\subsection{The equivariant index for the spindle}  
We consider the spindle $\spindle=\mathbb{WP}^1_{[n_+,n_-]}$. As in section \ref{sec:spindle} we take $v^1=n_+$, $v^2=-n_-$  and GLSM charges $Q=(n_-,n_+)$. We take $n_+$ and $n_-$ relatively prime. 

The character computes the equivariant index
\be \mathds{Z}(q, \Lambda_a) = \sum_{p=0}^1 (-1)^p {\rm Tr} \{ q | H^{(0,p)}(\spindle, O(\Lambda_+ D_++\Lambda_- D_-))\} \, ,\ee
where $q$ is the weight for the $\mathbb{T}$ action.
We recall that not all the $\mathbb{T}$-invariant divisors are inequivalent. In particular, for the spindle, \eqref{diveq} implies $n_+ D_+=n_- D_-$.

The fixed point formula 
\eqref{aa1} has contribution from two poles of the spindle. In each contribution we need to add an average over the local singularity $\ZZ_{d_A}$
\be\label{C2}  \mathds{Z}(\Lambda_a,q) = \frac{1}{n_+} \sum_{k=0}^{n_+-1} \frac{ \ex^{-2\pi i k \Lambda_+/n_+} q^{-\Lambda_+/n_+}}{1- \ex^{2\pi i k/n_+} q^{1/n_+}} + \frac{1}{n_-} \sum_{k=0}^{n_--1} \frac{ \ex^{-2\pi i k \Lambda_-/n_-} q^{\Lambda_-/n_-}}{1- \ex^{2\pi i k/n_-} q^{-1/n_-}} \, .\ee
Setting $q=\ex^{-\epsilon\hbar}$ and $\Lambda_\pm=-\lambda_{1,2}/\hbar$ and taking the limit $\hbar\rightarrow 0$ we recover the equivariant volume \eqref{equivspindle} as the coefficient of the leading pole as in \eqref{Klimit}.
Only the terms with $k=0$ contribute to the limit. 
The index can been resummed  using \cite{Inglese:2023wky}
\be\label{idre1}
\frac{1}{k}\sum_{\ell=0}^{k-1}\frac{\omega_k^{-\alpha \ell}}{1-\omega_k^\ell u}  = \frac{ u^{\alpha - k\lfloor \tfrac{\alpha}{k} \rfloor} } {1-u^k}\, , 
\ee
where $\omega_k=\mathrm{e}^{\frac{2\pi i}{k}}$,  $u$ is a complex number, $\alpha$ is an \emph{integer} number  and the floor function denotes the integer part.
This formula can be proved by expanding both sides in power series of $u$.
The result is
\begin{align}
\label{floorycharacter}
 \mathds{Z}(\Lambda_a,q)  =  \frac{q^{-\left\lfloor \frac{\Lambda_+}{n_+} \right\rfloor }}{1-q} +  \frac{q^{\left\lfloor \frac{\Lambda_-}{n_-} \right\rfloor }}{1-q^{-1}} = \frac{q^{-\left\lfloor \frac{\Lambda_+}{n_+} \right\rfloor }-
 q^{1+\left\lfloor \frac{\Lambda_-}{n_-} \right\rfloor }}{1-q}\, . 
\end{align}
Notice that, although there were fractional powers of $q$ in \eqref{C2}, the final formula contains only integer powers. We see  that,  for $\Lambda_\pm >0$, 
\be
\label{floorycharacter2}
 \mathds{Z}(\Lambda_a,q)  =q^{-\left\lfloor \frac{\Lambda_+}{n_+} \right\rfloor} +\ldots + q^{\left\lfloor \frac{\Lambda_-}{n_-}\right\rfloor} \, , 
\ee
the character has only positive signs and  counts the sections of 
\be H^0(\mathbb{WP}^1_{[n_+,n_-]}, O(\Lambda_+ D_++\Lambda_- D_-)) \, ,\ee
refined with respect to the $\mathbb{T}$ action. Geometrically, we see that the exponents are the integers 
\be -\frac{\Lambda_+}{n_+}\le m\le  \frac{\Lambda_-}{n_-}\, , \qquad m \in \mathbb{Z} \, ,\ee
corresponding to the integer points in the polytope \eqref{pol}
\be \Delta(\Lambda_a)=\{m \in \mathbb{Z} |  v^1 m\ge -\Lambda_+\, , v^2 m\ge -\Lambda_-\} \, .\ee

The Molien-Weyl formula \eqref{a2} reads
\be   \mathds{Z}_{MW}(T, \bar q_a)   = -\int \frac{ \dd z}{2\pi i z^{1+T}}\frac{1}{(1- z^{n_-} \bar q_1) (1- z^{n_+} \bar q_2)} \, .\ee
The relation between the two formulas is as in \eqref{ccc}
 \be\label{ccc0} \mathds{Z}_{MW}\left ( T= n_+ \Lambda_- +n_- \Lambda_+, \bar q_a\right ) = \prod_{a=\pm} \bar q_a^{\Lambda_a} \mathds{Z}\left (\Lambda_\pm , q = \bar q_+^{n_+} \bar q_-^{-n_-} \right) \, .\ee 
This can be checked  by computing explicitly the residue  related to $\bar q_i$. One obtain two sums over $z_{k_-}= \bar q_1^{-1/n_-} \ex^{2 \pi i k_-/n_-}$ and $z_{k_+}= \bar q_2^{-1/n_+} \ex^{2 \pi i k_+/n_+}$ for $k_\pm=0,\ldots n_\pm -1$  that match  the fixed point formula.\footnote{We assume that $n_+$ and $n_+$ are relatively prime.} 
On the other hand, we can also evaluate the integral in a simpler way. We define the grand-canonical partition function 
\be   \sum_{T=0}^\infty \mathds{Z}_{MW}(T, \bar q_a)  w^T = -\int \frac{ \dd z}{2\pi i }\frac{1}{(1- z^{n_-} \bar q_1) (1- z^{n_+} \bar q_2)(z-w)}  = \frac{1}{(1- w^{n_-} \bar q_1) (1- w^{n_+} \bar q_2)}\, ,\ee
where we evaluated the integral by deforming the contour circling $z_{k_\pm}$ into a contour around $z=w$, which is the only other singularity of the integrand.
From this expression we see that $\mathds{Z}_{MW}(T, \bar q_a)$ for $T>0$ counts the monomials in $(\bar q_1,\bar q_2)$ of charge $T$ under the rescaling with weights $Q=(n_-,n_+)$. These are precisely the holomorphic sections 
\be H^0(\mathbb{WP}^1_{[n_+,n_-]}, O(T)) \, , \ee
where the line bundle $O(T)$ corresponds to Chern class $c_1=\left [ \frac{F}{2\pi}\right ]$ with Chern number $\frac{1}{2\pi}\int F = \frac{T}{n_+ n_-}$. 

\subsection{The equivariant index for $\mathbb{WP}^2_{[N_1,N_2,N_3]}$}  For simplicity, we only consider the case of the ``non-minimal" fan 
\be v^1=(n_3,n_3)\, ,\qquad  v^2=(-n_1,0)\, , \qquad v^3=(0,-n_2) \, ,
\ee
 and  $Q= (N_1,N_2,N_3) =(n_1 n_2,n_2 n_3,n_1 n_3) $. The orbifold is  $\Morb_4=\mathbb{W}\mathbb{P}^2_{[N_1,N_2,N_3]}$ if the $n_a$ are coprime.  There are only labels and not extra orbifold singularities in the sense that the corresponding primitive vectors $\hat v^a$, obtained by dividing $v^a$ by the label $n_a$, satisfy $\hat d_{a,a+1}=\det (\hat v^a, \hat v^{a+1})=1$.

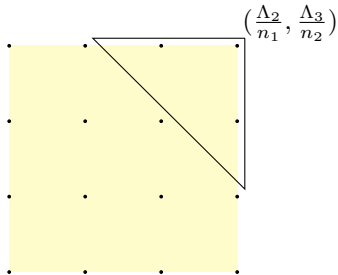
\begin{figure}[h!!!!!!]\begin{center}
\begin{tikzpicture}[font = \footnotesize, rotate=180]
\fill [yellow, opacity=0.2] (0,0)--(3,0)--(3,3)--(0,3)--(0,0);
\node[draw=none] at (-0.7,-0.3) {$(\frac{\Lambda_2}{n_1},\frac{ \Lambda_{3}}{n_2})$}; 
\draw[-] (-0.1,-0.1)--(1.9,-0.1) --(-0.1,1.9)--(-0.1,-0.1) ;
\filldraw (0,0) circle (0.5pt) ;
\filldraw (1,0) circle (0.5pt) ;
\filldraw (2,0) circle (0.5pt) ;
\filldraw (3,0) circle (0.5pt) ;

\filldraw (0,1) circle (0.5pt) ;
\filldraw (1,1) circle (0.5pt) ;
\filldraw (2,1) circle (0.5pt) ;
\filldraw (3,1) circle (0.5pt) ;

\filldraw (0,2) circle (0.5pt) ;\filldraw (1,2) circle (0.5pt) ;\filldraw (2,2) circle (0.5pt) ;\filldraw (3,2) circle (0.5pt) ;
\filldraw (0,3) circle (0.5pt) ;\filldraw (1,3) circle (0.5pt) ;\filldraw (2,3) circle (0.5pt) ;\filldraw (3,3) circle (0.5pt) ;

\end{tikzpicture}
\end{center}
\caption{The integer points counted by \eqref{pol1}.}\label{fig:pol2}
\end{figure}

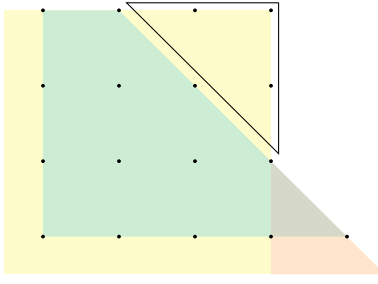
\begin{figure}[h!!!!!!]\begin{center}
\begin{tikzpicture}[font = \footnotesize, rotate=180]
\fill [yellow, opacity=0.2] (0,0)--(3.5,0)--(3.5,3.5)--(0,3.5)--(0,0);
\fill [cyan, opacity=0.2] (2,0)--(3,0)--(3,3)--(-1,3)--(2,0);
\fill [orange, opacity=0.2] (0,2)--(0,3.5)--(-1.5,3.5)--(0,2);
\draw[-] (-0.1,-0.1)--(1.9,-0.1) --(-0.1,1.9)--(-0.1,-0.1) ;
\filldraw (0,0) circle (0.5pt) ;
\filldraw (1,0) circle (0.5pt) ;
\filldraw (2,0) circle (0.5pt) ;
\filldraw (3,0) circle (0.5pt) ;

\filldraw (0,1) circle (0.5pt) ;
\filldraw (1,1) circle (0.5pt) ;
\filldraw (2,1) circle (0.5pt) ;
\filldraw (3,1) circle (0.5pt) ;

\filldraw (0,2) circle (0.5pt) ;\filldraw (1,2) circle (0.5pt) ;\filldraw (2,2) circle (0.5pt) ;\filldraw (3,2) circle (0.5pt) ;
\filldraw (0,3) circle (0.5pt) ;\filldraw (1,3) circle (0.5pt) ;\filldraw (2,3) circle (0.5pt) ;\filldraw (3,3) circle (0.5pt) ;
\filldraw (-1,3) circle (0.5pt) ;

\end{tikzpicture}
\end{center}
\caption{The power series contributing to  \eqref{VNCP123Molien:final}. The yellow and orange power series enter with a plus sign, while the green with a minus sign.
The result is just the set of integer points in the triangle. }\label{fig:pol3}
\end{figure}

The fixed point formula \eqref{aa1} requires averaging over the local orbifold singularity. For example, we can consider the contribution 
from the fixed point associated with the cone $(v^2,v^3)$. The local data are $d_{2,3}=n_1 n_2$ and $u_1=(-n_2,0)$ and $u^2=(0,-n_1)$. The local orbifold action is determined by $J_2 n_1=s, J_3 n_2 =p$ where $s,p$ are integers:
\be ( \ex^{2\pi i J_2}, \ex^{2\pi i J_3})=(\ex^{2\pi i \frac{s}{n_1}}, \ex^{2\pi i \frac{ p}{n_2}})\, .\ee
The fixed point contribution is
\bea\label{pol1} \evol &=\frac{1}{n_1 n_2} \sum_{s=0}^{n_1-1} \sum_{p=0}^{n_2-1} \frac{(\ex^{2\pi i \frac{s}{n_1}} q_1^{-1/n_1})^{-\Lambda_2} (\ex^{2\pi i \frac{p}{n_2}} q_2^{-1/n_2})^{-\Lambda_3}}{(1-\ex^{2\pi i \frac{s}{n_1}} q_1^{-1/n_1}) (1-\ex^{2\pi i \frac{p}{n_2}} q_2^{-1/n_2}) } \\
&= \frac{  q_1^{\lfloor \frac{\Lambda_2}{n_1}\rfloor} q_2^{\lfloor \frac{\Lambda_3}{n_2}\rfloor}}{(1- 1/q_1) (1- 1/q_2) } \, , \eea
where we use the identity \eqref{idre1}.
The contributions of the other fixed points can be computed similarly. Combining the three contributions we
find
\begin{align}
\label{VNCP123Molien:final}
\evol (\Lambda_a , q_i)  =\frac{q_1^{\left\lfloor \frac{\Lambda_2}{n_1}\right\rfloor} q_2^{\left\lfloor \frac{\Lambda_3}{n_2}\right\rfloor}}{(1-1/q_1)(1-1/q_2)}
+ \frac{\left(\frac{q_1}{q_2}\right)^{-\left\lfloor \frac{\Lambda_3}{n_2}\right\rfloor} q_1^{-\left\lfloor \frac{\Lambda_1}{n_3}\right\rfloor}}{(1-\frac{q_1}{q_2})(1-q_1)}
+  \frac{\left(\frac{q_2}{q_1}\right)^{-\left\lfloor \frac{\Lambda_2}{n_1}\right\rfloor} q_2^{-\left\lfloor \frac{\Lambda_1}{n_3}\right\rfloor}}{(1-\frac{q_2}{q_1})(1-q_2)}\, ,
\end{align}
in agreement with the general formula \eqref{aaa1}. One can check that 
\be \evol (\Lambda_a , q_i)  = \sum_{\underline{m}\in \Delta(\Lambda_a)} \underline{q}^{\underline{m}} \, \ee
counts (with a $\mathbb{T}^2$ weight) the integer points in 
the polytope
  \be \Delta(\Lambda_a) =\{ m\cdot v^a \ge - \Lambda_a \} \, ,\qquad m\in \mathbb{Z}^2 \, .\ee
This can be proved with elementary methods. By expanding in power series \eqref{pol1} in $1/q_1$ and $1/q_2$ we obtain all the integer points in an infinite cone, as in figure \ref{fig:pol2}.
By similarly expanding  the other power series in \eqref{VNCP123Molien:final} for $|q_i|>1$ and $|q_2/q_1|>1$, we find three competing contributions which cancel except for the points 
inside the polytope as depicted in figure \ref{fig:pol3}.

The Molien-Weyl formula  gives instead
\be \mathds{Z}_{MW} (T , \bar q_a) =  -   \int \frac{\dd z}{2\pi i z^{1+T}} \frac{1}{(1- z^{N_1} \bar q_1)(1- z^{N_2} \bar q_2)(1- z^{N_3} \bar q_3)} \, .\ee
This can be explicitly evaluated by taking the residues at $z^{N_a}\bar q_a=1$ and one recovers formula \eqref{ccc}:
 \be\label{ccc3} \mathds{Z}_{MW}\left ( T_A= \sum_a Q^A_a \Lambda_a, \bar q_a\right ) = \prod_{s=1}^d \bar q_a^{\Lambda_a} \mathds{Z}\left (\Lambda_a , q_i =\prod_a \bar q_a^{v^a_i} \right) \, ,\ee  
where the gauge invariant quantities are explicitly given by
\be q_1=\bar q_1^{n_3}\bar q_2^{-n_1}\, ,\qquad q_2=\bar q_1^{n_3}\bar q_3^{-n_2}\, ,\qquad T= n_1 n_2 \Lambda_1+n_2 n_3\Lambda_2+n_3 n_1 \Lambda_3\, .\ee
One can check that the three set of residues at $z^{N_a}\bar q_a=1$ precisely correspond to the three fixed point contributions.
For example, the residue associated with $\bar q_1$ reads
\bea  \frac{1}{n_1 n_2} \sum_{k=0}^{n_1 n_2-1} \frac{\bar q_1^{T/N_1} \omega_{N_1}^{-k T}}{(1- \omega_{N_1}^{k N_2}  \bar q_2 \bar q_1^{-N_2/N_1})(1- \omega_{N_1}^{k N_3}  \bar q_3 \bar q_1^{-N_3/N_1})} \\
= \frac{1}{n_1 n_2} \sum_{k=0}^{n_1 n_2-1}\frac{\bar q_1^{T/n_1 n_2} \omega_{n_1 n_2}^{-k T}}{(1- \omega_{n_1}^{k n_3}  \bar q_2 \bar q_1^{-n_3/n_1})(1- \omega_{n_2}^{k n_3}  \bar q_3 \bar q_1^{-n_3/n_2})}\\
= \frac{1}{n_1 n_2} \sum_{k=0}^{n_1 n_2-1}\frac{\bar q_1^{\Lambda_1+\frac{n_3 \Lambda_2}{n_1} +\frac{n_3 \Lambda_3}{n_2}} \omega_{n_1}^{-k n_3 \Lambda_2} \omega_{n_2}^{-k n_3 \Lambda_3} }{(1- \omega_{n_1}^{k n_3}  \bar q_2 \bar q_1^{-n_3/n_1})(1- \omega_{n_2}^{k n_3}  \bar q_3 \bar q_1^{-n_3/n_2})}\\
=\bar q_1^{\Lambda_1} \bar q_2^{\Lambda_2}\bar q_3^{\Lambda_3}\frac{(\bar q_2^{n_1} \bar q_1^{-n_3})^{-\lfloor \frac{\Lambda_2}{n_1}\rfloor} (\bar q_3^{n_2} \bar q_1^{-n_3})^{-\lfloor \frac{\Lambda_3}{n_2}\rfloor} }{(1-   \bar q_2^{n_1} \bar q_1^{-n_3})(1- \bar q_3^{n_2} \bar q_1^{-n_3})}\\
=\bar q_1^{\Lambda_1} \bar q_2^{\Lambda_2}\bar q_3^{\Lambda_3}\frac{q_1^{\lfloor \frac{\Lambda_2}{n_1}\rfloor} q_2^{\lfloor \frac{\Lambda_3}{n_2}\rfloor} }{(1-  1/q_1)(1- 1/q_2)}\, , 
\eea
reproducing the contribution \eqref{pol1}. In the sum  we replaced $k$ with $n_3 k$ using the fact that the $n_i$ are coprime. We also used the identity 
\be  \frac{1}{n_1 n_2} \sum_{k=0}^{n_1 n_2-1} \frac{ \omega_{n_1}^{-k \Theta_1}\omega_{n_2}^{-k \Theta_2}}{(1- \omega_{n_1}^{k} u_1)(1- \omega_{n_2}^{k} u_2)}=\frac{ u_1^{\Theta_1-n_1\lfloor \frac{\Theta_1}{n_1}\rfloor} u_2^{\Theta_2-n_2 \lfloor \frac{\Theta_2}{n_2}\rfloor}}{(1-u_1^{n_1}) (1-u_2^{n_2})}\, ,\ee
which can be proved by expanding both sides in power series of $u$. 

The  grand-canonical partition function 
\be   \sum_{T=0}^\infty \mathds{Z}_{MW}(T, \bar q_a)  w^T = -\int \frac{ \dd z}{2\pi i }\frac{1}{(1- z^{N_1} \bar q_1)(1- z^{N_2} \bar q_2)(1- z^{N_3} \bar q_3)(z-w)}  \ee
can be also computed deforming the contour to encircle $z=w$ which gives
\be  \sum_{T=0}^\infty \mathds{Z}_{MW}(T, \bar q_a)  w^T = \frac{1}{(1- w^{N_1} \bar q_1)(1- w^{N_2} \bar q_2)(1- w^{N_3} \bar q_3)}\, .\ee
We see that $\mathds{Z}_{MW}(T, \bar q_a)$ for $T>0$ counts the monomials in $(\bar q_1,\bar q_2, \bar q_3)$ of charge $T$ under the rescaling with weights $Q=(N_1,N_2,N_3)$. These are precisely the holomorphic sections 
\be H^0(\mathbb{WP}^2_{[N_1,N_2,N_3]}, O(T)) \, , \ee
of the line bundle $O(T)$. In the familiar case of $\mathbb{WP}^2$, where $N_i=1$, these are just the homogeneous polynomials of degree $T$.

In the presence of $\hat d_{a,a+1}=\det (\hat v^a, \hat v^{a+1})\neq 1$ we would need  more general formulas for resumming the averages over the orbifold action and we will not discuss  this  case.

\bibliographystyle{JHEP}
\bibliography{biblio}

\end{document}